\documentclass[twocolumn,showpacs,aps,prd,Here]{revtex4}

\usepackage{graphicx}
\usepackage{dcolumn}
\usepackage{amsmath}
\usepackage{epsfig}
\usepackage{hhline}
\usepackage[usenames,dvipsnames]{color}

\input babarsym

\newcommand{\BaBarType}     {PUB}
\newcommand{\BaBarYear}     {11}
\newcommand{\BaBarNumber}     {013}
\newcommand{\SLACPubNumber}     {14521}
\begin{document}

\begin{flushleft}
{\babar-\BaBarType-\BaBarYear/\BaBarNumber} \\
{SLAC-PUB-\SLACPubNumber} \\
\end{flushleft}

\title
{
Branching Fraction Measurements of the Color-Suppressed Decays \\
{${\Bzb}$} to ${\Dstze \piz}$, ${\Dstze \eta}$,
${\Dstze \omega}$, and ${\Dstze \eta^{\prime}}$ \\
and Measurement of the Polarization in the Decay ${\Bzb}\ra {\Dstarz}\omega$}


%
\author{J.~P.~Lees}
\author{V.~Poireau}
\author{V.~Tisserand}
\affiliation{Laboratoire d'Annecy-le-Vieux de Physique des Particules (LAPP), Universit\'e de Savoie, CNRS/IN2P3,  F-74941 Annecy-Le-Vieux, France}
\author{J.~Garra~Tico}
\author{E.~Grauges}
\affiliation{Universitat de Barcelona, Facultat de Fisica, Departament ECM, E-08028 Barcelona, Spain }
\author{M.~Martinelli$^{ab}$}
\author{D.~A.~Milanes$^{a}$}
\author{A.~Palano$^{ab}$ }
\author{M.~Pappagallo$^{ab}$ }
\affiliation{INFN Sezione di Bari$^{a}$; Dipartimento di Fisica, Universit\`a di Bari$^{b}$, I-70126 Bari, Italy }
\author{G.~Eigen}
\author{B.~Stugu}
\affiliation{University of Bergen, Institute of Physics, N-5007 Bergen, Norway }
% ...
\author{(The \babar\ Collaboration)}
\noaffiliation

\begin{abstract}
{We report updated branching fraction
measurements of the color-suppressed decays $\Bzb\ra\Dz\piz$,
$\Dstarz\piz$, $\Dz\eta$, $\Dstarz\eta$, $\Dz\omega$, $\Dstarz\omega$,
$\Dz\etapr$, and $\Dstarz\etapr$. We measure the branching fractions ($\times 10^{-4}$):
$\BF(\Bzb\ra\Dz\piz)= 2.69\pm 0.09  \pm 0.13 $,
$\BF(\Bzb\ra\Dstarz\piz)=3.05\pm 0.14  \pm 0.28$,
$\BF(\Bzb\ra\Dz\eta)=2.53\pm 0.09 \pm 0.11 $,
$\BF(\Bzb\ra\Dstarz\eta)=2.69\pm 0.14  \pm 0.23$,
$\BF(\Bzb\ra\Dz\omega)=2.57\pm 0.11 \pm 0.14 $,
$\BF(\Bzb\ra\Dstarz\omega)=4.55\pm 0.24  \pm 0.39$,
$\BF(\Bzb\ra\Dz\etapr)=1.48\pm 0.13 \pm 0.07 $,
 and $\BF(\Bzb\ra\Dstarz\etapr)=1.49\pm 0.22  \pm 0.15 $.
We also present the first measurement of the longitudinal polarization fraction of the
decay channel $\Dstarz\omega$, $f_L$=$(66.5\pm 4.7\pm 1.5)\%$. In
the above, the first uncertainty is statistical and the second is systematic.
The results are based on a sample of $(454\pm 5)\times 10^{6}$ $\BB$
pairs collected at the $\Upsilon(4S)$ resonance,
with the $\babar$ detector at the \pep2 storage rings at SLAC.
The measurements are the most precise determinations of these
quantities from a single experiment. They are compared
to theoretical predictions obtained by factorization, Soft Collinear
Effective Theory (SCET) and perturbative QCD (pQCD). We find that the
presence of final state interactions is favored  and the
measurements are in better agreement with SCET
than with pQCD.}

\end{abstract}


\maketitle

\section{INTRODUCTION}
\label{sec:Introduction}
Weak decays of hadrons provide direct access to the parameters
of the Cabibbo-Kobayashi-Maskawa (CKM) matrix and thus to the study
of \CP violation. Strong  interaction scattering in the final
state~\cite{ref:ChangChuaSoni} (Final State Interactions, or FSI)
can modify the decay dynamics and must be well understood.
The two-body hadronic $B$ decays with a  charmed final state,
$B\ra \Dst h$,  where $h$ is a light meson, are of great help in studying
strong-interaction physics related to the confinement
of quarks and gluons in hadrons.

\begin{figure}[h]
\begin{center}
\includegraphics[width=0.850\linewidth]{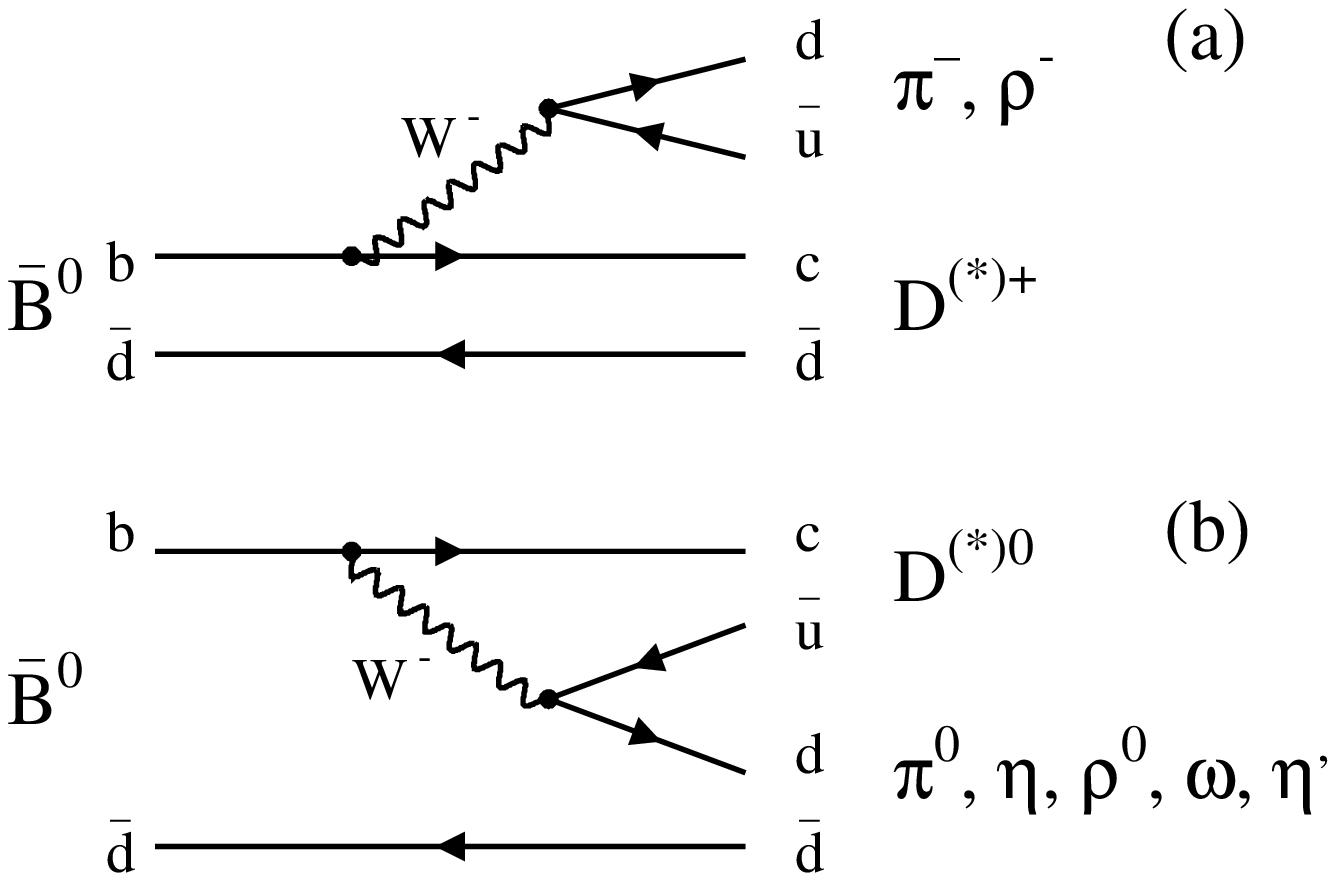}
 \caption{External (a) and internal (b) tree diagrams for $\Bzb\ra \Dst h$ decays.}
 \label{fig:feynman}
\end{center}
\end{figure}

The decays $B\ra \Dst h$ can proceed through the emission of a $W^{\pm}$
boson following three possible diagrams: external, internal (see
Fig.~\ref{fig:feynman}), or by a $W^{\pm}$ boson exchange
whose contribution to the decay rate  is expected to be much smaller than the external
and internal amplitudes~\cite{ref:exchangeW}. The neutral  $\Bzb\ra\Dstze\hz$ decays proceed
through the internal diagrams~\cite{ref:Neubert}. Since mesons are color singlet objects,
 the quarks from the $W^{\pm}$
decay are constrained to have the anti-color of the spectator quark,
which induces a suppression of internal diagrams.
For this reason, internal diagrams are called {\it color-suppressed} and external ones
are called {\it color-allowed}.

We already discussed factorization models~\cite{ref:Neubert,ref:BauerStechWirbel,ref:NeubertPetrov,ref:Deandrea_0}
in our previous publication~\cite{ref:Babar2004}. Within that approach
the non-factorizable interactions in the final state by soft
gluons are neglected. The matrix element in the effective weak
Hamiltonian of the decay $B\ra \Dst h$ is then factorized into a
product of asymptotic states. Factorization appears to be
successful in the description of the color-allowed
decays~\cite{ref:HSW}.

The color-suppressed $b \to c$ decays $\Bzb\ra\Dstze\piz$ were first observed by 
Belle~\cite{ref:Belle2002}  and CLEO~\cite{ref:Cleo2002}  with
 $23.1\times 10^6$ and $9.67\times 10^6$~$\BB$  pairs  respectively.
Belle has also observed the decays $\Dz\eta$ and $\Dz\omega$ and put upper limits on the branching fraction ($\BF$)
of $\Dstarz\eta$ and $\Dstarz\omega$~\cite{ref:Belle2002}. The branching fraction of the color-suppressed decays
$\Bzb\ra\Dstze\piz$, $\Dstze\eta$, $\Dstze\omega$, and $\Dz\etapr$
were measured by $\babar$~\cite{ref:Babar2004} with
88$\times 10^6~\BB$ pairs and an upper limit was set on
$\BF (\Bzb \ra \Dstarz\etapr)$. Belle  updated with 152$\times
10^6~\BB$ pairs the measurement of $\BF(\Bzb\ra\Dstze\hz)$, $\hz=\piz$, $\eta$,
$\omega$ \cite{ref:Belle2005}, and $\etapr$ \cite{ref:Belle2006} and
studied the decays $\Bzb\ra\Dz\rho^0$ with 388$\times
10^6~\BB$ pairs~\cite{ref:Belle2007}.
In an alternative approach,  $\babar$~\cite{ref:BabarKpipiz}
used the charmless neutral  $B$ to $K^\pm \pi^\mp\piz$
Dalitz plot analysis with 232$\times 10^6~\BB$ pairs, and found  $\BF(\Bzb\ra\Dz\piz)$
to be in excellent agreement with earlier experimental results.  $\babar$
has also performed a preliminary
Dalitz-plot analysis of the mode $\Bz  \ra  \Dzb \pi^+ \pi^-$ with 471$\times
10^6~\BB$ pairs~\cite{ref:BaBar2010D0pipi}.

Many of these branching fraction measurements are significantly larger than predictions
obtained within the factorization approximation~\cite{ref:Chua,ref:Neubert}.
But, while the initial various experimental results demonstrated overall good consistency,
the most recent measurements published by Belle \cite{ref:Belle2005,ref:Belle2006}  have moved
on average towards lower  $\BF$  for the color-suppressed $\Bzb\ra\Dstze\hz$ decays,
closer to factorization predictions. However, it has been demonstrated~\cite{ref:LegangerEeg} that
non-factorizable contributions are mostly dominant for the color-suppressed charmed $\Bzb\ra\Dz\piz$
decay and therefore cannot be neglected.

Stronger experimental constraints are therefore needed to distinguish between
the different models of the color-suppressed dynamics like
pQCD ({\it perturbative QCD})~\cite{ref:pQCD_1,ref:pQCD_2} or
SCET ({\it Soft Collinear Effective Theory})~\cite{ref:SCET_1,ref:SCET_2,ref:SCET_3}.
Finally, we emphasize the need for accurate measurements of
hadronic color-suppressed $\Bzb\ra\Dstze\hz$ decays to constrain the theoretical predictions
on $\bar{B}_{u,d,s}$ decays to $D^{(*)}P$ and $\bar{D}^{(*)}P$ states, where $P$ is  a light pseudoscalar meson
such as  a pion or a kaon~\cite{ref:ChuaHou}. Using flavor $SU(3)$ symmetry,
the comparison of $B_d$ and $B_s$ decays offers new possibilities to determine the decay constant ratio
$f_s/f_d$ \cite{ref:ChuaHou}. These decays are and will be employed to extract the CKM-angle
$\gamma$ and other angles~\cite{ref:GronauFleischer},
  especially in the context of the $B$-physics program at the LHC.

 This paper reports improved branching fraction measurements of eight
color-suppressed decays $\Bzb\ra\Dstze\piz$, $\Dstze\eta$, $\Dstze\omega$, and $\Dstze\etapr$
 with 454$\times 10^6~\BB$ pairs and presents for the first time the measurement
of the longitudinal polarization for the decay mode to two vector mesons
$\Bzb\ra\Dstarz\omega$, which also constrains  QCD models and
challenges  {\it Heavy Quark Effective Theory}  (HQET) (see Sec.~\ref{se:fL}).

\section{THE \babar\ DETECTOR and DATA SAMPLE}
\label{sec:babar}

The data used in this analysis were collected with the $\babar$
detector at the \pep2 asymmetric-energy $\epem$ storage rings operating
at SLAC. The $\babar$ detector is described in detail in
Ref.~\cite{ref:detector}. Charged particle tracks are
reconstructed using a five-layer silicon vertex tracker (SVT) and
a 40-layer drift chamber (DCH) immersed in a 1.5~T magnetic field.
Tracks are identified as pions or kaons (particle identification
or PID) based on likelihoods constructed from energy loss
measurements in the SVT and the DCH and from Cherenkov radiation
angles measured in the detector of internally reflected Cherenkov light (DIRC).
Photons are reconstructed from showers measured in the CsI(Tl)
crystal electromagnetic calorimeter (EMC). Muon and neutral hadron
identification is performed with the instrumented flux return
(IFR).

The results presented are based on a data sample  of an integrated luminosity of
413~$\invfb$ recorded from 1999 to 2007 at the $\Upsilon(4S)$ resonance with a
$e^+e^-$ center-of-mass (CM) energy of 10.58~$\gev$, corresponding to
$(454\pm 5)\times 10^6~\BB$ pairs. The equal production rate of $\Bz\Bzb$ and $\Bp\Bm$ at
that resonance is assumed in this paper, as suggested by the Particle Data Group (PDG) \cite{ref:PDG}.
A data sample of 41.2~$\invfb$ with a CM energy of 10.54~$\gev$, below the $\BB$ threshold, is used
to study background contributions from continuum events
$\epem\ra\qqbar$ ($q=u$, $d$, $s$, $c$). We call that latter dataset
{\it off-peak events} in what follows.

Samples of simulated Monte Carlo (MC) events are used to
determine signal and background characteristics, to optimize
selection criteria and to evaluate efficiencies. Simulated events
$\epem\ra\Upsilon(4S)\ra\BpBm,~\Bz\Bzb$, $\epem\ra\qqbar$ ($q=u$,
$d$, $s$) and $\epem\ra\ccbar$ are generated with {\tt EvtGen}~\cite{ref:evtgen},
which interfaces to {\tt Pythia}~\cite{ref:pythia} and {\tt Jetset}~\cite{ref:jetset}.
Separate samples of exclusive $\Bzb\ra\Dstze\hz$ decays are
generated to study the signal features and to quantify the signal selection efficiencies.
We also use high statistics control samples of exclusive decays $\Bm\ra D^{(*)0}\pi^-$ and
 $\Dstze\rho^-$ for specific selection and background studies. We study these control
 samples both in data and in the MC, using the same selection criteria.
All MC samples include simulation of the $\babar$ detector response generated
through {\tt Geant4}~\cite{ref:GEANT4}. The equivalent integrated luminosity
of the MC samples is about three times that of the data  for
$\BB$, one time for $\epem\ra\qqbar$ ($q=u$, $d$, $s$) and twice
for $\epem\ra\ccbar$ respectively. The equivalent integrated luminosities of the exclusive $B$ decay mode
simulations range from 50 to 2500  times the dataset. 
\section{ANALYSIS METHOD}
\label{sec:Analysis}

\subsection{General considerations}

The color-suppressed $\Bzb$ meson decay modes are
reconstructed from $\Dstze$ meson candidates that
are combined with light neutral-meson candidates $\hz$
($\piz$, $\eta$, $\omega$, and $\etapr$). The $D^{(*)0}$ and $\hz$ mesons
are detected in various possible decay channels. In total,
we consider 72 different $\Bzb\ra\Dstze\hz$ decay modes.

We perform a blind analysis:  the optimization of the various event selections,
the background characterizations and rejections, the efficiency calculations,
and most of the systematic uncertainty computations
are based on studies done with MC simulations, data sidebands,
or data control samples. The fits to data, including the various signal regions,
are only performed after all analysis procedures are fixed
and systematic uncertainties are studied.

Intermediate particles of the decays $\Bzb\ra\Dstze\hz$ are
reconstructed by combining tracks and$/$or photons for the
decay channels with the highest decay rate and detection efficiency.
Vertex constraints are applied to charged daughter particles
 before computing their invariant masses. At each step in the decay
 chain we require that the candidate mesons
have masses consistent with their assumed particle type.
If daughter particles are produced in the decay of a parent meson
with a natural width that is small relative to the reconstructed width,
we constrain the mass of this  meson to its nominal value, except for
the $\omega$ and the $\rho^0$ \cite{ref:PDG}. The $\Bzb$ mass is
computed using the constraint of the beam energy (see Sec.~\ref{ref:BKine}).
This fitting technique improves the resolution of the energy and the
momentum of the $\Bzb$ candidates as they are calculated from
improved energies and momenta of the $\Dstze$ and $\hz$.

Charged particle tracks are reconstructed from measurements in
the SVT and$/$or the DCH, and they are assigned various particle identification
probabilities  by the PID algorithms.
Extrapolated tracks must be in the vicinity of the
$\epem$ interaction point, $i.e.$ within 1.5~cm in the plane
transverse to the beam axis and 2.5~cm along the beam axis. The charged
tracks used for the reconstruction of $\eta\ra\pip\pim\piz$ and  $\etapr \ra
\pip\pim\eta(\ra\gg)$  must
in addition have a transverse momentum $p_T$ larger than 100~$\mevc$ and at least
12 hits in the DCH. When a PID positive identification is required for a track, the track polar angle
$\theta$ must be in the DIRC fiducial region
$25.78^{\circ}<\theta<146.10^{\circ}$. Photons are defined as
single clusters of energy deposition in the EMC crystals not matched to a track, and with
shower lateral shape consistent with photons. Because of the high machine background in
the very forward part of the EMC, we reject photons detected in the region $\theta<21.19^{\circ}$.
We assume that the production point of the  photons
is the reconstructed primary vertex of each $e^+e^-$ collision.

The selections applied to each meson
($\piz$, $\eta$, $\omega$, $\etapr$, $\Dz$, and $\Dstarz$) are
optimized by maximizing the figure of merit
$S/\sqrt{S+B}$, where $S$ is the number of signal and $B$ is the
number of background events. The numbers $S$ and $B$ are computed
from simulations, and the branching ratios  used to evaluate $S$ are the present world
average values of color-suppressed decay modes~\cite{ref:PDG}.
Each particle mass distribution is fitted with a set of Gaussian functions or
a so-called {\it modified Novosibirsk} empirical function~\cite{ref:bukin},
which is composed of a Gaussian-like peaking
part with two tails at low and high values. Particle candidates are then
required to have a mass within $\pm 2.5\ \sigma$ around the fitted
mass central value, where $\sigma$ is the resolution of the mass
distribution obtained by the fit. For the decays
$\Dz\ra\Km\pip\piz$ and $\Dstarz\ra\Dz\g$, the lower bound is
extended to $3 \ \sigma$ because of the photon energy losses in front of
and between the EMC crystals, which
makes the mass distribution asymmetric with a tail at low values.

\subsection{Selection of intermediate particles}

\subsubsection{$\piz$ selection}\label{pizselection}

The $\piz$ mesons are reconstructed from photon pairs. Each
photon energy $E\gamma$ must be greater than 85~$\mev$ for $\piz$
produced directly from $\Bz$ decays, and greater than 60~$\mev$ for $\piz$
from $\eta$, $\omega$, or $\Dz$ meson decays. Slow neutral pions originating from
$\Dstarz\ra\Dz\piz$ decays must satisfy $E \gamma>30~\mev$.
The $\piz$ reconstructed mass  resolution ranges $6.5-7.0~\mevcc$ for
$\piz$  from $\eta$, $\omega$, and $\Dz$ mesons decays, and  $7.0-7.5~\mevcc$
for $\piz$ produced in $\Dstarz$  or $\Bz$ decays.

\subsubsection{$\eta$ selection}\label{etaselection}

The $\eta$ mesons are reconstructed in the $\gg$ and
$\pip\pim\piz$ decay modes, accounting for
about $62\%$ of the total decay rate~\cite{ref:PDG}, and may originate from
$\Bzb\ra\Dstze\eta$ or $\etapr\ra\pip\pim\eta$ decays.

The $\eta\ra\gg$ candidates are reconstructed by combining two
photons that satisfy $E \gamma>200~\mev$ for $\Bzb$ daughters and
$E \gamma>180~\mev$ for $\etapr$ daughters. As photons originating from high momentum
$\piz$ mesons may fake a $\eta\ra\gg$ signal, a veto is applied. The $\eta\ra\gamma\gamma$ candidate is rejected
if either photon  combined with any other photon in the event
with $E \gamma >200~\mev$   has an invariant mass between 115 and 150~$\mevcc$.
Such a veto retains $93\%$ of the signal while reducing the background of fake $\eta$ mesons candidates by a
factor of two. The resolution of the
$\eta\ra\gg$ mass distribution is  approximately $15~\mevcc$, dominated by the resolution on the photon
energy measurement in the EMC.

For $\eta$ candidates reconstructed in the decay channel $\pip\pim\piz$, the $\piz$
is required to satisfy the conditions described in
Sec.~\ref{pizselection}. The mass resolution is about $3~\mevcc$,
which is better than for the mode $\eta\ra\gg$, thanks to the relatively better resolution of the
tracking system and the various vertex and mass constraints applied
to the $\eta$ and $\piz$ candidates.

\subsubsection{$\omega$ selection}\label{omegaselection}

The $\omega$ mesons are reconstructed in the $\pip\pim\piz$ decay
mode. This mode accounts for approximately $89\%$ of the total decay rate. The $\piz$
is required to satisfy the conditions described in
Sec.~\ref{pizselection} and the transverse momenta of the
charged pions must be greater than 200~$\mevc$.
The natural width of the
$\omega$ mass distribution $\Gamma = 8.49~\mev$~\cite{ref:PDG}
is comparable to the experimental resolution $\sigma\sim7~\mevcc$,
therefore the $\omega$ mass is not constrained to its nominal
value. We define a total width
$\sigma_{\rm tot}=\sqrt{\sigma^2+\Gamma^2/c^4}\simeq  11~\mevcc$ and require
the $\omega$ candidates to satisfy
$|m_\omega-{\mu_{m_\omega}}|<2.5 \ \sigma_{tot}$ (where $\mu_{m_\omega}$ is the mean of the
$\omega$ mass distribution).

\subsubsection{$\rho^0$ selection}\label{rhoselection}

The $\rho^0$ mesons originate from $\etapr\ra\rho^0\g$ and are
reconstructed in the $\pip\pim$ decay mode. The charged tracks must
satisfy $p_T(\pipm)>100~\mevc$, where  $p_T$ is the transverse component of the momentum
with respect to the beam axis. We define the helicity angle
$\theta_{\rho^0}$ as the angle between the direction of the momentum of one of the two pions
and that of the $\etapr$ both evaluated in the
$\rho^0$ center-of-mass frame. Because the $\rho^0$ is a vector meson,
the angular distribution is proportional to
$\sin^2\theta_{\rho^0}$ for signal, and is flat for
background. The $\rho^0$ candidates with
$|\cos \theta_{\rho^0}|>0.73$ are rejected. Due to the large  $\rho^0$
natural width $\Gamma=149.1~\mev$~\cite{ref:PDG},
the mass of the $\rho^0$ candidate must lie within 160~$\mevcc$
around the nominal mass value and no mass constraint is applied.

\subsubsection{$\etapr$ selection}

The $\etapr$ mesons are reconstructed in the $\pip\pim\eta(\ra\gg)$ and
$\rho^0\g$ decay modes. These modes account for approximately $46.3\%$
of the total decay rate.

Only the $\eta\ra\gg$ sub-mode is used in the $\pip\pim\eta$ reconstruction due
to its higher efficiency.  The selection is described in Sec.~\ref{etaselection}.
For candidates reconstructed in the $\rho^0\g$ decay channel we select
$\rho^0$ candidates as described in Sec.~\ref{rhoselection}, and the photons must have an energy
larger than 200~$\mev$. As photons coming from $\piz$ decays may fake
signal, a veto as described in Sec.~\ref{etaselection} is applied.
The $\etapr$ mass resolution is about $3~\mevcc$ for $\pip\pim\eta$ and
8~$\mevcc$ for $\rho^0\g$.

\subsubsection{$\KS$ selection}\label{KSselection}

The $\KS$ mesons are reconstructed through their decay to two
charged pions ($\pim\pip$) which must originate from a common
 vertex, with  a $\chi^2$ probability of the vertex fit that must be larger than
$0.1\%$. We define the flight significance as the ratio $L/\sigma_L$, where
$L$ is the $\KS$ flight length in the plane transverse to the beam axis and
$\sigma_L$ is the resolution on $L$ determined from the vertex fit.
The combinatorial background is rejected by requiring a flight significance
larger than 5. The reconstructed $\KS$ mass resolution is about 2~$\mevcc$
for a core Gaussian part corresponding to about $70\%$ of the candidates and
5~$\mevcc$ for the remaining part, depending on the transverse position of the
$\KS$ decay within the tracking system (SVT or DCH).

\subsubsection{$\Dz$ selection}\label{Dzselection}

The $\Dz$ mesons are reconstructed in the $\Km\pip$, $\Km\pip\piz$, $\Km\pip\pim\pip$,
and $\KS\pip\pim$ decay modes. These modes account for about $29\%$ of the total decay rate.
All $\Dz$ candidates must satisfy $p^*(\Dz)>1.1~\gevc$, where $p^*$ refers to the value of the momentum
computed in the $\Upsilon(4S)$ rest frame. That requirement is loose enough so that  various sources
of background can populate the sidebands of the signal region.

For the decay modes reconstructed only with tracks, we require
that the  charged pions originating from the
$\Dz$ candidates fulfill $p_T(\pipm)>400~\mevc$ for
$\Km\pip$, $p_T(\pipm)>100~\mevc$ for
$\Km\pip\pim\pip$, and $p_T(\pipm)>120~\mevc$ for
$\KS\pim\pip$.

The charged tracks  must originate from a common
 vertex, therefore the $\chi^2$ probability of the vertex fit
 must be larger than $0.1\%$ for the decay channel $\Km\pip$ and larger than
$0.5\%$ for the other modes with more abundant background. Because of the increasing level of background
present for the various decay modes, the kaon candidates must satisfy from looser to tighter PID criteria
for  the modes  $\Km\pip$,  $\Km\pip\pim\pip$,  and $\Km\pip\piz$ respectively.
For $\KS\pip\pim$, the $\KS$ candidates must satisfy the selection criteria described in
Sec.~\ref{KSselection}.

For the decay $\Dz \ra \Km\pip\piz$ the combinatorial background can
significantly be reduced by using the parametrization of the $\Km\pip\piz$
Dalitz-plot distribution as provided by the Fermilab E691 experiment~\cite{ref:E691}. This
distribution is dominated by the two $\Kstar$ resonances ($\Kstarz\ra\Km\pip$ and $\Kstarm\ra\Km\piz$)
and by the $\rho^+(\pip\piz)$ resonance. Therefore we select only $\Dz$ candidates that fall in the enhanced
region of the Dalitz plot as determined by the above parametrization.
The $\piz$ must satisfy the selections described in Sec.~\ref{pizselection}.

The reconstructed $\Dz$ mass resolution is about 5, 5.5,  6.5, and
11~$\mevcc$ for the decay modes $\Km\pip\pim\pip$, $\KS\pip\pim$, $\Km\pip$, and
$\Km\pip\piz$ respectively.

\subsubsection{$\Dstarz$ selection}

The $\Dstarz$ mesons are reconstructed in the $\Dz\piz$ and
$\Dz\gamma$ decay modes. The $\piz$ and $\Dz$ candidates are requested
to satisfy the selections described in Sec.~\ref{pizselection}
 and \ref{Dzselection} respectively. The
photons from $\Dstarz\ra\Dz\g$ must fulfill the additional condition  $E\gamma >130~\mev$ and
must pass the  $\piz$ veto as described in Sec.~\ref{etaselection}.

The resolution of the mass difference $\Delta m\equiv
m_{\Dstarz}-m_{\Dz}$ is about 1.3~$\mevcc$ for $\Dz\piz$ and
7~$\mevcc$ for $\Dz\g$.

\subsection{Selection of $B$-meson candidates}

The $B$ candidates are reconstructed by combining a $\Dstze$ with
an $\hz$, with the $\Dstze$ and $\hz$ masses constrained to their
nominal values (except when $\hz$ is an $\omega$). One needs to
discriminate between true $B$ signal candidates and  fake $B$ candidates.
The fake $B$ candidates  originate  from combinatorial backgrounds,
from other specific $B$ modes, or from the cross feed  events between reconstructed
color-suppressed signals.

\subsubsection{$B$-mesons kinematic variables}
\label{ref:BKine}

Two kinematic variables are commonly used in $\babar$ to select $B$
candidates: the energy-substituted mass $\mes$ and the energy
difference $\DeltaE$. These two variables use  constraints from
the precise knowledge of the beam energies and from energy
conservation in the two-body decay $\Upsilon(4S)\ra\BB$. The
quantity $\mes$ is the invariant mass of the $B$ candidate where
the $B$ energy is set to the beam energy in the CM frame:
\begin{equation}
\mes =
\sqrt{{\left( {\frac {{s / 2}
+\vec{p}_0.\vec{p}_B} {E_0}}  \right)^2}-\vert \vec{p}_B \vert^2}.
\end{equation}
The variable $\DeltaE$ is the energy difference between the reconstructed
$B$ energy and the beam energy in the CM frame:

\begin{equation}
\DeltaE =E_{D^{(*)}}^{\ast} + E_h^{\ast} - \sqrt{s}/2,
\end{equation}
where $\sqrt{s}$ is the $\epem$ center-of-mass energy. The small
variations of the beam energy over the duration of the run are
corrected when calculating \mes. For the momentum
$\vec{p}_i$ ($i=0,B$) and the energy $E_0$, the subscripts $0$ and
$B$ refer to the $\epem$ system and the reconstructed $B$ meson,
respectively. The energies $E_{D^{(*)}}^{\ast}$ and $E_h^{\ast}$ are
calculated from the measured $\Dstze$ and $\hz$ momenta.

For the various decay channels  of the $B$ signal events, the $\mes$ distribution peaks at the
$B$ mass with a resolution of $2.6-3$~$\mevcc$,  dominated by the
beam energy spread, whereas $\DeltaE$ peaks near zero with a
resolution of $15-50~\mev$ depending on the number of photons in
the final state.

\subsubsection{Rejection of $\epem\ra\qqbar$ background}

The continuum background $\epem\ra\qqbar$, where  $q$ is a  light quark $u$, $d$, $s$, or
$c$, creates high momentum mesons $\Dstze$, $\piz$,
$\eta^{(')}$, $\omega$ that can fake the signal mesons originating
from the two body decays $\Bzb\ra\Dstze\hz$. That background is dominated
by $c\bar{c}$ processes and to a lesser extent by $s\bar{s}$ processes.
Since the $B$ mesons are produced almost at rest in the $\Upsilon(4S)$
frame, the $\Upsilon(4S)\ra\BB$ event shape is isotropically distributed. By
comparison, the $\qqbar$ events have a back-to-back jet-like shape.
The $\qqbar$ background is therefore discriminated by  employing event shape variables.
The following set of variables was found
to be optimal among various tested configurations:

\begin{itemize}

\item The thrust angle $\theta_T$ defined as the angle between the
thrust axis of the $B$ candidate and the thrust axis of the rest
of event, the thrust axis being the axis on which the sum of projected
momentum is maximal. The distribution of $|\cos\theta_T|$ is flat for signal and
peaks at $1$ for continuum background.

\item Event shape monomials $L_0$ and $L_2$ defined as
\begin{equation}
L_0 = \sum_i |\overrightarrow{p}^*_i | \\ ;\ L_2 = \sum_i |\overrightarrow{p}^*_i|\cos^2\theta^*_i,
\end{equation}

with $\overrightarrow{p}_i^*$ being the CM momentum of the
particle $i$  that does not come from the $B$ candidate, and
$\theta_i^*$ is the angle between $\overrightarrow{p}_i^*$ and the thrust axis of
the $B$ candidate.

\item The polar angle $\theta^*_B$ between the $B$ momentum in the
$\Upsilon(4S)$ frame and the beam axis. With the $\Upsilon(4S)$ being
vector and the $B$ mesons being pseudoscalar, the
angular distribution is proportional to $\sin^2\theta^*_B$ for
signal and roughly flat for background.

\end{itemize}

These  four variables are combined into a Fisher discriminant built
with the {\tt TMVA}~\cite{ref:TMVA} toolkit package.  An alternate approach
employing a multi-layer perceptron artificial neural network with two hidden
layers within the same framework was tested and showed marginal relative
gain, therefore the Fisher discriminant is used.

The Fisher discriminant $\mathcal{F}_{\rm shape}$ is trained with signal MC events and
off-peak data events. In order to maximize the number of off-peak
events all the $\Bzb\ra\Dstze\hz$ modes are combined. We retain signal MC events with
 $\mes$ in the signal region $5.27-5.29$~$\mevcc$ and off-peak data events with $\mes$
 in the range  $5.25-5.27$~$\mevcc$, accounting for half of the 40~$\mev$ CM energy-shift
 below the $\Upsilon(4S)$ resonance. The training and testing of the multivariate classifier are performed
with non-overlapping data samples of equal size obtained from  a cocktail of $20,000$ MC simulation signal events
 and from $20,000$ off-peak events. The obtained Fisher formula is
\begin{eqnarray}
\mathcal{F}_{\rm shape} = 2.36-1.18 \times |\cos\theta_T|+\nonumber\\
0.20\times  L_0 - 1.01\times L_2 - 0.80\times |\cos\theta_B^*|.
\end{eqnarray}
The $\qqbar$ background is reduced by applying a selection cut on
$\mathcal{F}_{\rm shape}$. The selection is optimized for each of the 72 signal decay channels by
maximizing the statistical significance with signal MC against generic MC $\epem\ra\qqbar$, $q\ne b$.
This requirement for the various decay modes retains between about  $30\%$ and $97\%$ of $B$ signal events, while rejecting
between about $98\%$ and $35\%$ of the background from light $\qqbar$.

\subsubsection{Rejection of other specific
backgrounds}\label{bruit_BB}

The $\omega$ mesons in $\Bzb\ra\Dz\omega$ decays are
longitudinally polarized. We define the angle
$\theta_\omega$~\cite{ref:Babar2004,ref:BermanJacob} as the
the angle between the normal to the plane of the three daughter
pions in the $\omega$ frame and the line-of-flight
of the $\Bzb$ meson in the $\omega$  rest  frame. This definition is the
equivalent of the two-body helicity angle for the three-body
decay.  To describe the three-body decay distribution of
$\omega\ra\pip\pim\piz$, we define the {\it Dalitz angle}
$\theta_D$~\cite{ref:Babar2004} as the angle between the $\piz$
momentum in the $\omega$ frame and the $\pip$ momentum in the
frame of the pair of charged pions.

The signal distribution is proportional to $\cos^2\theta_\omega$ and $\sin^2\theta_D$, while
the combinatorial background distribution is roughly flat as
a function of  $\cos\theta_\omega$ and $\cos\theta_D$. These
two angles are combined into a Fisher discriminant
$\mathcal{F}_{\rm hel}$ built from signal MC events and generic
$\qqbar$ and $\BB$ MC events:
\begin{equation}
\mathcal{F}_{\rm hel}=-1.41-1.01\times|\cos\theta_D|+3.03\times|\cos\theta_\omega|.
\end{equation}
We require $\Bzb\ra\Dz\omega$ candidates to satisfy
$\mathcal{F}_{\rm hel}>-0.1$, to obtain an efficiency (rejection) on signal
(background) of about $85\%$ ($62\%$).

We also exploit the angular distribution properties in the decay
$\Dstarz\ra\Dz\piz$ to reject combinatorial background.
We define the helicity angle $\theta_{D^*}$ as the angle between the line-of-flight
of the $\Dz$ and that of the $\Bzb$, both evaluated in the $\Dstarz$ rest frame.
The angular distribution is proportional to $\cos^2\theta_{D^*}$
for signal and roughly flat for combinatorial background. Although
in principle such a behavior could be employed for $\Bzb\ra\Dstarz\piz$,
$\Dstarz\eta$, and  $\Dstarz\etapr$, a selection on $|\cos\theta_{D^*}|$ significantly improves the
statistical significance for the $\Bzb\ra\Dstarz\piz$ mode only.
Therefore  $\Dstarz$ candidates coming from the decay
$\Bzb\ra\Dstarz\piz$ are required to satisfy $|\cos\theta_{D^*}|>0.4$
with an efficiency (rejection) on signal (background) of about $91\%$
($33\%$).

A major $\BB$ background contribution in the analysis
of the $\Bzb\ra\Dstze\piz$ decay channel comes from the color-allowed decay
$\Bm\ra\Dstze\rho^-$. If the charged pion (mostly slow)  from the decay
$\rho^-\ra\pim\piz$ is omitted in the reconstruction of the $\Bzb$
candidate, $\Bm\ra\Dstze\rho^-$ events can mimic the $\Dstze\piz$
signal. Moreover, the decay modes $\BF(\Bm\ra\Dstze\rho^-)$ are $30-50$
times larger than those of the  $\Bzb\ra\Dstze\piz$ modes, and are poorly known:
$\Delta\BF/\BF=13.4\%-17.3\%$~\cite{ref:PDG}.  A veto is applied to reduce this background.
For each $\Bzb \ra \Dstze\piz$ candidate, we combine any remaining negatively
charged track in the event to reconstruct a $\Bm$ candidate in the decay mode
 $\Dstze\rho^-$. If the reconstructed $\Bm$ candidate satisfies
$\mes(\Bm)>5.27~\gevcc$, $|\DeltaE(\Bm)|<100~\mev$, and
$|m_{\rho^-}-m_{\rho^-}^{\textrm{PDG}}|<250~\mevcc$, then the initial
$\Bzb$ candidate is rejected. For the analysis of the decay mode
$\Bzb\ra\Dz\piz$ ($\Bm\ra\Dstarz\piz$), the veto retains about $90\%$ ($82\%$)
of signal and rejects about $67\%$  ($56\%$) of $\Bm\ra\Dz\rho^-$ and
$44\%$  ($66\%$) of $\Bm\ra\Dstarz\rho^-$ background.

\subsubsection{Choice of the ``best" $B$ candidate in the event}

The average number of $\Bzb\ra\Dstze\hz$ candidates per event after all selections
ranges between 1 and 1.6 depending on the complexity of the sub-decays. We perform
all the 72 $\Bzb\ra\Dstze\hz$ decay mode analyzes in parallel. Such that each possible decay channel is selected
with a dedicated analysis, for a given decay  mode one $B$ candidate only is kept per event. 
The chosen $B$ is that with the smallest value of
\begin{equation}
\chi^2_B = \left(\frac{m_{{\Dz}}-\mu_{m_{\Dz}}}{\sigma_{m_{\Dz}}}\right)^2 + \left(\frac{m_{\hz}-\mu_{m_{\hz}}}{\sigma_{m_{\hz}}}\right)^2,
\end{equation}
for ${\Dz}{\hz}$ modes, and
\begin{eqnarray}
\chi^2_B &=& \left(\frac{m_{\Dz}-\mu_{m_{\Dz}}}{\sigma_{m_{\Dz}}}\right)^2 + \left(\frac{m_{\hz}-\mu_{m_{\hz}}}{\sigma_{m_{\hz}}}\right)^2 \nonumber \\
&+&\left(\frac{\Delta m-\mu_{\Delta m}}{\sigma_{\Delta m}}\right)^2,
\end{eqnarray}
for the $\Dstarz{\hz}$ modes. The quantities $\sigma_{m_{{\Dz}}}$ and
$\sigma_{m_{{\hz}}}$ ($\mu_{m_{\Dz}}$ and $\mu_{m_{\hz}}$) are the resolution (mean)
of the mass distributions. The quantities $\mu_{\Delta m}$ and $\sigma_{\Delta m}$
are respectively the mean and resolution of the $\Delta m$   distributions.
 These quantities are obtained from fits of the mass distribution of  simulated
candidates selected from signal MC simulations.

The probability of choosing the true $\Bzb$ candidate in the event
according to the above criteria ranges from 71 to
$100\%$. The cases with lower probabilities correspond to the $\Dstze{\hz}$ modes with high
neutral multiplicity.

\subsubsection{Selection efficiencies}\label{sec:MCcorrection}

The branching fractions of the $\Bzb\ra\Dstze\hz$ decays is computed as

\begin{equation}\label{eq:BF}
\BF(\Bzb\ra\Dstze\hz)=\frac{N_S}{N_{B\bar{B}}\cdot{\cal E}\cdot\BF_{\rm sec}},
\end{equation}
where $\BF_{\rm sec}$ is the product of the branching fractions associated with the secondary
decays of the $\Dstze$ and $\hz$ mesons for the each of the 72 decay channels
considered in this paper~\cite{ref:PDG}. $N_{B\bar{B}}$ is the number of $\BB$ pairs in
data and $N_S$ is the number of signal events remaining after all the selections.
The quantity ${\cal E}$ is the total  signal efficiency including reconstruction
(detector and trigger acceptance) and analysis selections. It is computed from
each of the 72 exclusive high statistics MC simulation samples.

The selection efficiency from MC simulation is slightly different from the efficiency
in data. The MC efficiency and its systematic uncertainty therefore has to be adjusted
according to control samples. For the reconstruction of $\piz/\g$, the efficiency corrections are obtained
from detailed studies performed with a high statistics and high purity control sample
of $\piz$ mesons produced  in $\tau\ra\rho(\pi\piz)\nu_{\tau}$ decays normalized to
$\tau\ra\pi\nu_{\tau}$, to unfold tracking effects. Such corrections are validated against studies
 performed on the relative ratio of the number of detected $\Dz$  mesons in the decays
  $\Dz \ra K^- \pip  \piz$ and $\Dz \ra K^- \pip$, and produced in the decay
of $D^{*+}$ mesons from $\epem\ra c \bar{c}$ events. The relative data/simulation
efficiency measurements for charged tracks are similarly based on studies of
track mis-reconstruction using $\epem\ra \tau^+ \tau^-$ events.
On one side the events are tagged from a lepton in the
decay $\tau^- \ra l^- \bar{\nu}_l \nu_{\tau}$ and on the other side one reconstructs
two or three tracks from the decay $\tau^+ \ra \pip\pim  h^+ \bar{\nu}_{\tau}$.
The simulated efficiency of charged particle identification is compared
to the efficiency computed in data with control samples of kaons from $D^{*+}\ra\Dz(\Km\pip)\pip$
produced in $\epem\ra c \bar{c}$ events. The efficiency for $\KS$ candidates is modified using a data sample
of $\KS$, mainly arising from the continuum processes $\epem$ into $\qqbar$.

The efficiency corrections for the selection criteria applied to $\Dstze$
candidates and on the Fisher discriminant ($\mathcal{F}_{\rm hel}$)
for the continuum $\qqbar$  ($q\ne b$) rejection are obtained from studies of a  $\Bm\ra\Dstze\pim$ control sample.
This abundant control sample is chosen  for its kinematic similarity with
$\Bzb\ra\Dstze\hz$. The corrections are computed from the
ratios ${\cal E}_{\rm rel.}({\rm data})/{\cal E}_{\rm rel.}({\rm MC})$, where the relative
efficiencies ${\cal E}_{\rm rel.}$  are computed with the
signal yields as obtained from fits to $\mes$ distributions
of $\Bm\ra\Dstze\pim$  candidates in data and MC simulation,
before and after applying the various selections. The obtained
results are checked with the color-allowed control sample
$\Bm\ra\Dstze\rho^-$, which has slightly different kinematics
due to the relatively higher mass of the $\rho^-$, and therefore validates
those corrections for the modes such as $\Dstze\etapr$.

The reconstruction efficiency of  $\Bzb\ra\Dstarz\omega$
depends on the angular distribution, which is not yet
known ($f_L\sim 0.5-1$). To evaluate this efficiency we combine a set of properly
weighted fully longitudinally and fully transversely polarized MC samples,
according to the fraction of longitudinal polarization that we measure in
this paper (see Sec.~\ref{se:fL}).

\subsection{Fit procedure and data distributions}
\label{FITProc}

\begin{figure*}[htb]
\begin{center}
\includegraphics[width=0.325\linewidth]{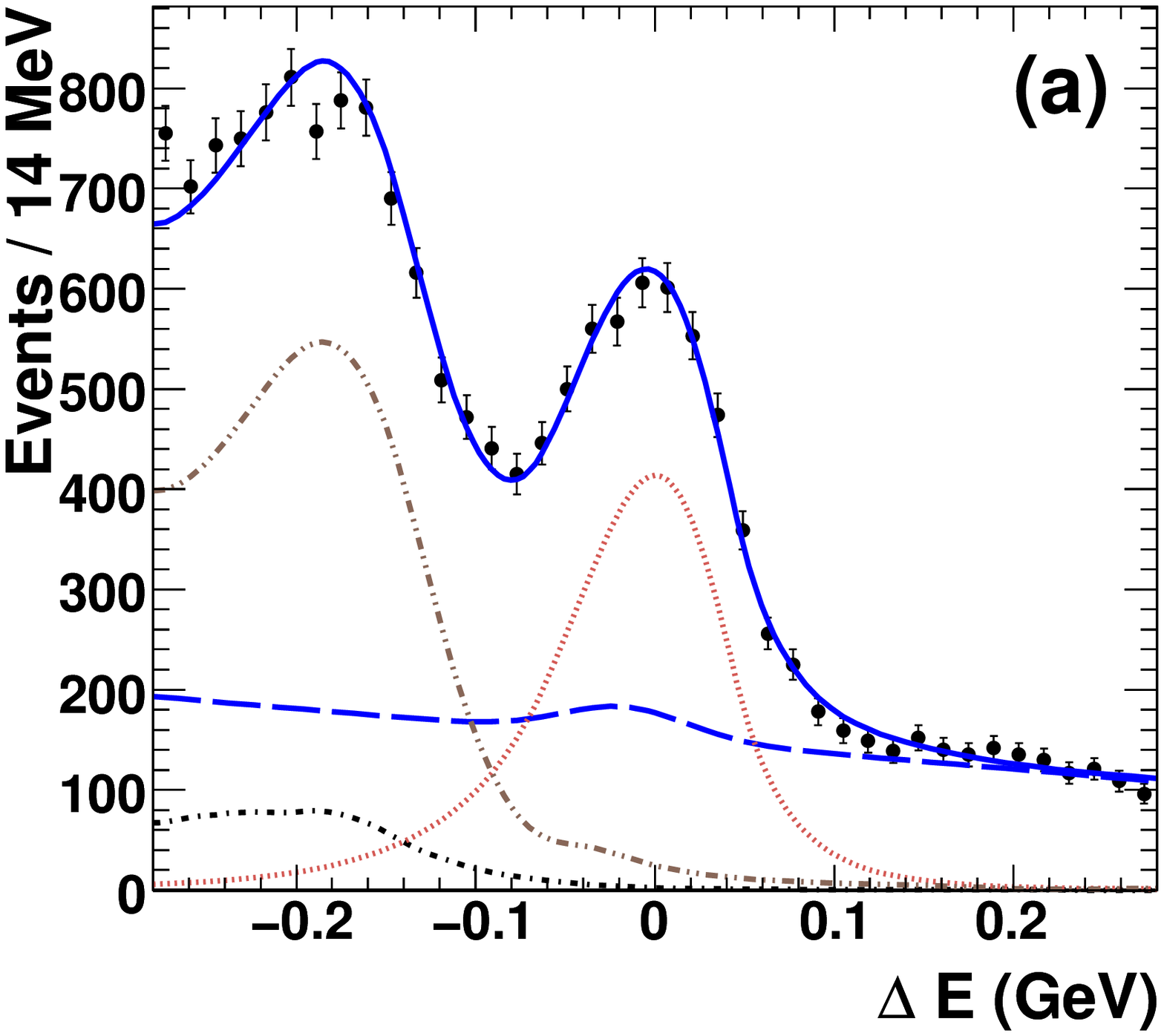}
\includegraphics[width=0.325\linewidth]{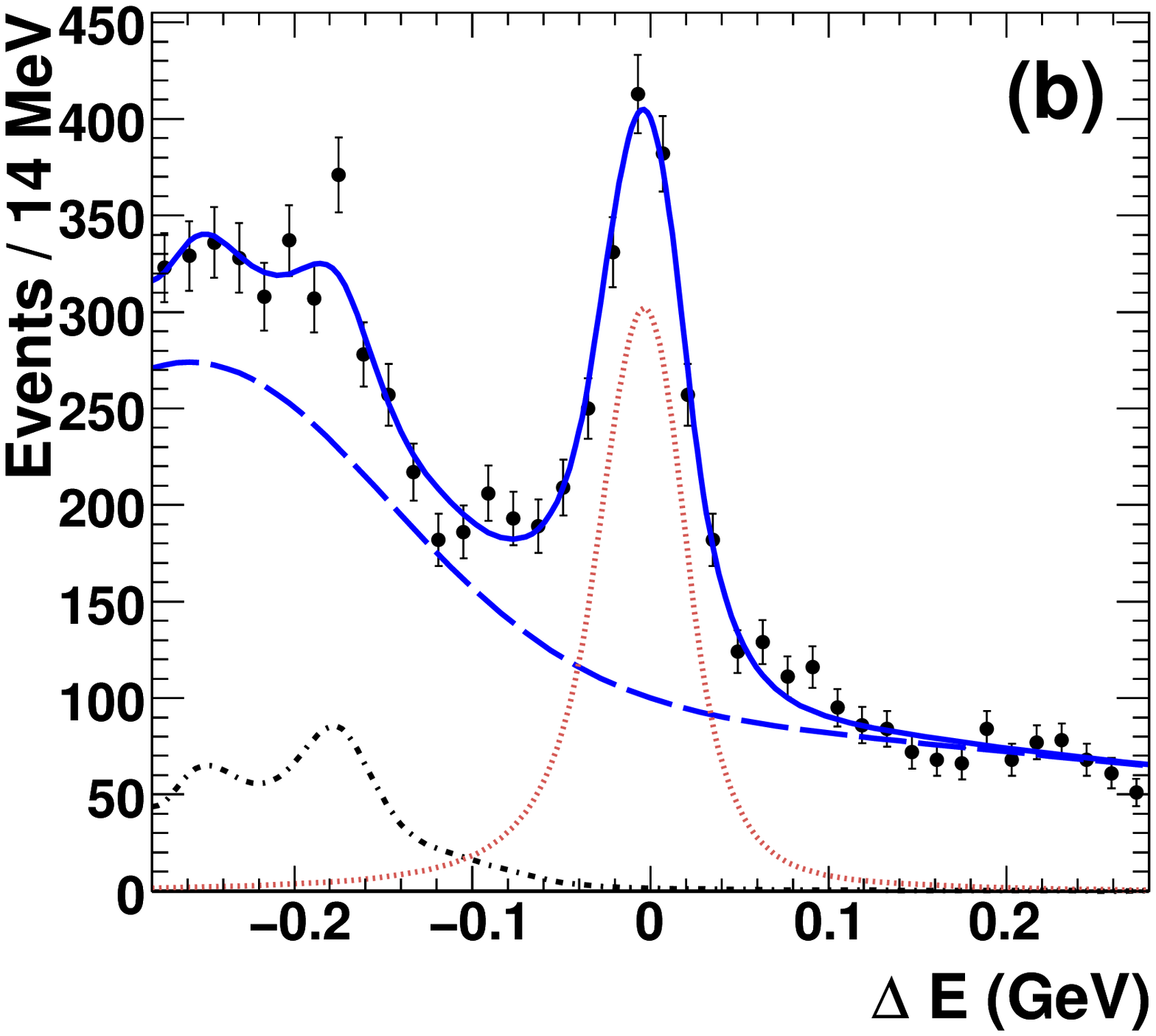}
\includegraphics[width=0.325\linewidth]{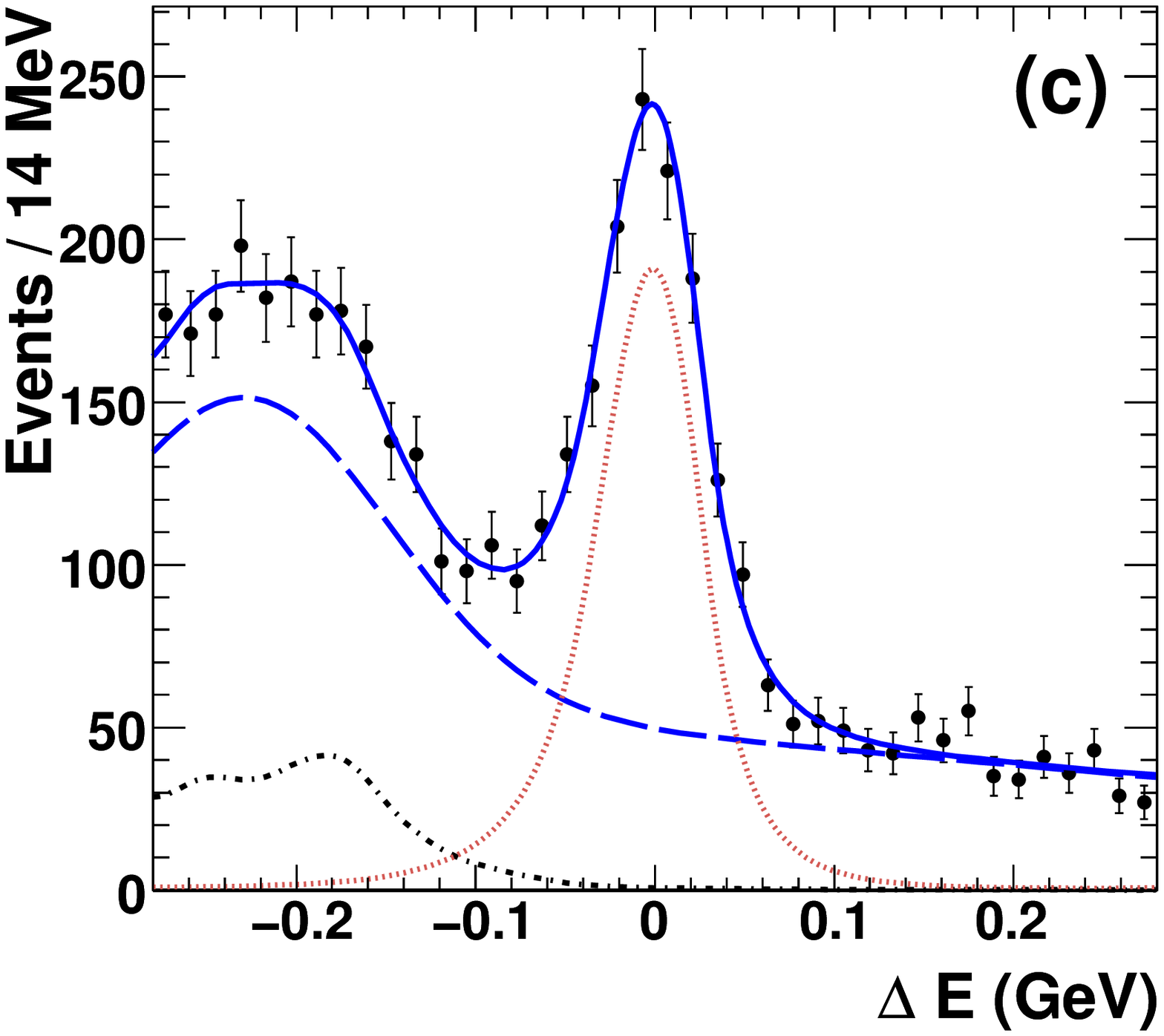}\\
\includegraphics[width=0.325\linewidth]{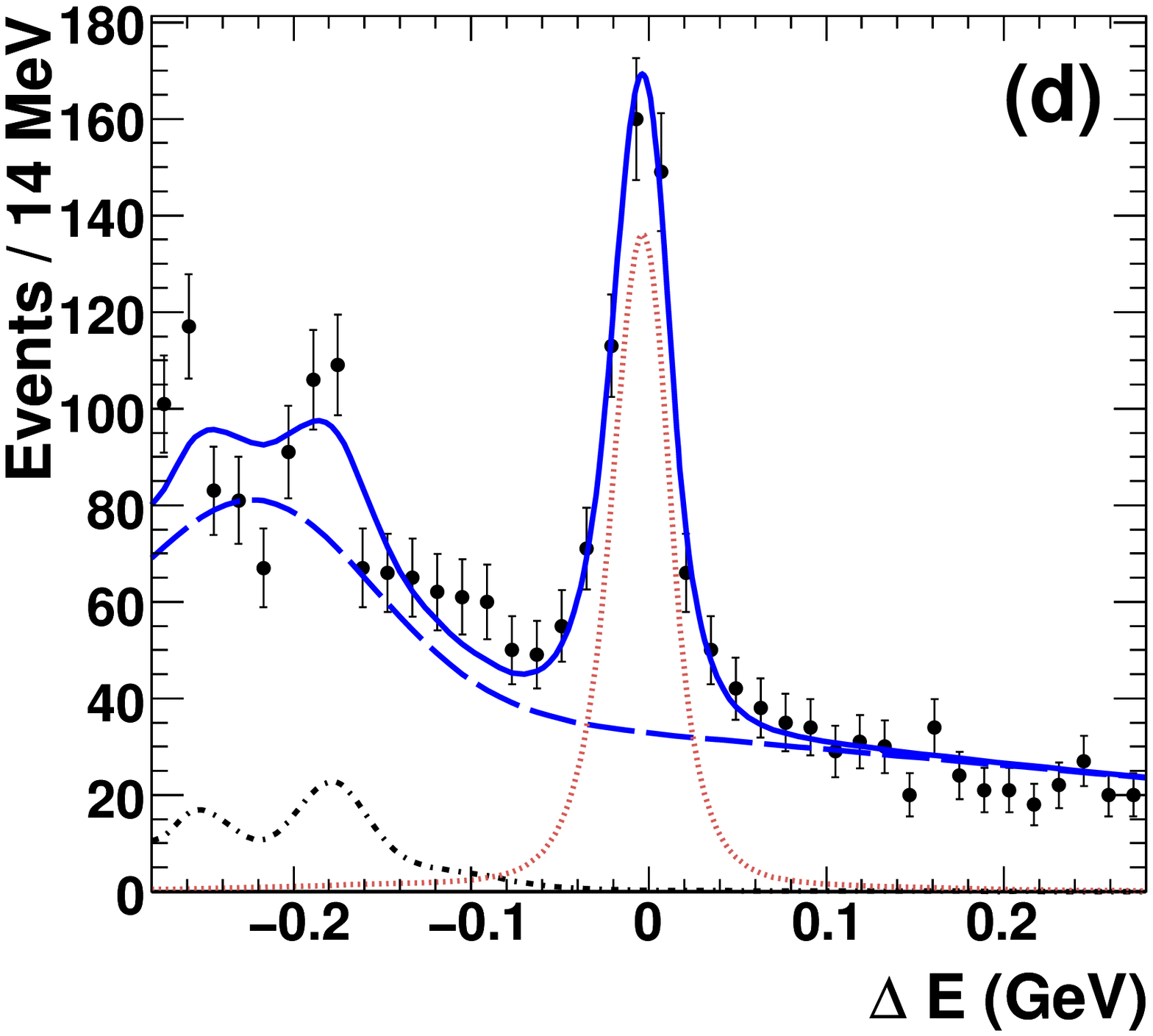}
\includegraphics[width=0.325\linewidth]{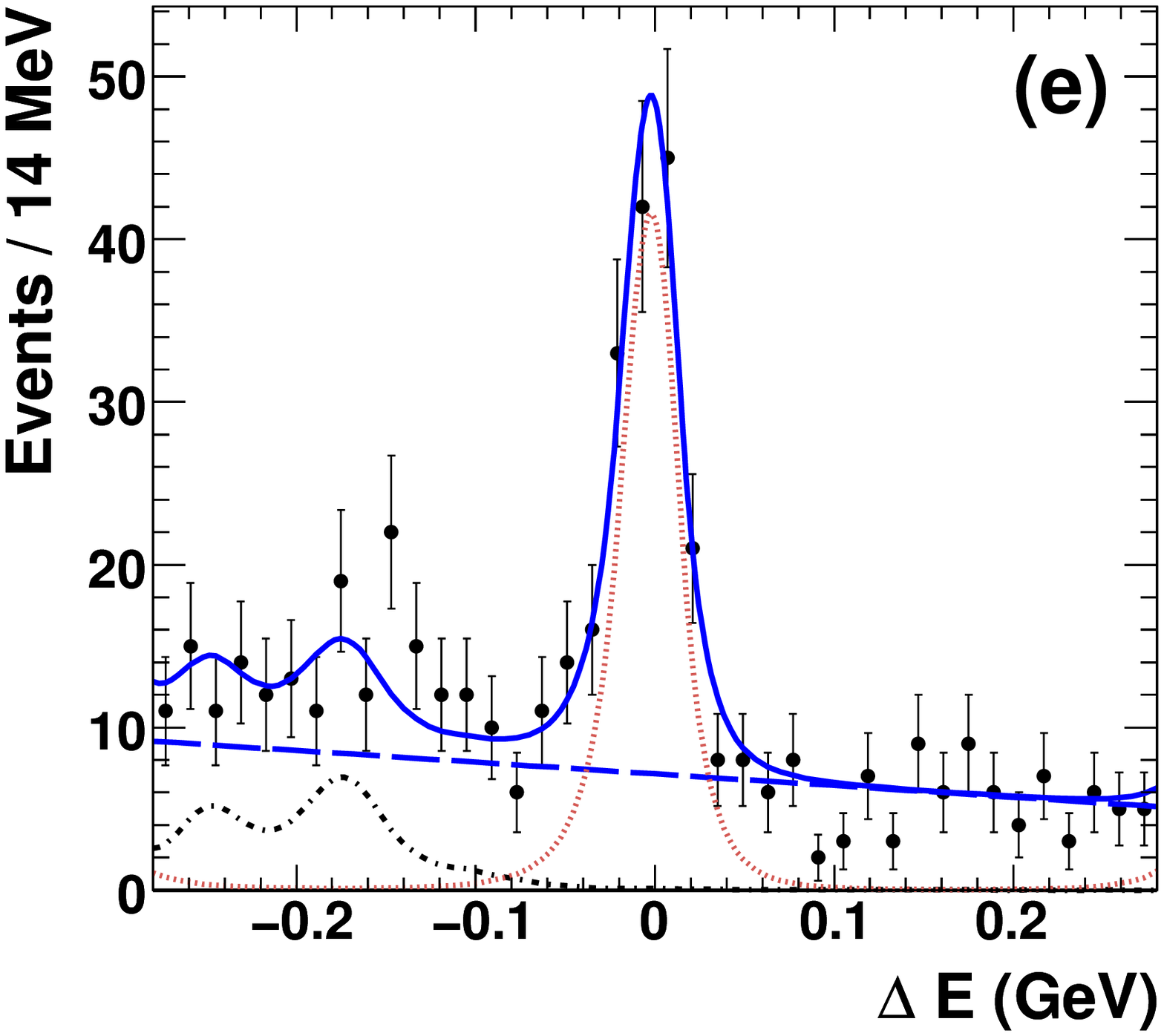}
\includegraphics[width=0.325\linewidth]{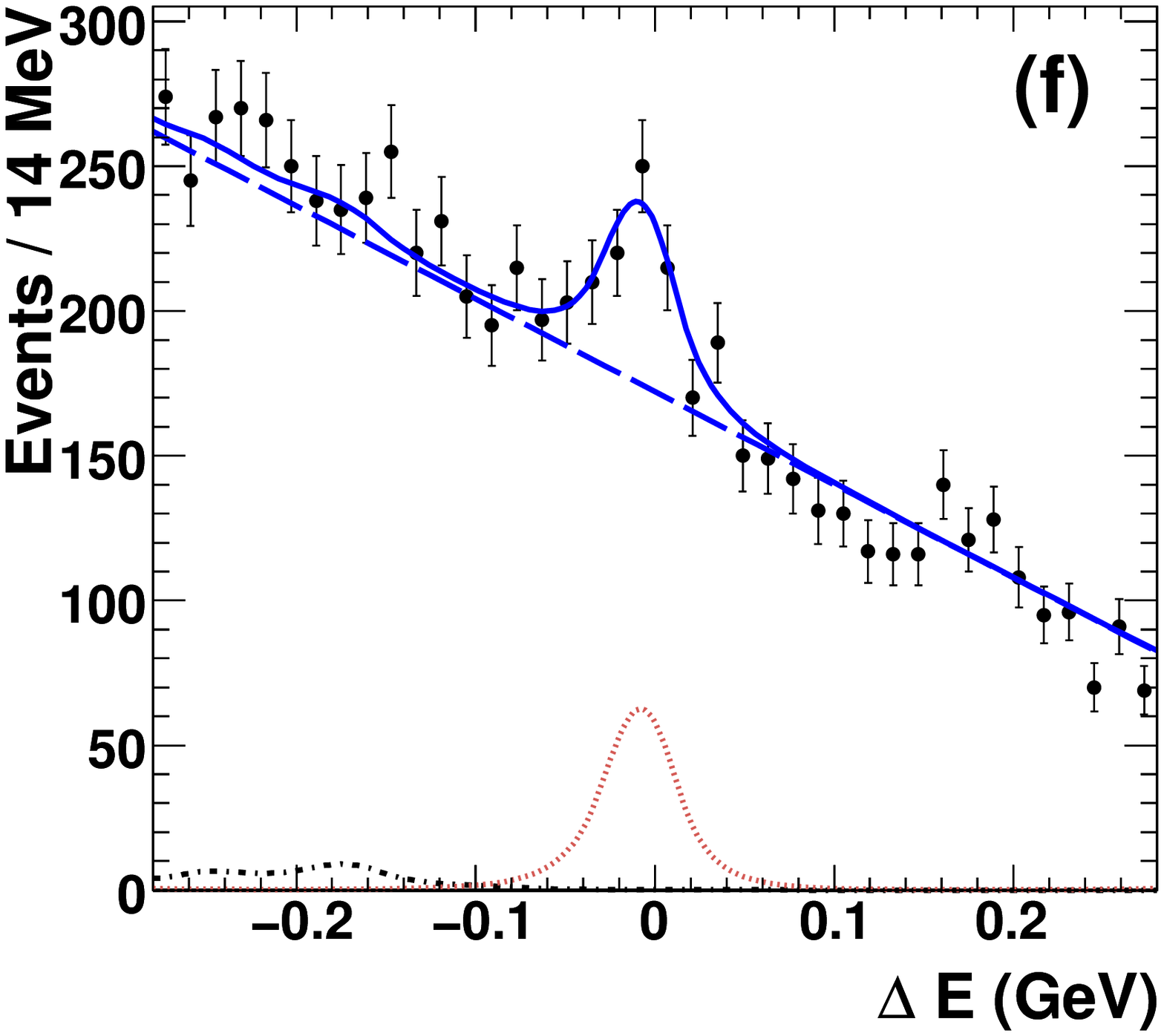}
\caption{\label{fig:fitData_1} Fits of $\DeltaE$ distributions in data for modes $\Bzb\ra\Dz\piz$ (a), $\Bzb\ra\Dz\omega$
(b), $\Bzb\ra\Dz\eta(\gg)$ (c),  $\Bzb\ra\Dz\eta(\pi\pi\piz)$ (d), $\Bzb\ra\Dz\etapr(\pi\pi\eta)$
(e), and $\Bzb\ra\Dz\etapr(\rho^0\g)$ (f). The data points with error bars are measurements in data,
the curves are the various PDF components: the  solid (blue) fitted total PDF, the dotted (red) signal PDF, the dotted-dashed (black)
cross-feed PDF, the double dotted-dashed (brown) $\Bm\ra\Dstze\rho^-$ PDF, and the long dashed (blue) combinatorial background PDF.}
\end{center}
\end{figure*}
\begin{figure*}[htb]
\begin{center}
\includegraphics[width=0.325\linewidth]{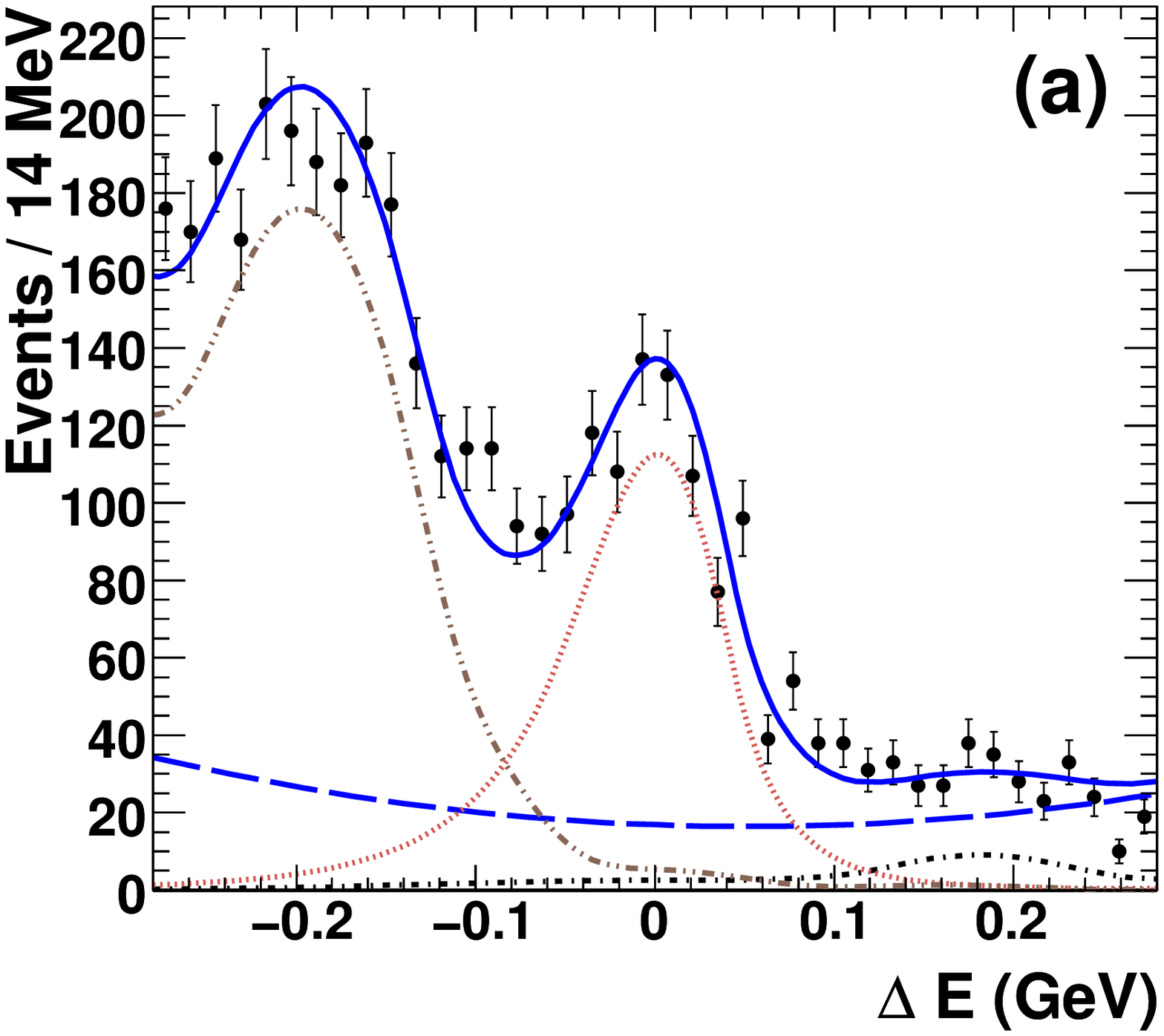}
\includegraphics[width=0.325\linewidth]{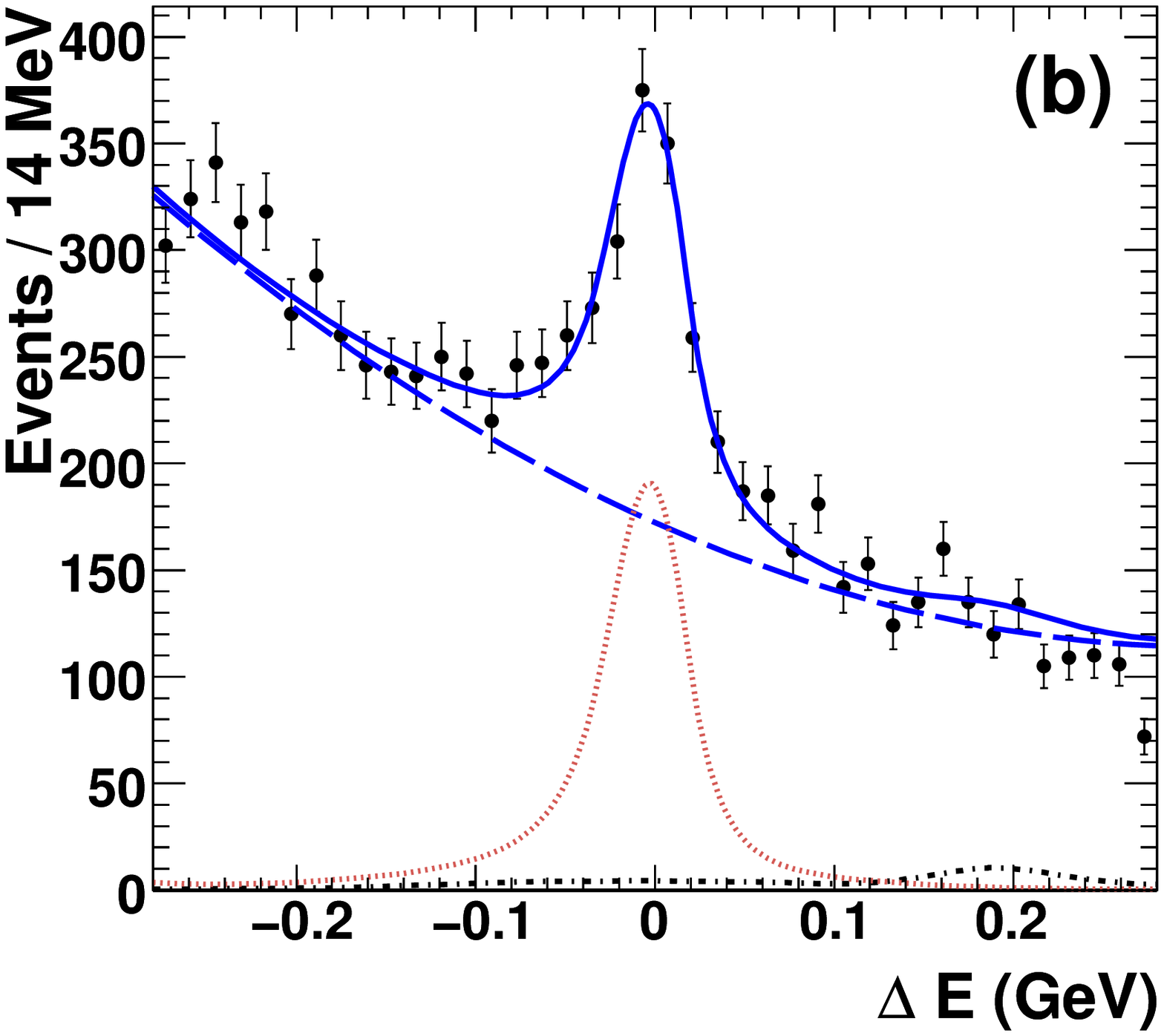}
\includegraphics[width=0.325\linewidth]{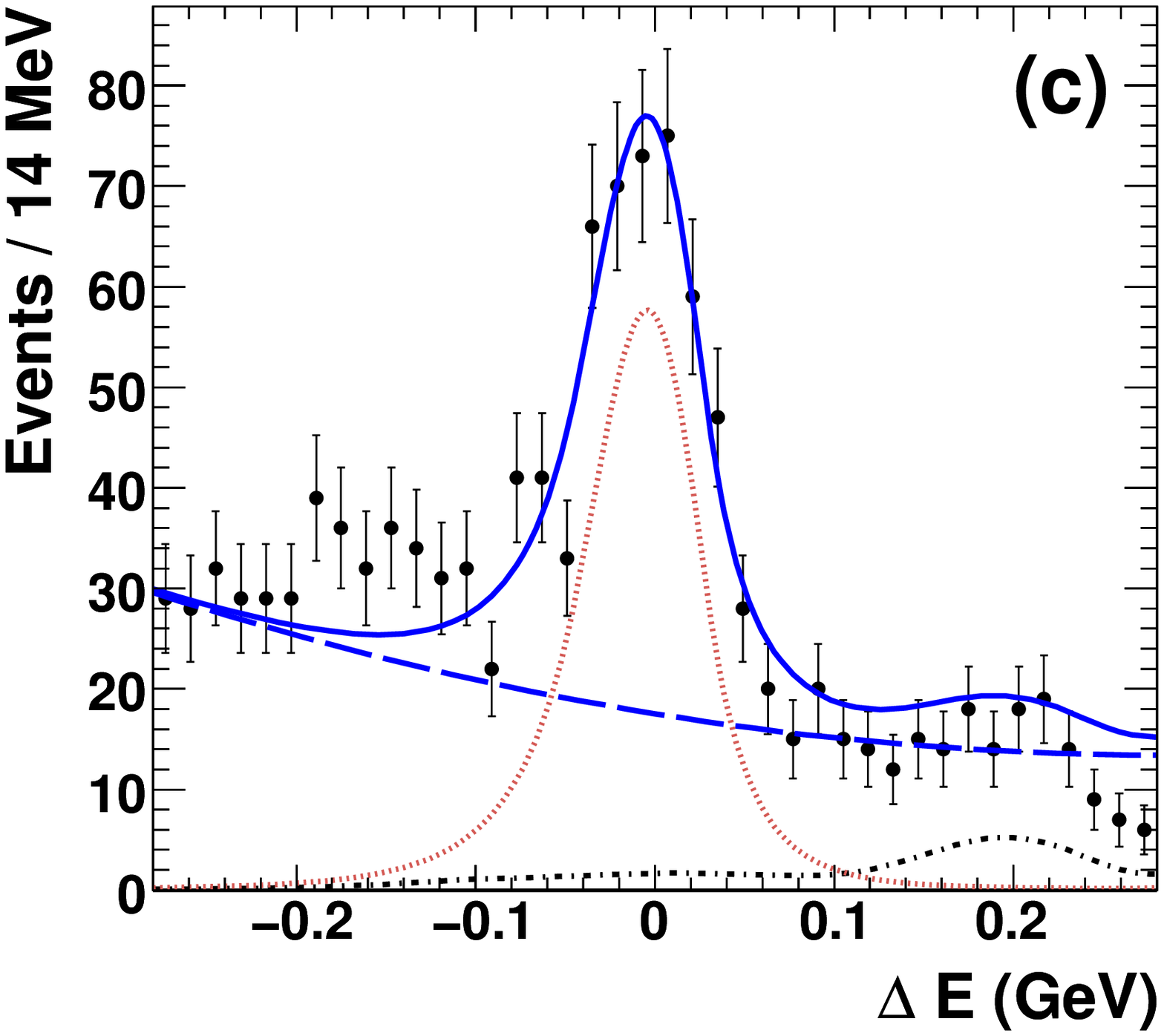}\\
\includegraphics[width=0.325\linewidth]{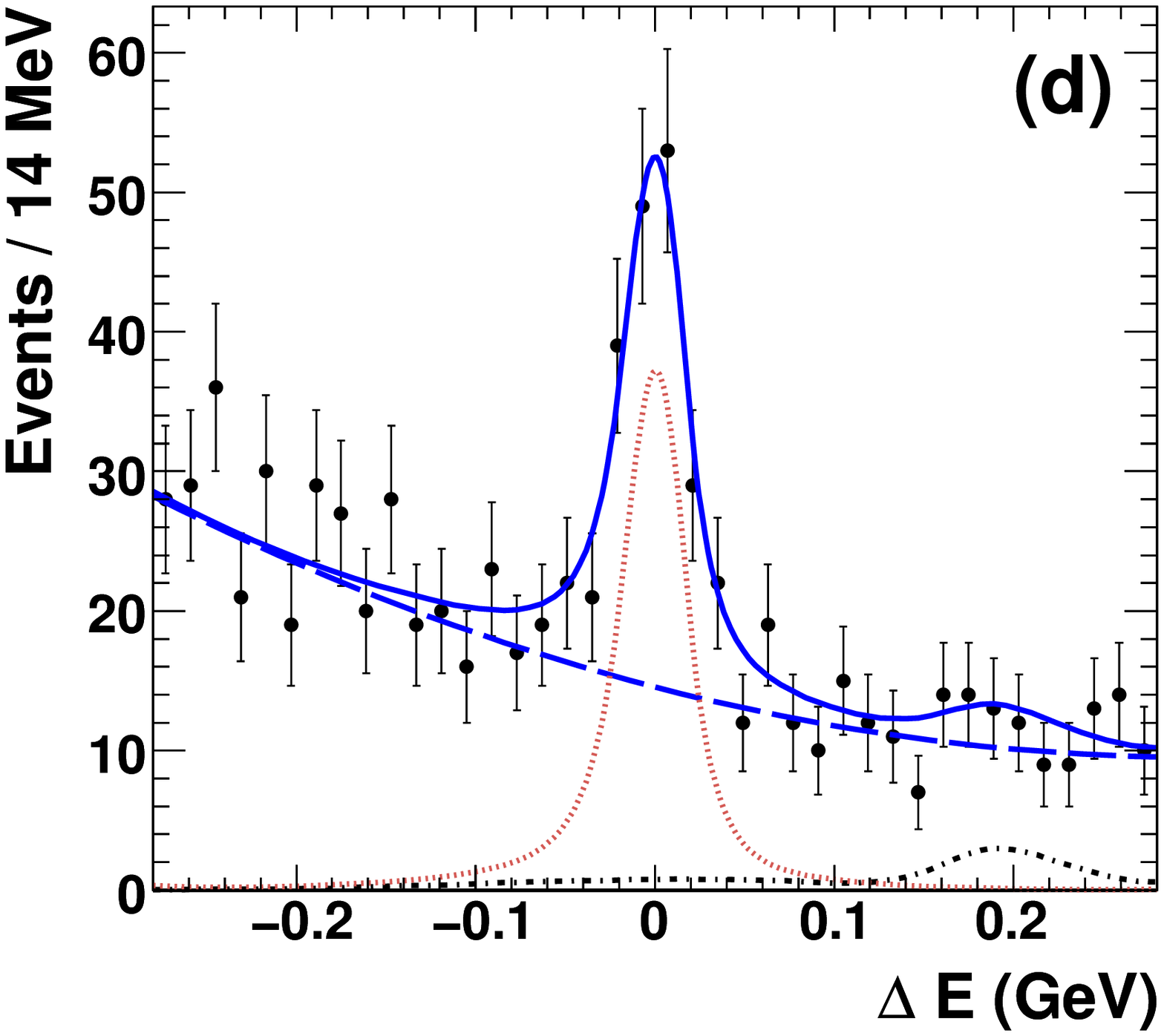}
\includegraphics[width=0.325\linewidth]{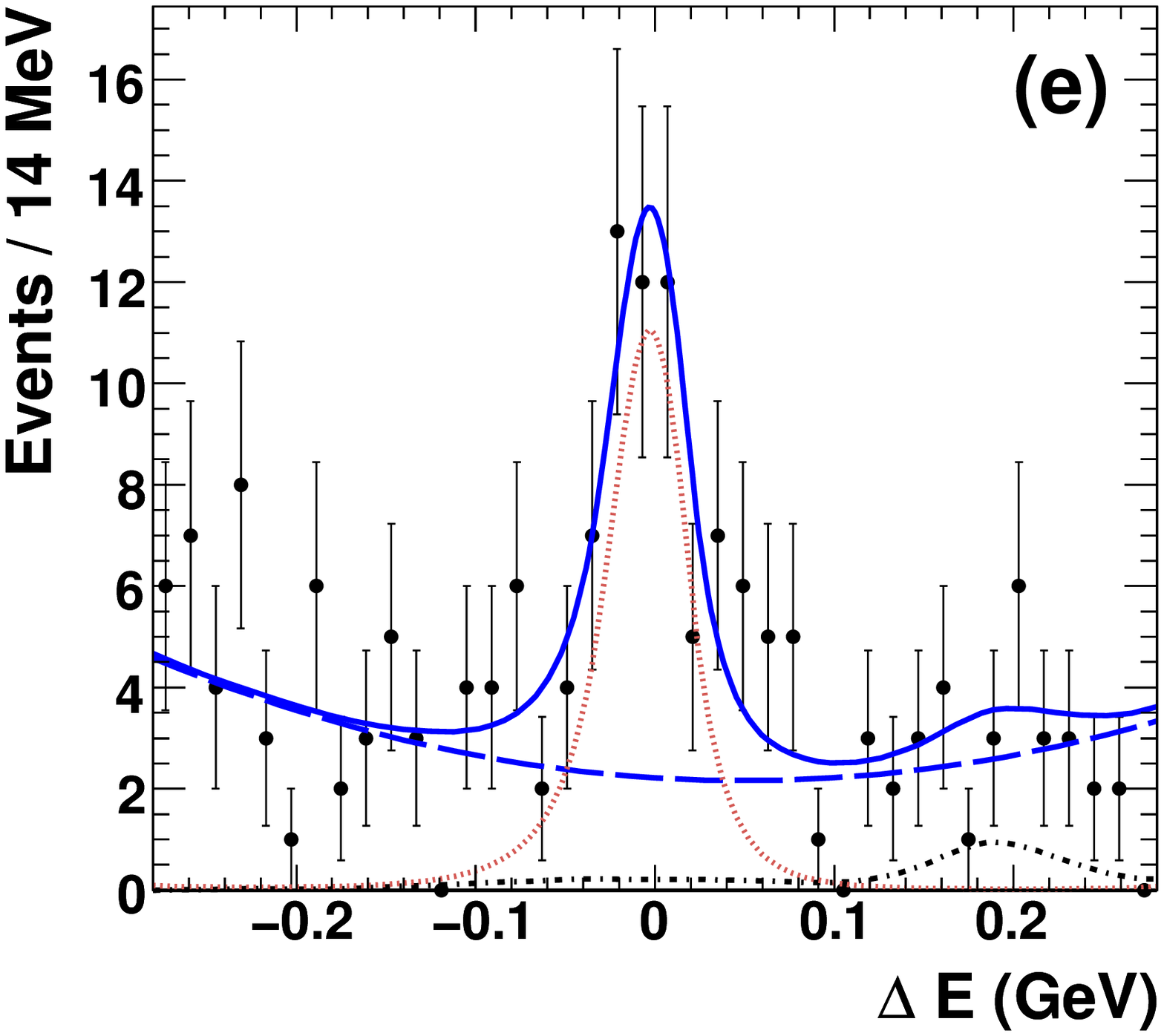}
\includegraphics[width=0.325\linewidth]{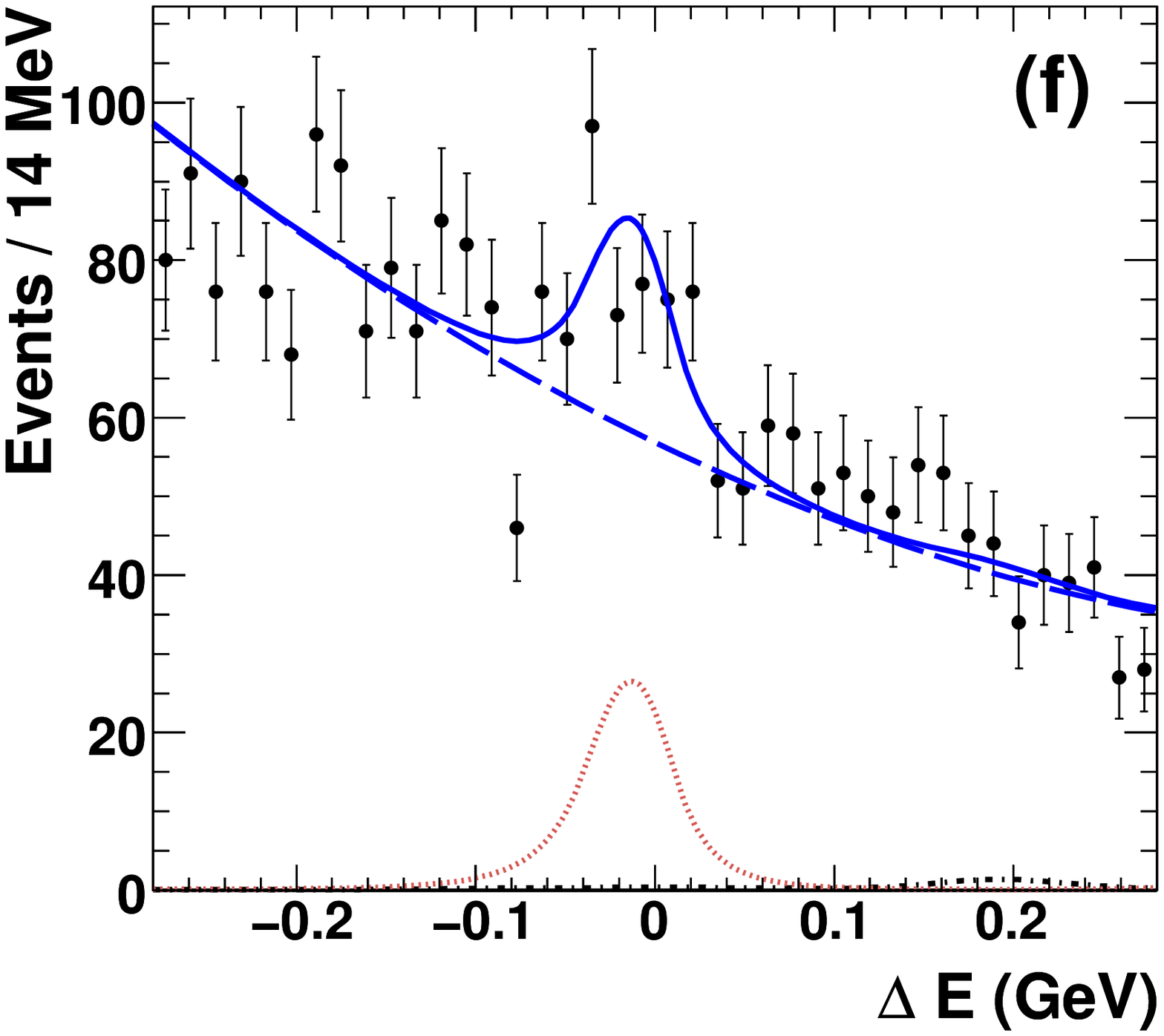}
\caption{\label{fig:fitData_2} Fits of $\DeltaE$ distributions in
data for modes $\Bzb\ra\Dstarz\piz$ (a), $\Bzb\ra\Dstarz\omega$ (b), $\Bzb\ra\Dstarz\eta(\gg)$ (c),
$\Bzb\ra\Dstarz\eta(\pi\pi\piz)$ (d), $\Bzb\ra\Dstarz\etapr(\pi\pi\eta)$ (e), and $\Bzb\ra\Dstarz\etapr(\rho^0\g)$ (f),
where the $\Dstarz$ mesons decay into the signal mode $\Dz\piz$. A detailed legend
is provided in the caption of Fig.~\ref{fig:fitData_1}.}
\end{center}
\end{figure*}

\begin{figure*}[htb]
\begin{center}
\includegraphics[width=0.325\linewidth]{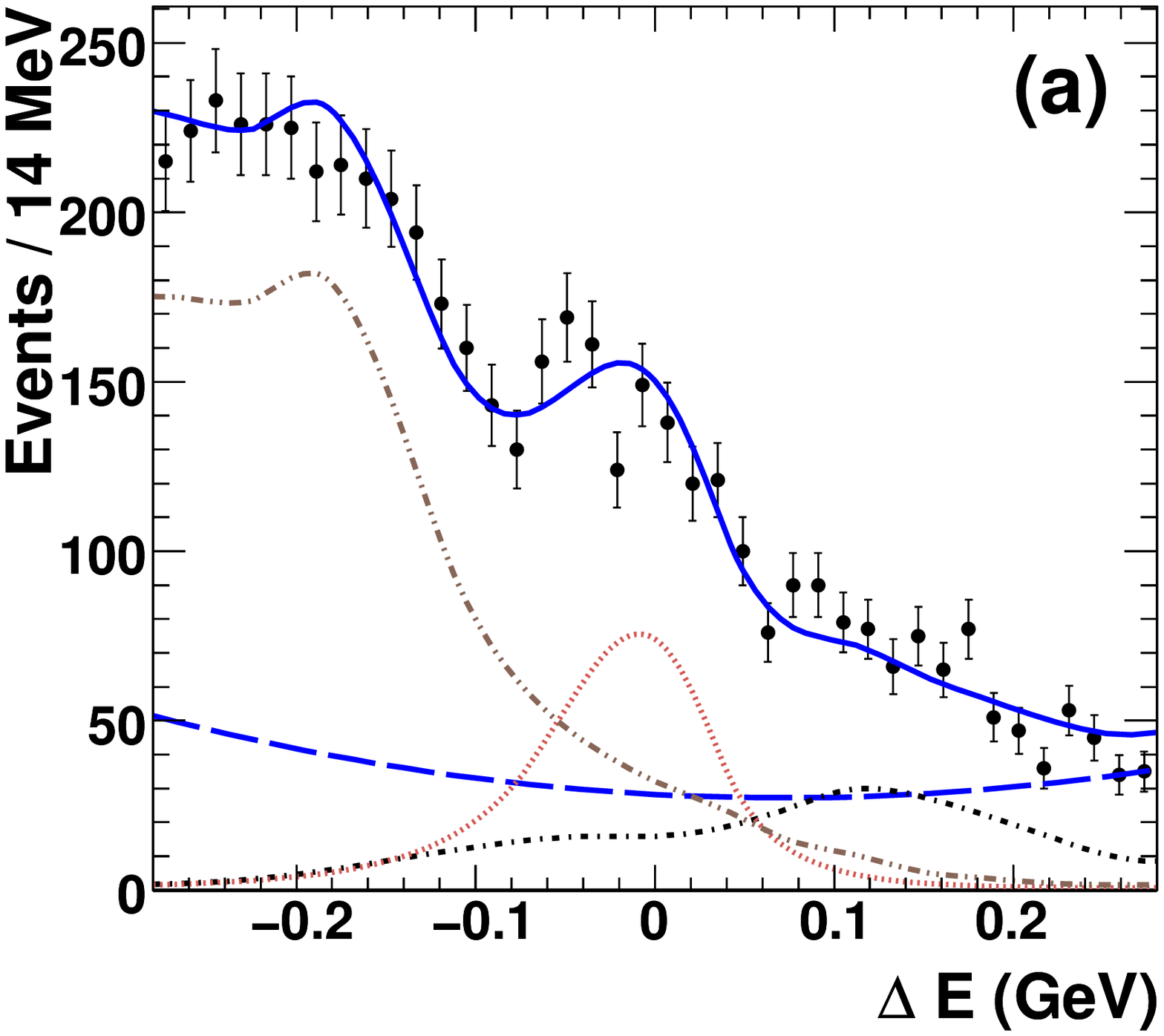}
\includegraphics[width=0.325\linewidth]{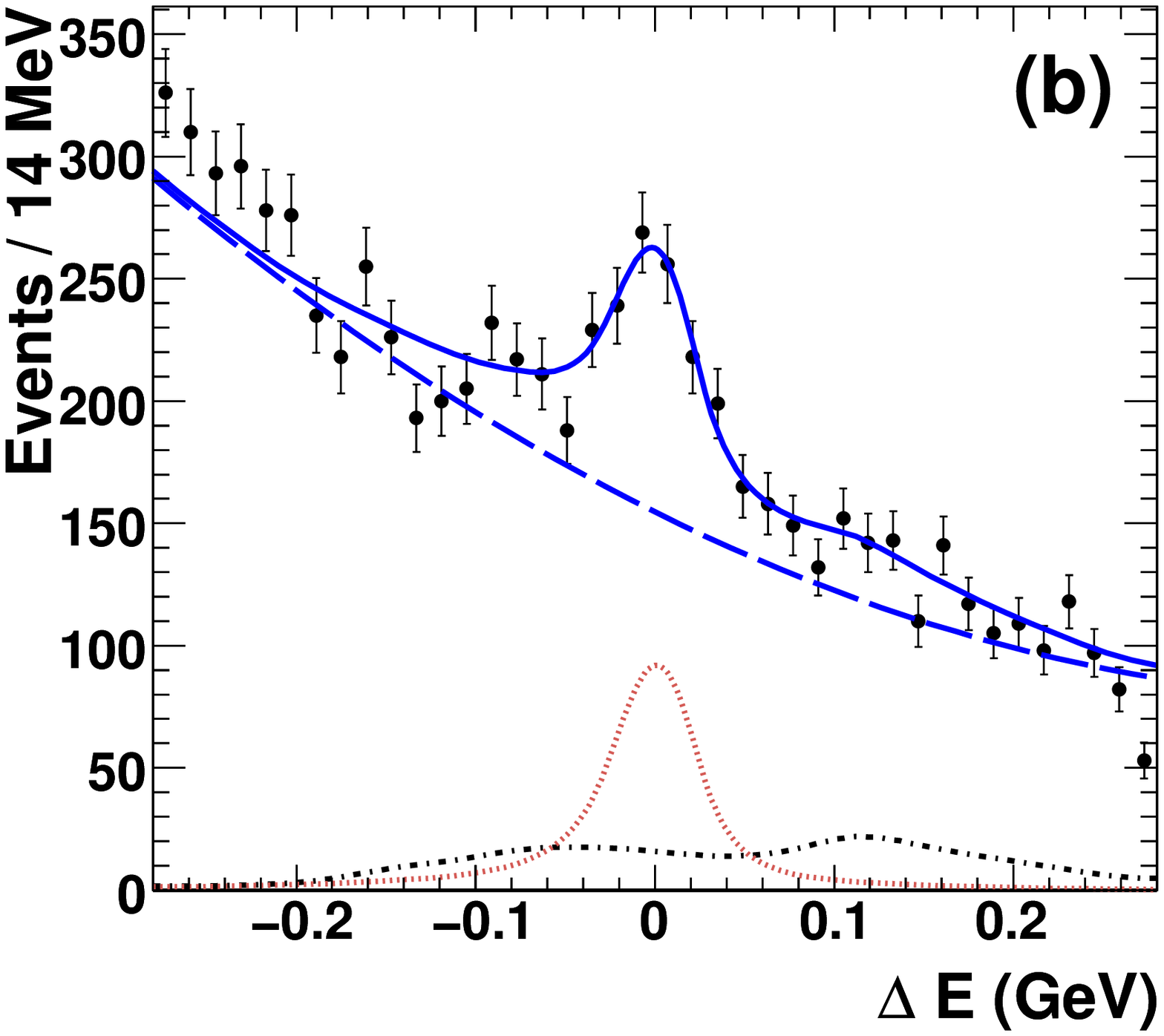}
\includegraphics[width=0.325\linewidth]{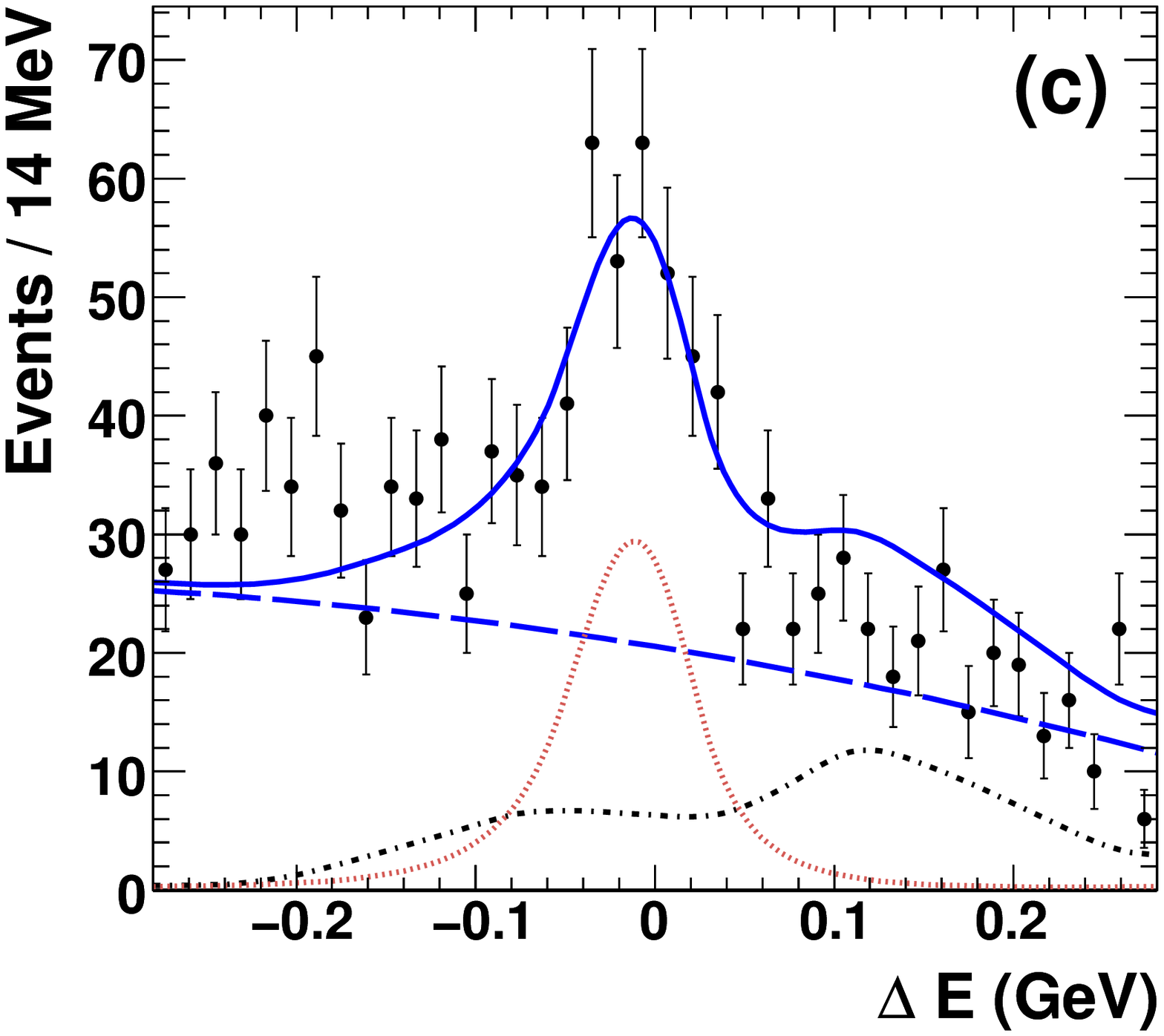}\\
\includegraphics[width=0.325\linewidth]{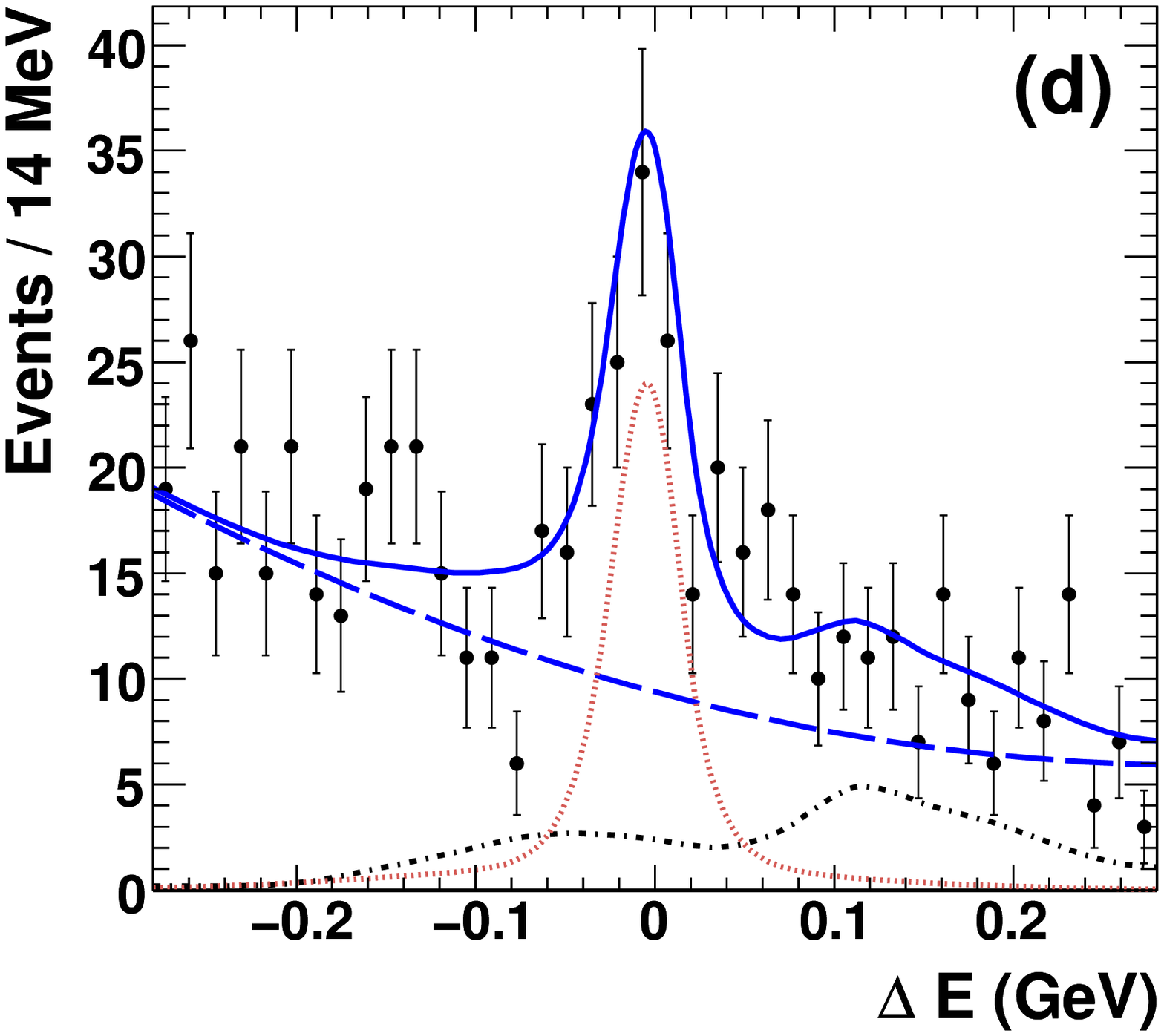}
\includegraphics[width=0.325\linewidth]{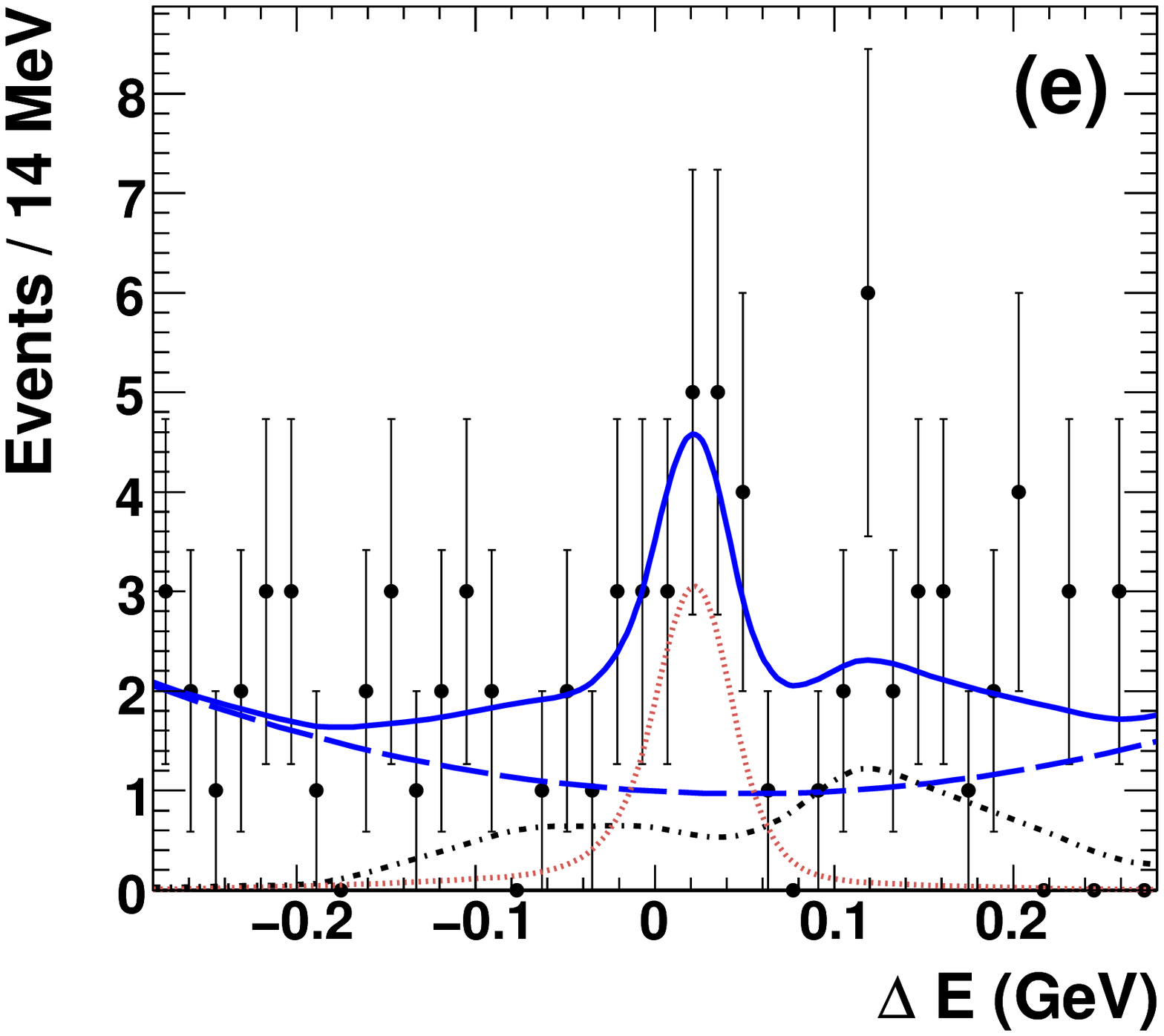}
\includegraphics[width=0.325\linewidth]{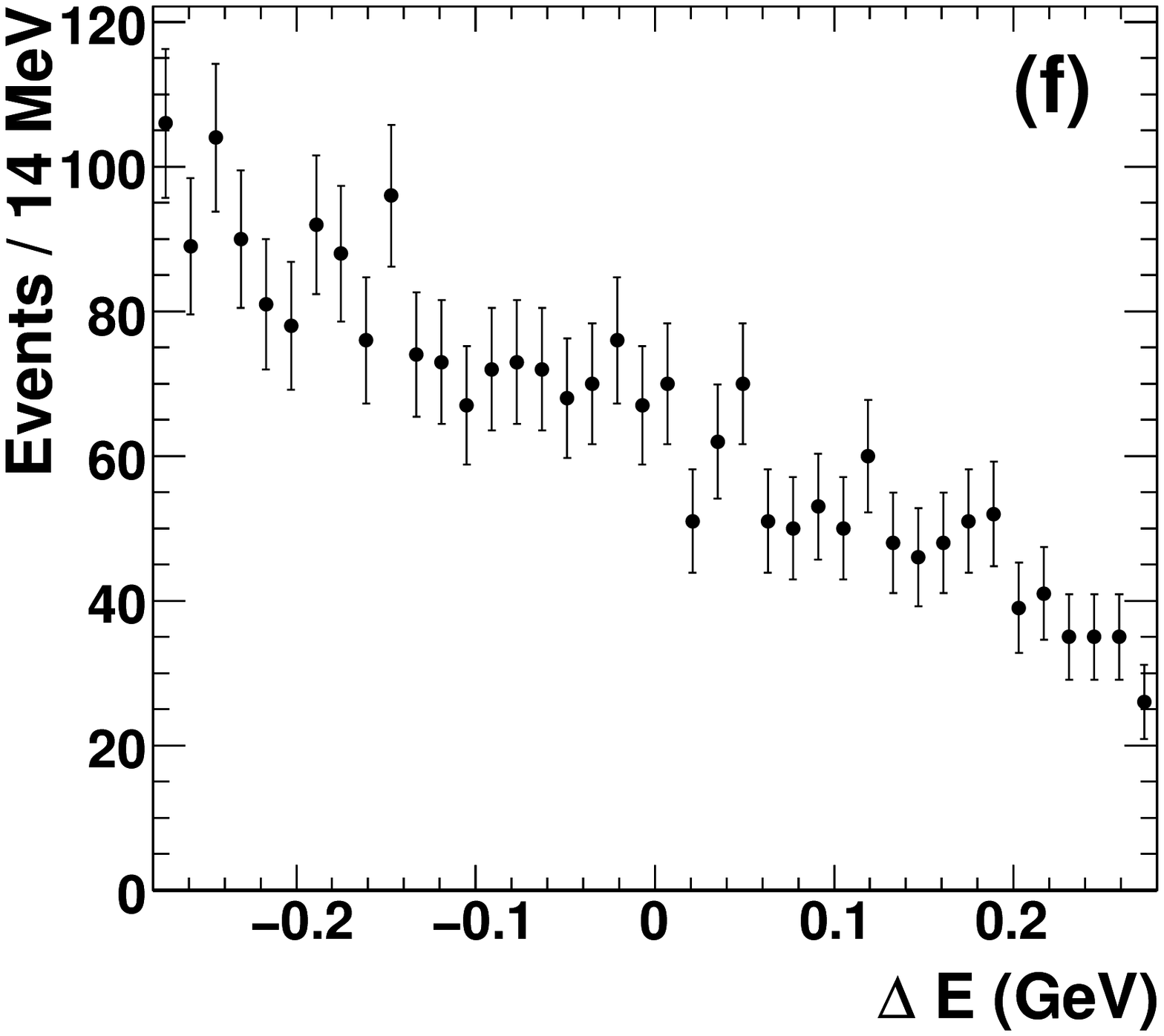}
\caption{\label{fig:fitData_3} Fits of $\DeltaE$ distributions in
data for modes $\Bzb\ra\Dstarz\piz$ (a), $\Bzb\ra\Dstarz\omega$ (b), $\Bzb\ra\Dstarz\eta(\gg)$ (c),
$\Bzb\ra\Dstarz\eta(\pi\pi\piz)$ (d),  and $\Bzb\ra\Dstarz\etapr(\pi\pi\eta)$ (e), where the $\Dstarz$ mesons
decay into the signal mode $\Dz\g$. The unfitted $\Delta E$ distribution of $\Bzb\ra\Dstarz(\Dz\gamma)\etapr(\rho^0\g)$
candidates is also displayed (f). A detailed legend is provided in the caption of Fig.~\ref{fig:fitData_1}.}
\end{center}
\end{figure*}

We present the fits used to extract the branching fractions $\BF$.
For each of the 72 possible  $\Bzb\ra\Dstze\hz$ sub-decay modes,
using an iterative procedure (discussed in Sec.~\ref{IterProc}), we fit  the $\DeltaE$ distribution in the
range $|\DeltaE|<280~\mev$ for $\mes>5.27~\gevcc$ to get the
signal ($N_S$) and background yields. The fit of the  $\DeltaE$ distribution
allows us to model and adjust  the complex non-combinatoric $B$
background structure without relying completely on simulation.

The data samples corresponding to each $\Bzb$ decay mode are
disjoint and the fits are performed independently for each
mode. According to their physical origin, four categories
of events with differently shaped  $\DeltaE$ distributions are
considered:  signal events, cross-feed
events, peaking background events, and combinatorial background events.
The event (signal and background) yields are obtained from unbinned
extended maximum likelihood (ML)  fits. We write the extended likelihood $\cal L$ as
\begin{equation}
{\cal L} = \frac{e^{-n}}{N!} n^{N} \prod_{j=1}^N
f(\DeltaE_j|\theta,n),
\label{eq:LH}
\end{equation}
where $\theta$ indicates the set of parameters which are fitted  from the data.
$N$ is the total number of signal and background events for each sub-decay mode, and
$n=\sum_i N_i$ is the expectation value for the total number of events. The sum runs
over the different expected number $N_i$ of signal and background events in the various
$i$ categories. The total probability density function (PDF) $f(\DeltaE_j|\theta,n)$ is written as
the sum over the different signal and  background categories,
\begin{equation}
f(\DeltaE_j |\theta,n) = \frac{\sum_i N_i f_i(\DeltaE_j|\theta)}{n},
\label{eq:PDF}
\end{equation}
where $f_i(\DeltaE |\theta)$ is the PDF of category $i$ (signal or background component).
Some of the PDF component parameters are fixed from the MC simulation (see details in the following sections).

 The individual corresponding branching ratios are computed and then
combined as explained in Sec.~\ref{se:results}.

\subsubsection{Signal contribution}

All of the 72 possible reconstructed $\Bzb$ decay channels contain at least one photon.
Due to the possible energy losses of early showering photons in the detector material before the
EMC, the $\DeltaE$ shape for signal is modeled by the {\it modified Novosibirsk} PDF~\cite{ref:bukin}.
A Gaussian PDF is added to the modes with a  large $\DeltaE$ resolution to describe mis-reconstructed
events. The signal shape parameters are estimated from a ML fit to the distributions of simulated signal
events in the high statistics exclusive decay modes.

\subsubsection{Cross-feed contribution}

We call ``cross feed'' the events from all of the reconstructed $\Dstze\hz$
modes, except the one under consideration, that pass the complete
selection. The cross-feed events are a non-negligible part of
the  $\DeltaE$ peak in some of the modes, and the signal event
yield must be corrected for these cross-feed events.
As the various decay channels are studied in parallel, we use an iterative
procedure to account for those contributions in the
synchronous measurements (see Sec.~\ref{IterProc}).

The dominant cross-feed contribution to $\Bzb\ra\Dz\hz$ comes from
the companion decay channel $\Bzb\ra\Dstarz\hz$, when the $\piz/\g$ from the
$\Dstarz$ decay is not reconstructed. Such cross-feed events are
shifted in $\DeltaE$ by approximately the mass of the $\piz$ ($-135~\mev$),
with a long tail from $\Dstarz(\ra\Dz\g)\hz$ leaking into the signal region.
Similarly, the decay channel   $\Bzb\ra\Dstarz\hz$ receives a cross-feed contribution from the
associated decay mode $\Bzb\ra\Dz\hz$ and there is a cross-contamination
between the  $\Dstarz\ra \Dz\piz$ and $\Dstarz\ra \Dz\g$ decay channels.

The remaining cross-feed contributions,  {\it i.e.} from other  $\Bzb\ra\Dstze\hz$ color-suppressed decay modes,
were studied with the generic MC simulation. They were found to be highly negligible
and  represent at most 1~$\%$ of the signal, in the region $|\Delta E|<100~\mev$. Therefore
 they are accounted for by the generic MC simulation,
 whose generated branching fractions were taken from the PDG~\cite{ref:PDG}.

\subsubsection{Peaking $\BB$ background contributions}
\label{PeakBBbgd}

The major background in the reconstruction of $\Bzb\ra\Dstze\piz$ comes from the
decays $\Bm\ra\Dstze\rho^-$ (see Sec.~\ref{bruit_BB}). Their contribution is modeled by histogram-based PDFs
built from the high statistics exclusive signal MC
simulation samples. The individual distributions of the two
backgrounds $B^-\ra\Dz\rho^-$ and $B^-\ra\Dstarz\rho^-$ that pass  the $\Bzb\ra\Dstze\piz$  selections,
including the specific veto requirement as described in Sec.~\ref{bruit_BB},
cannot be distinguished. As a consequence, given the large uncertainty on their
branching fractions, the overall normalization of $B^-\ra\Dstze\rho^-$ PDF is left floating but the
relative ratio $N(B^-\ra\Dstarz\rho^-)/N(B^-\ra\Dz\rho^-)$ of the PDF normalization
is fixed. The value of this  ratio is extracted directly from the data by reconstructing exclusively
each of the $B^-\ra\Dstze\rho^-$ modes rejected by the veto requirements.  Those fully reconstructed
$B^-$ mesons differ from the   $\Bm\ra\Dstze\rho^-$, that pass  all the $\Bzb\ra\Dstze\piz$  selections,
by the additional selected slow charged $\pi$ originated from the $\rho^-$ meson. The relative correction on that ratio for
events surviving the veto selection is then computed using  the MC simulation for generated $B^-\ra\Dstze\rho^-$ decays.
A systematic uncertainty on that assumption is  assigned (see Sec.~\ref{sec:Systematics}).

In the cases of $\Bzb\ra\Dstze\omega/\eta(\ra \pip\pim\piz)$ modes, additional
contributions come from the $B$ decay modes $D^{(*)} n \pi \pi^{(0)}$ ,
where $n=1, \ 2, {\rm or} \ 3$, and through intermediate resonances such as
$\omega$ and $\rho^{-}_3(1690) (\ra \omega \pi^-)$.
These peaking backgrounds are modeled by a first-order polynomial PDF plus a
Gaussian PDF determined from the generic $\BB$ MC simulation.
The relative normalization of that Gaussian PDF component is left floating in the fit, since some of the branching fractions of  the $B$
decay modes $D^{(*)} n \pi \pi^{(0)}$ are not precisely known~\cite{ref:PDG}.

\subsubsection{Combinatorial background contribution}

The shape parameters of the combinatorial background PDFs are
obtained from ML fits to the generic $\BB$ and continuum MC, where all
signal, cross feed and peaking $\BB$ background  events have been removed.
The combinatorial background from $\BB$ and
$\qqbar$  ($q\ne b$) are summed and modeled by a second-order polynomial PDF.

\subsubsection{Iterative fitting procedure}
\label{IterProc}

We fit the $\DeltaE$ distribution using the PDFs for the signal,
for the cross feed, for the peaking background, and
for the combinatorial background as detailed in the previous
sections. The normalization of the signal, the peaking $\BB$ backgrounds,
and the combinatorial background components are allowed to float in the fit.
The mean of the signal PDF is left floating for the sum of
$\Dstze$ sub-decays. For each $\Dz$ sub-mode, the signal mean PDF is
fixed to the value obtained from the fit to the sum of $\Dz$
sub-modes. Those free parameters are extracted by maximizing the
unbinned extended likelihood to the $\DeltaE$ distribution defined in
Eqs.~(\ref{eq:LH}) and (\ref{eq:PDF}). Other PDF parameters are fixed from
fit results obtained with MC simulations, when studying separately
each of the signal and background  categories.

In the global event yield extraction of all the various $\Bzb\ra\Dstze\hz$
color-suppressed signals studied in this paper, a given mode can be signal and
cross feed to other modes at the same time. In order to use the
$\BF$ computed in this analysis, the yield extraction is performed
through an iterative fit  on $\Dstarz\hz$ and
$\Dz\hz$. The normalization of cross-feed contribution from
$\Dstze\hz$ is then fixed to the $\BF$ measured in the previous
fit iteration. For the cross-feed contributions, the
 PDG branching fraction~\cite{ref:PDG} values are used as starting points.
This iterative method converges quickly to stable
$\BF$ values, with a variation of less than $10\%$ of the  statistical uncertainty,
in less than 5 iterations.

We check the absence of biases  in our fit procedure by studying
pseudo-experiments with a large number of different data-sized
samples for the various signals.
The extraction procedure is applied to these samples where
background events are generated and added from the fitted PDFs. The
signal samples are assembled from non-overlapping samples
corresponding to the exclusive high statistics MC signals, with
yields corresponding to the MC-generated value of the branching
fraction. No significant biases are found.

\subsubsection{Data distributions and
event yields from summed sub-decay modes}
\label{DataDistAndYield}

\begin{table*}[htb]
{\footnotesize \caption{\label{tab:nbEvtFitDATA}
Numbers of signal events ($N_S$), combinatorial background ($N_{\textrm{combi}}$), cross feed ($N_{\textrm{cf}}$), and
$\Bm\ra\Dstze\rho^-$ ($N_{\textrm{D}\rho}$) events computed from the $\DeltaE$ fits to
data and counted in a signal box $|\Delta E|<2.5\ \sigma$, together with the statistical significances in numbers of standard deviations ${\cal S}_{\rm stat}$  (see text).
The quoted uncertainties are statistical only.}
\begin{center}
\begin{tabular}{l c rcr c rcr c rcr c rcr c c}
\hline\hline \\
$\Bzb\ra$ & &  \multicolumn{3}{c}{$N_S$} & & \multicolumn{3}{c}{$N_{\textrm{combi}}$} & &
\multicolumn{3}{c}{$N_{\textrm{cf}}$} & & \multicolumn{3}{c}{$N_{\textrm{D}\rho}$}  & & Statistical \\
(decay channel) & & \multicolumn{3}{c}{ } & &  \multicolumn{3}{c}{ } & & \multicolumn{3}{c}{ }  & &  \multicolumn{3}{c}{ }
& & significance ${\cal S}_{\rm stat}$ \\ \hline \\
$\Dz\piz$   & & 3429 & $\pm$ & 123 & &  2625 & $\pm$ & 75 & &  97 & $\pm$ & 3 & &  700 & $\pm$ & 14 & & 41 \\
$\Dz\eta(\gg)$   & & 1022 & $\pm$ & 55 &  & 532 & $\pm$ & 14 &&   13 & $\pm$ & 1 &&    & - &  & & 36  \\
$\Dz\eta(\pi\pi\piz)$   & & 411 & $\pm$ & 29 &&   191 & $\pm$ & 6 & &  2 & $\pm$ & 0 &&    & - & & & 23 \\
$\Dz\omega$   & & 1374 & $\pm$ & 120 &  & 886 & $\pm$ & 25 & &  18 & $\pm$ & 2 & &   & - &  && 38\\
$\Dz\etapr(\pi\pi\eta(\gg))$ &  &  122 & $\pm$ & 13 & &  41 & $\pm$ & 3 & &    & - &  &&   & -&&  & 14 \\
$\Dz\etapr(\rho^0\gamma)$ &  &  234 & $\pm$ & 40 & &  1253 & $\pm$ & 17 & &  1 & $\pm$ & 0 &  & &-&& & 7.4 \\
$\Dstarz(\Dz\piz)\piz$  & &  883 & $\pm$ & 40 & &  268 & $\pm$ & 21 &  & 39 & $\pm$ & 2 & &  175 & $\pm$ & 5 & & 34 \\
$\Dstarz(\Dz\gamma)\piz$ &  &  622 & $\pm$ & 47 & &  469 & $\pm$ & 33 &  & 295 & $\pm$ & 23 &  & 602 & $\pm$ & 20 && 17 \\
$\Dstarz(\Dz\piz)\eta(\gg)$  & &  338 & $\pm$ & 25 & &  201 & $\pm$ & 9 &   &17 & $\pm$ & 1 & &  &-& & & 19\\
$\Dstarz(\Dz\gamma)\eta(\gg)$  & &  187 & $\pm$ & 24 &&   254 & $\pm$ & 12 & &  85 & $\pm$ & 11 &&   & -& && 8.7 \\
$\Dstarz(\Dz\piz)\eta(\pi\pi\piz)$  & &  123 & $\pm$ & 15 &  & 90 & $\pm$ & 4 & &  5 & $\pm$ & 1 &&   &-& && 11\\
$\Dstarz(\Dz\gamma)\eta(\pi\pi\piz)$ &  &  88 & $\pm$ & 14 &  & 65 & $\pm$ & 4 &  & 16 & $\pm$ & 3 & & & -& & & 7.6 \\
$\Dstarz(\Dz\piz)\omega$ &  &  806 & $\pm$ & 48 & &  1365 & $\pm$ & 18 & &  33 & $\pm$ & 2 & &  &-& && 20 \\
$\Dstarz(\Dz\gamma)\omega$ &  &  414 & $\pm$ & 44 & &  1290 & $\pm$ & 19 & &  132 & $\pm$ & 14 & & & -&& & 10 \\
$\Dstarz(\Dz\piz)\etapr(\pi\pi\eta)$ &  &  45 & $\pm$ & 8 &  & 18 & $\pm$ & 2 &&   2 & $\pm$ & 0 &  & &-& & & 8.5\\
$\Dstarz(\Dz\gamma)\etapr(\pi\pi\eta)$ &  &  12 & $\pm$ & 5 &  & 8 & $\pm$ & 1 & &  5 & $\pm$ & 2 & &   &-& & & 3.2\\
$\Dstarz(\Dz\piz)\etapr(\rho^0\gamma)$ &  &  115 & $\pm$ & 25 & &  487 & $\pm$ & 11 &&   3 & $\pm$ & 1 &&  & -& && 5.4\\
\hline\hline
\end{tabular}
\end{center}
}
\end{table*}

The fitting procedure is applied to data at the very last stage of the blind analysis.
Though the event yields and $\BF$  measurements are performed separately for each of the
72 considered sub-decay modes, we illustrate  here, in a compact manner,
the magnitude of the signal and background component yields,
and the statistical significances, of the various decay channels  $\Bzb\ra\Dstze\hz$,
summing  together all the  $\Dz$ sub-modes.
The fitted $\DeltaE$ distributions, for the sum of $\Dz$ sub-modes, are given
in Figs. \ref{fig:fitData_1}, \ref{fig:fitData_2}, and \ref{fig:fitData_3}, for, respectively, the $\Bzb\ra\Dz\hz$,
$\Dstarz(\ra\Dz\piz)\hz$, and $\Dstarz(\ra\Dz\g)\hz$ modes.

The signal and background yields obtained from the fit to the summed sub-mode
data for the $\Bzb\ra\Dstze\hz$ are presented  in Table~\ref{tab:nbEvtFitDATA}, with the corresponding statistical
significances. The  signal and background yields are
computed in the signal region $|\Delta E|<2.5\ \sigma$ (where $\sigma$ is the
signal resolution). In the same range we  calculate the statistical significance of the various signals
from the cumulative Poisson probability $p$ to have a background statistical fluctuation reaching the observed data yield,
\begin{equation}
p =  \sum_{k=N_{\rm cand}}^{+\infty} \frac{e^{-\nu}}{k!} \nu^{k},
\end{equation}
where $N_{\rm cand}$ is the total number of selected candidates in the signal region and
$ \nu$ the mean value of the total expected background, as
extracted from the fit.  This probability is then converted into
a number of equivalent one-sided standard deviations ${\cal S}_{\rm stat}$,
\begin{equation}
{\cal S}_{\rm stat}=\sqrt{2}~{\tt erfcInverse}(p/2).
\end{equation}
The function ${\tt erfcInverse}$ is the inverse of the complementary error function
  (see statistics review in~\cite{ref:PDG}).

The majority of the decay channels present clear and significant signals. In particular, the
modes $\Dz\etapr(\pi\pi\eta(\gg))$ and $\Dz\etapr(\rho^0\gamma)$ are
observed for the first time.

Before performing  the final unblinded fits on data, among the various 72 initial possible decay channels,
several sub-decay modes have been discarded. The decision to remove those sub-modes
have been taken according to analyses performed on MC simulation,  as no
significant signals are expected. The eliminated decay channels are:   $\Bzb\ra\Dstze\etapr$ and  $\Dstarz(\Dz\gamma)\eta(\pi\pi\piz)$,
where $\Dz\ra \KS\pipi$, $\Dstarz(\Dz\gamma)\etapr(\pi\pi\eta)$, where $\Dz\ra K^-\pi^+\pi^-\pi^+$,
as well as the whole decay channel $\Dstarz(\Dz\gamma)\etapr(\rho^0\gamma)$. These are sub-modes
with poor signal efficiency, caused by large track multiplicity or modest $\Dz$
secondary branching fractions, such that the expected signal yields are very low. In addition, they
have large  background contributions. We concluded that adding
such decay channels in the global combinations  would degrade  the $\BF$ measurements.
These choices based on a Monte Carlo simulation
 only  studies  have been confirmed in data (see for example Fig.~\ref{fig:fitData_3} (bottom right)).

\section{SYSTEMATIC UNCERTAINTIES ON BRANCHING FRACTIONS}
\label{sec:Systematics}

\begin{table*}[htb]
{\footnotesize
\caption{\label{tab:SystCombi3}Combined contributions to the
branching fraction $\BF(\Bzb\ra\Dstze\hz)$  relative systematic uncertainties ($\%$).}
\begin{center}
\begin{tabular}{ l c c c c c c c c c c }
\hline\hline \\
Sources &  \multicolumn{10}{c}{$\Delta \BF/\BF(\%)$ for the $\Bzb$ decay} \\
  & $\Dz\piz$ & $\Dz\eta(\gamma\gamma)$ & $\Dz\eta(\pi\pi\piz)$ & $\Dz\omega$ &
  $\Dz\etapr(\pi\pi\eta)$ &  $\Dz\etapr(\rho^0\gamma)$ & $\Dstarz\piz$ & $\Dstarz\eta$ & $\Dstarz\omega$ & $\Dstarz\etapr$ \\ \hline \\
$\piz/\gamma$ detection         & 3.5 & 3.5 & 3.6 & 3.6 & 3.7 & 2.3 & 6.2 & 5.5 & 5.7 & 5.8 \\
Tracking                        & 0.9 & 0.9 & 1.6 & 1.7 & 1.6 & 1.6 & 0.9 & 1.1 & 1.6 & 1.6 \\
Kaon ID                         & 1.0 & 1.1 & 1.1 & 1.1 & 1.2 & 1.1 & 1.1 & 1.1 & 1.1 & 1.2 \\
$\KS$ reconstruction            & 0.7 & 0.7 & 0.6 & 0.8 & 0.6 & 0.6 & 0.4 & 0.3 & 0.5 & - \\
Secondary $\BF$                 & 1.6 & 1.6 & 2.0 & 1.8 & 2.3 & 2.4 & 5.1 & 5.7 & 5.5 & 5.1 \\
$\BB$ counting                  & 1.1 & 1.1 & 1.1 & 1.1 & 1.1 & 1.1 & 1.1 & 1.1 & 1.1 & 1.1 \\
MC statistics                   & 0.1 & 0.2 & 0.3 & 0.2 & 0.4 & 0.4 & 0.2 & 0.2 & 0.3 & 0.3 \\
Particles selection            & 0.3 & 0.4 & 0.2 & 1.0 & 0.3 & 1.0 & 0.2 & 0.1 & 0.1 & 1.2 \\
$\Delta E$ fit                  & 2.1 & 2.2 & 1.3 & 2.1 & 1.1 & 2.1 & 1.0 & 0.6 & 1.4 & 0.5 \\
Combinatorial background      & 0.1 & 0.1 & 0.1 & 0.1 & 0.1 & 0.1 & 0.1 & 0.1 & 0.1 & 0.1 \\
$\Dstze\rho^-$ background       & 1.8 & - & - & - & - & - & 5.6 & - & - & - \\
$\Dstarz\omega$ polarization    & - & - & - & - & - & - & - & - & 1.4 & - \\
\hline \\
Total                           & 5.1 & 4.9 & 4.9 & 5.3 & 5.1 & 4.7 & 9.6 & 8.2 & 8.5 & 8.2 \\
\hline\hline
\end{tabular}
\end{center}
}
\end{table*}

\begin{table*}[htb]
{\footnotesize \caption{\label{tab:allBF}Branching fractions of the decay channels $\Bzb\ra\Dstze\hz$
measured in the different secondary decay modes. The first uncertainty is statistical
and the second is systematic. The cells with ``-" correspond to decay channels that  have been discarded after the analysis on simulation,
and confirmed with data, as no significant signal is expected or seen for them.}
\begin{center}
\begin{tabular}{ l cc  cc  cc   cc }
\hline\hline \\
$\BF(\Bzb\ra)$ ($\times 10^{-4}$) & & $\Dz\ra K\pi$ & & $\Dz\ra K3\pi$ & & $ \Dz\ra K\pi\piz$ & & $\Dz\ra \KS\pipi$ \\\hline \\
$\Dz\piz$ & & 2.49 $\pm$ 0.13 $\pm$ 0.16 & & 2.69 $\pm$ 0.15 $\pm$ 0.17 & & 2.97 $\pm$ 0.15  $\pm$ 0.25 & & 2.90 $\pm$ 0.28 $\pm$ 0.23  \\
$\Dz\eta(\gg)$ & & 2.46 $\pm$ 0.18 $\pm$ 0.14 & & 2.56 $\pm$ 0.19 $\pm$ 0.16 & & 2.37 $\pm$ 0.20  $\pm$ 0.20 & & 2.62 $\pm$ 0.37 $\pm$ 0.21  \\
$\Dz\eta(\pi\pi\piz)$ & & 2.59 $\pm$ 0.27 $\pm$ 0.12 & & 2.65 $\pm$ 0.30 $\pm$ 0.14 & & 2.48 $\pm$ 0.29  $\pm$ 0.20 & & 2.28 $\pm$ 0.54 $\pm$ 0.18  \\
$\Dz\omega$ & & 2.59 $\pm$ 0.18 $\pm$ 0.20 & & 2.34 $\pm$ 0.19 $\pm$ 0.15 & & 2.42 $\pm$ 0.20  $\pm$ 0.21 & & 3.17 $\pm$ 0.39 $\pm$ 0.24  \\
$\Dz\etapr(\pi\pi\eta(\gg))$ & & 1.40 $\pm$ 0.25 $\pm$ 0.07 & & 1.37 $\pm$ 0.26 $\pm$ 0.08 & & 1.34 $\pm$ 0.27  $\pm$ 0.11 & & 1.30 $\pm$ 0.50 $\pm$ 0.12  \\
$\Dz\etapr(\rho^0\gamma)$ &  & 1.58 $\pm$ 0.42 $\pm$ 0.09 & & 1.79 $\pm$ 0.57 $\pm$ 0.10 & & 1.91 $\pm$ 0.54  $\pm$ 0.15 & & 1.55 $\pm$ 0.89 $\pm$ 0.16  \\
$\Dstarz(\Dz\piz)\piz$ & & 2.95 $\pm$ 0.25 $\pm$ 0.30 & & 2.95 $\pm$ 0.29 $\pm$ 0.33 & & 3.52 $\pm$ 0.29  $\pm$ 0.43 & & 2.32 $\pm$ 0.56 $\pm$ 0.24  \\
$\Dstarz(\Dz\g)\piz$ & & 3.49 $\pm$ 0.40 $\pm$ 0.83 & & 2.25 $\pm$ 0.50 $\pm$ 0.63 & & 3.02 $\pm$ 0.50  $\pm$ 0.90 & & 3.53 $\pm$ 1.14 $\pm$ 0.99  \\
$\Dstarz(\Dz\piz)\eta(\gg)$ & & 2.52 $\pm$ 0.32 $\pm$ 0.26 & & 2.57 $\pm$ 0.33 $\pm$ 0.29 & & 2.41 $\pm$ 0.32  $\pm$ 0.32 & & 4.09 $\pm$ 0.74 $\pm$ 0.49  \\
$\Dstarz(\Dz\g)\eta(\gg)$ & & 2.62 $\pm$ 0.45 $\pm$ 0.33 & & 2.81 $\pm$ 0.49 $\pm$ 0.35 & & 2.87 $\pm$ 0.55  $\pm$ 0.39 & & 2.75 $\pm$ 0.78 $\pm$ 0.36  \\
$\Dstarz(\Dz\piz)\eta(\pi\pi\piz)$ & & 2.27 $\pm$ 0.50 $\pm$ 0.20 & & 2.60 $\pm$ 0.55 $\pm$ 0.24 & & 1.93 $\pm$ 0.46  $\pm$ 0.22 & & 1.21 $\pm$ 0.87 $\pm$ 0.13  \\
$\Dstarz(\Dz\g)\eta(\pi\pi\piz)$ & & 2.93 $\pm$ 0.71 $\pm$ 0.32 & & 2.55 $\pm$ 0.80 $\pm$ 0.29 &  & 1.94 $\pm$ 0.81  $\pm$ 0.24 & &- \\
$\Dstarz(\Dz\piz)\omega$ & & 5.07 $\pm$ 0.45 $\pm$ 0.47 & & 4.00 $\pm$ 0.49 $\pm$ 0.36 & & 4.38 $\pm$ 0.51  $\pm$ 0.51 & & 5.02 $\pm$ 0.98 $\pm$ 0.53  \\
$\Dstarz(\Dz\g)\omega$ & & 3.66 $\pm$ 0.64 $\pm$ 0.41 & & 4.46 $\pm$ 0.80 $\pm$ 0.56 & & 4.59 $\pm$ 0.87  $\pm$ 0.57 & & 4.28 $\pm$ 1.71 $\pm$ 0.57  \\
$\Dstarz(\Dz\piz)\etapr(\pi\pi\eta(\gg))$ & & 1.09 $\pm$ 0.38 $\pm$ 0.09 & & 1.67 $\pm$ 0.44 $\pm$ 0.15 & & 1.34 $\pm$ 0.49  $\pm$ 0.15 & &- \\
$\Dstarz(\Dz\g)\etapr(\pi\pi\eta(\gg))$ & & 0.75 $\pm$ 0.49 $\pm$ 0.24 & &-& & 1.19 $\pm$ 0.69  $\pm$ 0.39 & &- \\
$\Dstarz(\Dz\piz)\etapr(\rho^0\g)$ & & 2.10 $\pm$ 0.82 $\pm$ 0.23 & & 1.21 $\pm$ 0.90 $\pm$ 0.14 & & 1.45 $\pm$ 0.95  $\pm$ 0.18 & &- \\
$\Dstarz(\Dz\gamma)\etapr(\rho^0\g)$ &  &- & & - & & - & &- \\
\hline\hline
\end{tabular}
\end{center}
}
\end{table*}

There are several possible sources of systematic uncertainties
in this analysis, which  are summarized in Table~\ref{tab:SystCombi3}.

The categories ``$\piz/\gamma$ detection'' and ``Tracking'' account
respectively for the systematic uncertainties on the reconstruction of $\piz/\g$ and for charged particle
tracks, and are taken from  the efficiency
corrections computed in the studies of $\tau$ decays from $\epem\ra \tau^+ \tau^-$ events (see
Sec.~\ref{sec:MCcorrection}).

Similarly, the systematic uncertainties on kaon identification and on the reconstruction
of $\KS$ mesons are estimated from the MC efficiency corrections computed in the study
of pure samples of kaons and $\KS$ mesons compared to data
(see Sec.~\ref{sec:MCcorrection}).

The uncertainty on the secondary branching fraction results from our limited knowledge
of the $\Dstze$ and $\hz$  sub-mode branching ratios~\cite{ref:PDG} (including secondary decays into
detected stable particles). Correlations between the different decay channels were accounted for.

The uncertainty related to the number of $\BB$ pairs and
the limited available MC-sample statistics when computing the
efficiency of various selection criteria are also included.

Systematic uncertainties due to the intermediate particles mass selections are computed as the relative difference of
signal yield when the values of the mass means and mass resolutions are taken from a fit to the data.
Systematic effects from the $\qqbar$  ($q\ne b$) rejection and the $\Dstze$
selections are obtained from the study performed on the control
sample $\Bm\ra\Dstze\pim$ and are estimated as the limited confidence on
the efficiency correction ratio: ${\cal E}_{\rm rel.}({\rm data})/{\cal E}_{\rm rel.}({\rm MC})$,
including the correlations between the samples before and after selections (see
Sec.~\ref{sec:MCcorrection}). The effects of the cuts on $\rho^0$ and $\Dz\omega$ helicities are
obtained by varying the selection cut values by $\pm 10\%$ around the maximum of statistical significance.
All uncertainties on intermediate particle selections are combined into the category ``Particles selection''.

The uncertainty quoted for ``$\Delta E$ Fit'' gathers the changes due the limited knowledge on the shapes of signal and
background PDFs, and on the cross-feed branching fraction. For example, the modes $\Dstze\piz/\eta(\g\g)$  have high or low momentum
$\g$ in the final state. The difference between data and MC simulation in energy scale and  resolution
for neutrals is estimated from a study of the high statistics control sample $\Bm\ra\Dz(K^-\pi^+)\rho^-(\piz\pi^-)$,
which yields the difference between data and MC simulation, $\simeq 5.7~\mev$, for
 the mean and $\simeq 3.3~\mev$, for the resolution. This study was cross-checked against another one
performed with the high statistics and very pure control sample $\Bm\ra\Dz(K^-\pi^+\piz)\pi^-$.
With that control sample we find a difference between data and MC simulation of respectively  $\simeq 1.7~\mev$, for
 the mean and $\simeq 0.04~\mev$, for the resolution. For the modes $\Bzb\ra \Dstze\piz/\eta(\g\g)$,
 the uncertainty due to the signal shape is obtained by varying conservatively the signal PDF
mean by $\pm 5.7~\mev$ and the width by $\pm 3.3~\mev$.  For the other $\Bzb$ signal modes, each PDF parameter
is varied within  $\pm1 \ \sigma$ of its MC simulation precision, and the relative difference on the fitted event yield
is taken as  a systematic incertitude. The various parameters are varied one at a time, independently.
The relative differences while varying the $\DeltaE$ PDF parameters are
then summed up in quadrature. This sum is taken as the systematic doubt on  the $\DeltaE$ shapes.

The uncertainty on the combinatorial background shape (including both $\BB$  and $\qqbar$  ($q\ne b$)
events) is evaluated from the comparison of generic MC simulation and data in 
the $\mes$ sidebands:  $5.24<\mes<5.26~\gevcc$.
The observed difference between data and simulation has then been used as a systematics.
When a Gaussian is added to the combinatorial background shape, to model additional
peaking $\BB$ background contributions (see Sec.~\ref{PeakBBbgd}), the related effect
is computed by varying its means and resolution by $\pm 1 \ \sigma$.

We account for possible differences in the PDF shape of the $B^-\ra\Dstze\rho^-$
background that is modeled by a non-parametric PDF. As above, it is obtained by shifting
and smearing the PDF mean and resolution by $\pm5.7$~$\mev$ and $\pm 3.3$~$\mev$
respectively. The non-parametric PDF is  convoluted with  a Gaussian
with the previously defined mean and width values. The quadratic sum of the various
changes on the signal event yield is taken as the systematic uncertainty.

The relative ratio of the $B^-\ra\Dstarz\rho^-$ and $B^-\ra\Dz\rho^-$
backgrounds for the studies of the modes $\Bzb\ra\Dstze\piz$ has been
fixed to the selected $\B^-\ra\Dstze\rho^-$  events of the
 data control sample,  for rejected $B^-$ events with the veto described in
Sec.~\ref{bruit_BB}. The effect of such a veto on that ratio is then computed
from MC simulation. We assign as a conservative systematic
uncertainty half of the difference between the nominal result
and the result from the MC simulation assuming the
PDG branching ratios of $B^-\ra\Dstze\rho^-$~\cite{ref:PDG}.

The acceptance of $\Bzb\ra\Dstarz\omega$ is computed from the sum of purely
longitudinally  ($f_L=0$) and transversely ($f_L=1$) polarized MC simulation signals,
weighted by our measurement  of $f_L$ (see Sec.~\ref{se:fL}). The systematic limited
knowledge  of the efficiency due to the unknown fraction of $\Dstarz\omega$
longitudinal polarization is then estimated by varying $f_L$
by $\pm 1  \ \sigma$ in the estimation of the signal acceptance. This contribution
is slightly larger  than $1\%$, while it is expected to be about $10.5\%$
if the fraction $f_L$ were unknown. This is one of the motivations for measuring the polarization
of the decay channel $\Bzb\ra\Dstarz\omega$ (see Sec.~\ref{se:fL}).

The most significant sources of systematic uncertainties come from the $\piz/\g$
reconstruction, the $\Delta E$ fits,  and the uncertainties on the
world average branching fractions of the secondary decay channels.
In the case of the modes $\Bzb\ra\Dstze\piz$, the contributions from
$B^-\ra\Dstze\rho^-$ backgrounds are also not negligible.

\begin{table}[htb]
\caption{\label{tab:BFDataCombi}Branching fractions of decay channels $\Bzb\ra\Dstze\hz$, where
the branching fraction measured in each $\Dz$ modes are combined.  For the modes with $\hz=\eta, \ \etapr$,
we give the combination (comb.) of the branching fraction computed with each sub-modes of $\eta^{(')}$. The first
uncertainty is statistical and the second is systematics. The quality of the combination
is given through the value of $\chi^2/{\rm ndof}$, with the corresponding probability
($p$-value) given in parenthesis in percents.}
\begin{center}
\begin{tabular}{ l c c }
\hline\hline \\
$\Bzb$ mode &  $\BF(\times 10^{-4})$ & $\chi^2/{\rm ndof}$ \\
 & & ($p$-value $\%$) \\
 \hline \\
  $\Dz\piz$ &  2.69  $\pm$  0.09  $\pm$ 0.13 & 2.81/3 (42.2)\\
\hline \\
  $\Dz\eta(\gg)$ &  2.50  $\pm$  0.11 $\pm$ 0.12 & 0.45/3 (93.0)\\
  $\Dz\eta(\pi\pi\piz)$ &  2.56  $\pm$  0.16 $\pm$ 0.13 & 0.39/3 (94.2)\\
  $\Dz\eta$  (comb.) &  2.53  $\pm$  0.09 $\pm$ 0.11 & 0.95/7 (99.6)\\
\hline \\
  $\Dz\omega$ &  2.57  $\pm$  0.11 $\pm$ 0.14 & 3.19/3 (36.3)\\
\hline \\
  $\Dz\etapr(\pi\pi\eta(\gg))$ & 1.37  $\pm$  0.14 $\pm$ 0.07 & 0.05/3 (99.7)\\
  $\Dz\etapr(\rho^0\gamma)$ &  1.73  $\pm$  0.28 $\pm$ 0.08 & 0.27/3 (96.6)\\
  $\Dz\etapr$  (comb.) &  1.48 $\pm$  0.13 $\pm$ 0.07 & 1.55/7 (98.1)\\
\hline \\
  $\Dstarz\piz$ &  3.05 $\pm$ 0.14 $\pm$ 0.28 & 4.73/7 (69.3)\\
\hline \\
  $\Dstarz\eta(\gg)$ &        2.77  $\pm$  0.16 $\pm$ 0.25 & 4.20/7 (75.6)\\
  $\Dstarz\eta(\pi\pi\piz)$ &   2.40  $\pm$  0.25 $\pm$ 0.21 & 3.81/6 (70.2)\\
  $\Dstarz\eta$  (comb.) &               2.69  $\pm$  0.14 $\pm$ 0.23 & 10.48/14 (72.6)\\
  \hline \\
  $\Dstarz\omega$ &             4.55  $\pm$  0.24 $\pm$ 0.39 & 4.05/7 (77.4)\\
   \hline \\
  $\Dstarz\etapr(\pi\pi\eta(\gg))$ &      1.37  $\pm$  0.23 $\pm$ 0.13 & 2.30/4 (68.1)\\
  $\Dstarz(\Dz\piz)\etapr(\rho^0\gamma)$ &   1.81  $\pm$  0.42 $\pm$ 0.16 & 0.68/2 (71.2)\\
 $\Dstarz\etapr$  (comb.) &                          1.48  $\pm$  0.22 $\pm$ 0.13 & 3.78/7 (80.5)\\
\hline\hline
\end{tabular}
\end{center}
\end{table}

\begin{figure*}
\begin{center}
\includegraphics[width=0.315\linewidth,height=7cm]{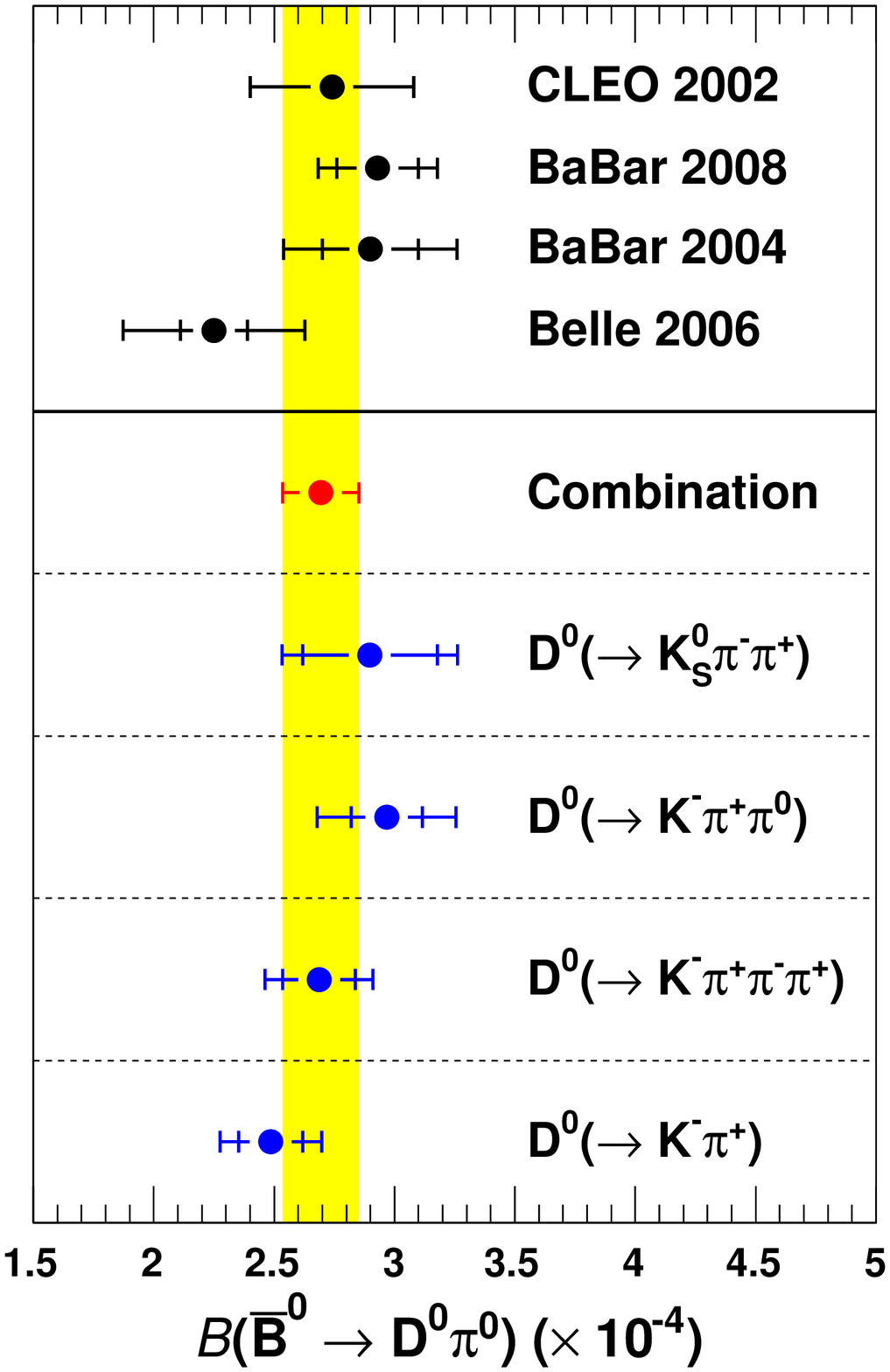}
\includegraphics[width=0.315\linewidth,height=7cm]{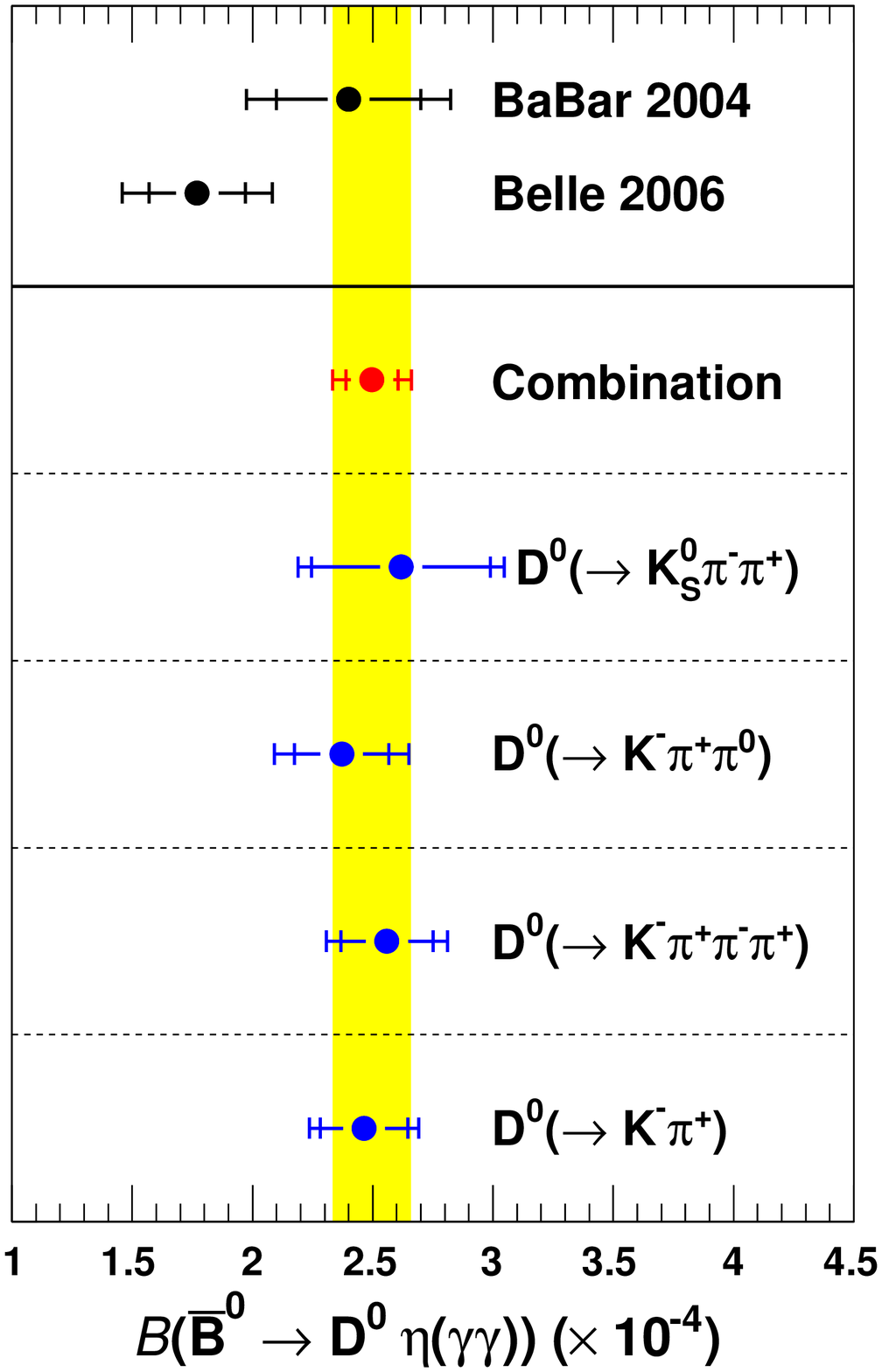}
\includegraphics[width=0.315\linewidth,height=7cm]{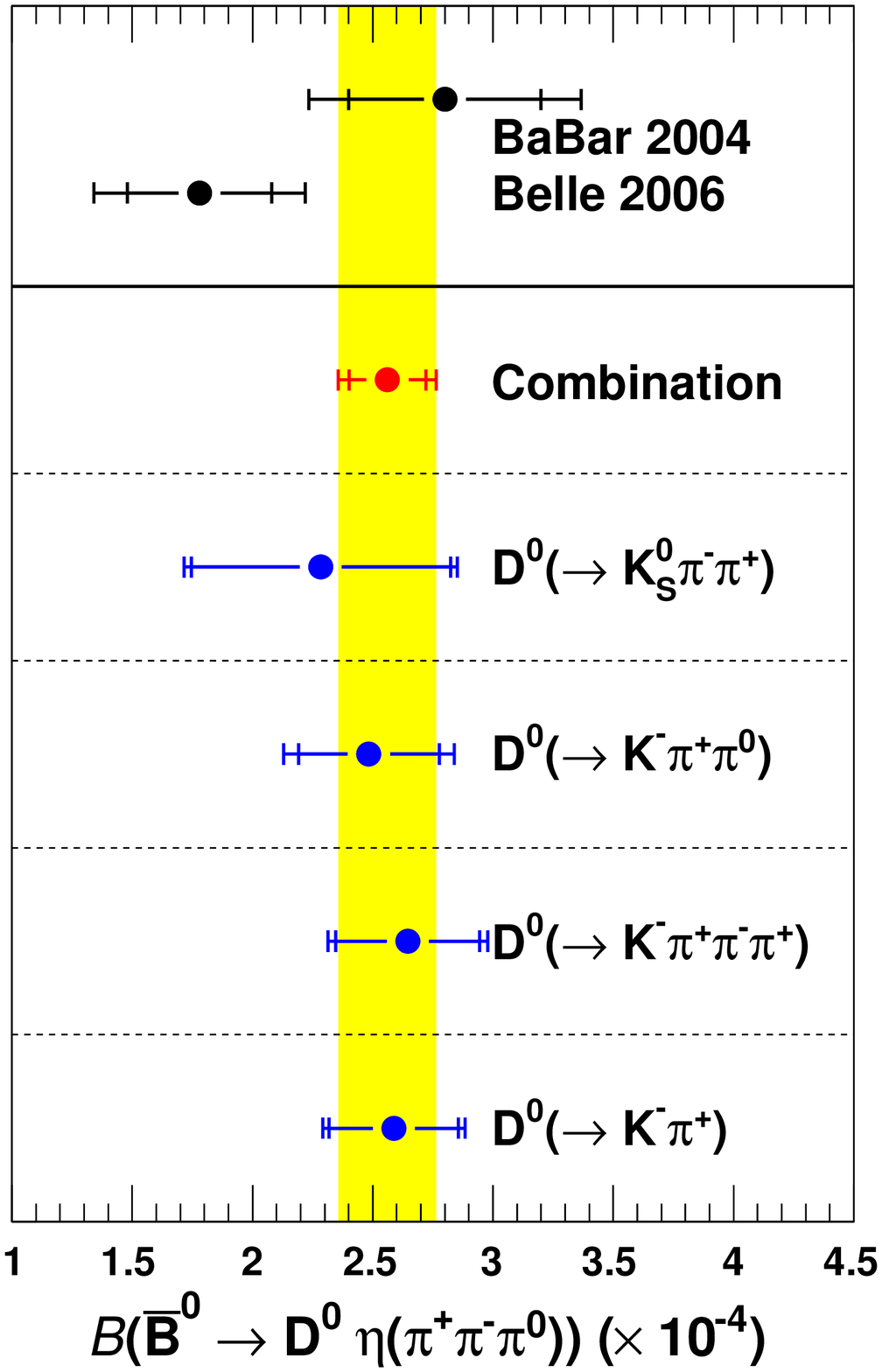} \\
\includegraphics[width=0.315\linewidth,height=7cm]{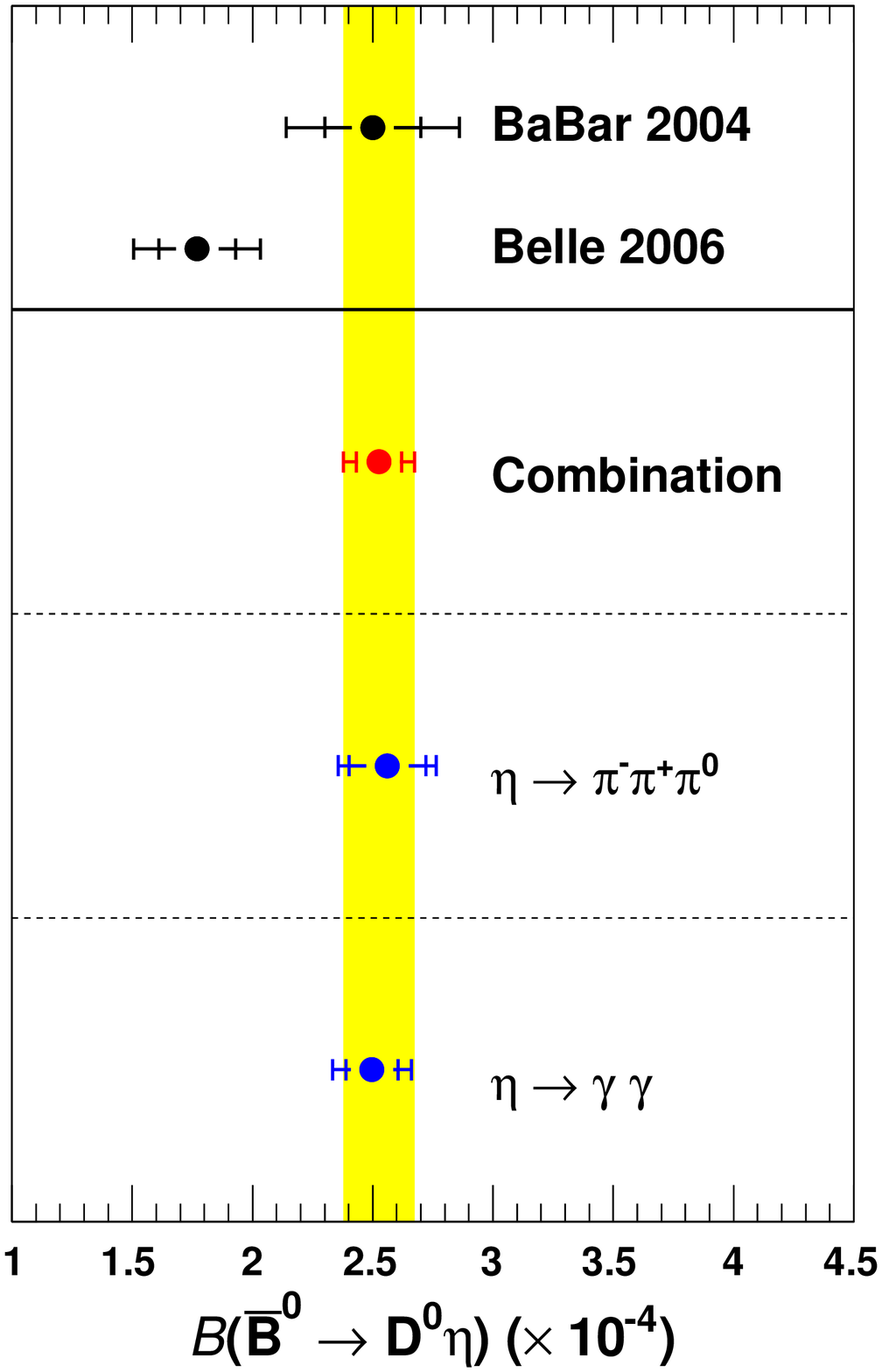}
\includegraphics[width=0.315\linewidth,height=7cm]{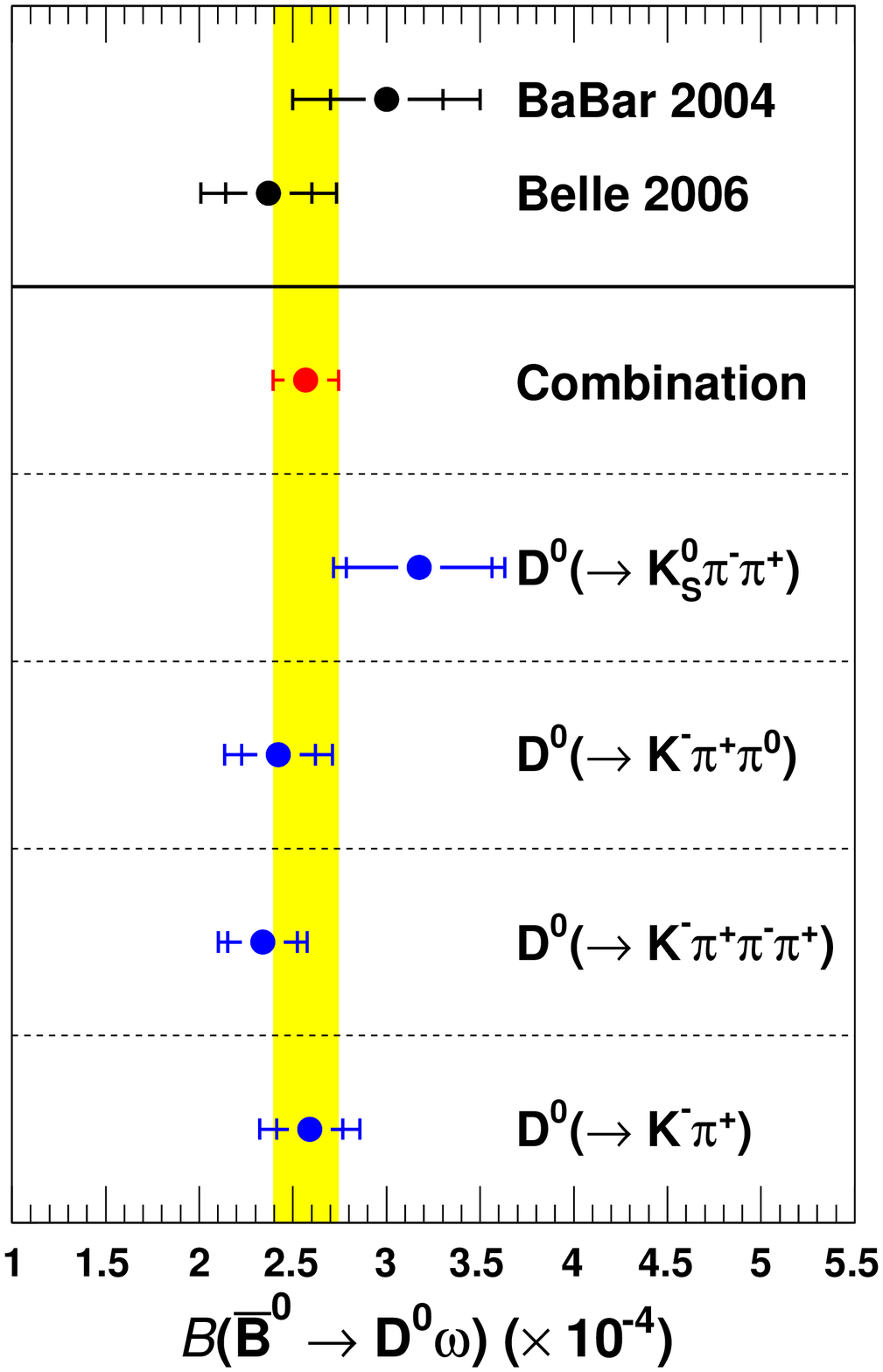}
\includegraphics[width=0.315\linewidth,height=7cm]{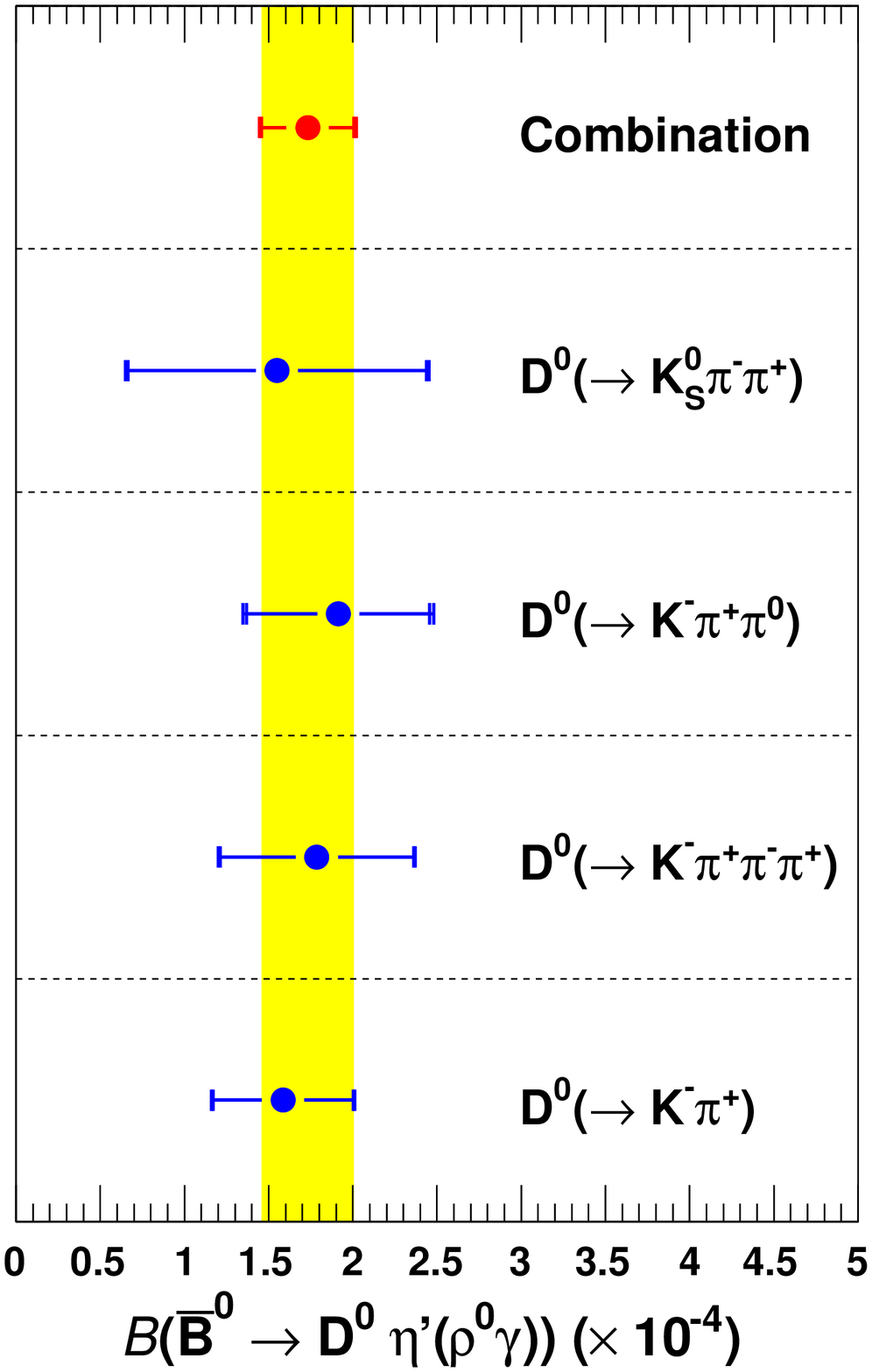} \\
\includegraphics[width=0.315\linewidth,height=7cm]{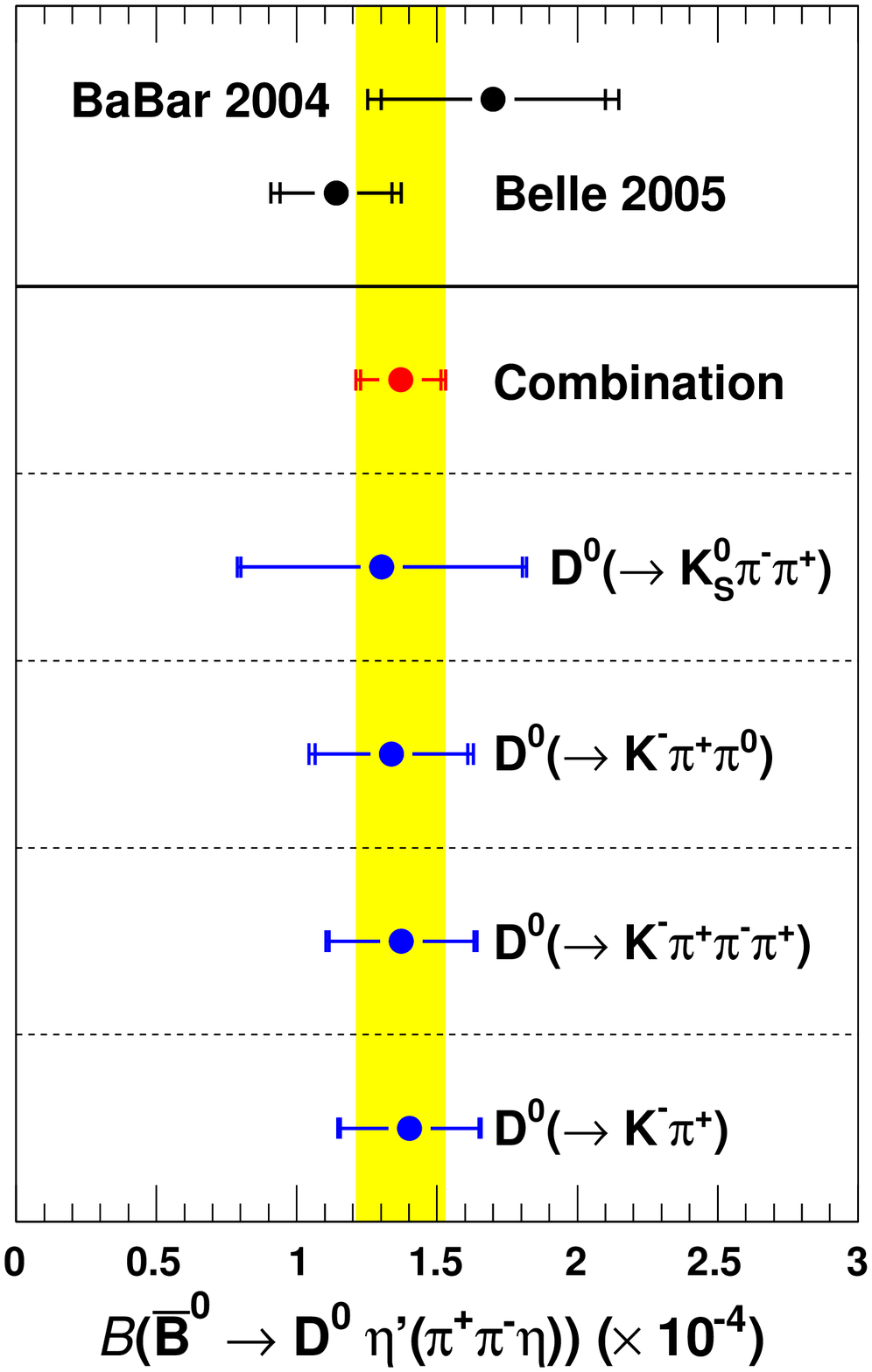}
\includegraphics[width=0.315\linewidth,height=7cm]{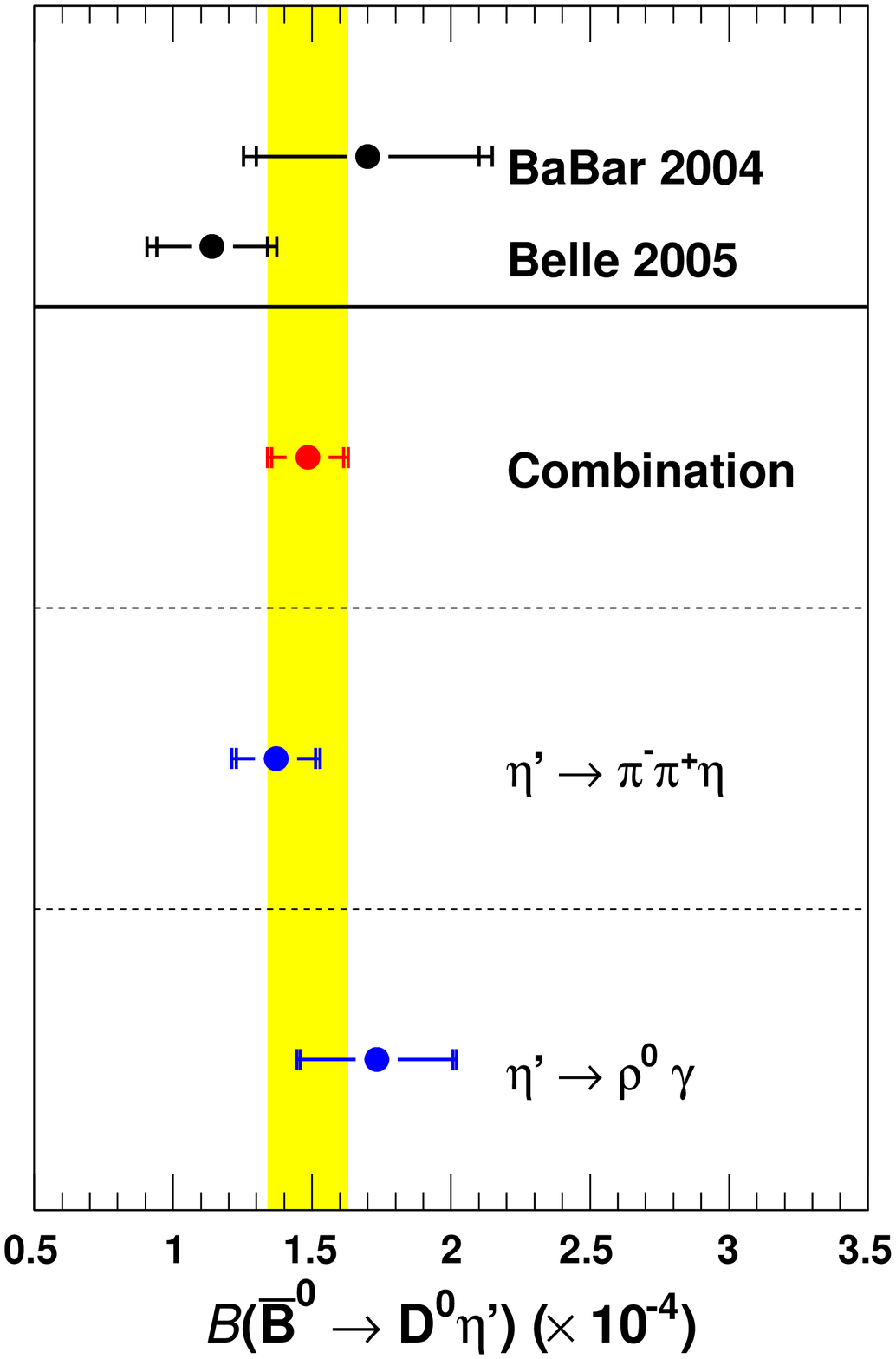}
\caption{\label{fig:ResumefitData_1}$\BF(\Bzb\ra \Dz\hz)$ ($\times10^{-4}$) for the individual
reconstructed $\Dz$ and $\hz$ decay channels (blue points) together with the {\it BLUE}
combination of this paper measurements (vertical yellow bands and the red points). The previous experimental results from
$\babar$~\cite{ref:Babar2004,ref:BabarKpipiz}, Belle~\cite{ref:Belle2005,ref:Belle2006}, and
CLEO~\cite{ref:Cleo2002} are also shown (black points). The horizontal bars represent the statistical
contribution alone and the quadratic sum of the  statistical and systematic uncertainty contributions. The width of the vertical yellow band
corresponds to $\pm1 \ \sigma$ of the combined measurement, where the  statistical and systematic uncertainties are summed in quadrature.}
\end{center}
\end{figure*}

\begin{figure*}
\begin{center}
\includegraphics[width=0.315\linewidth,height=7cm]{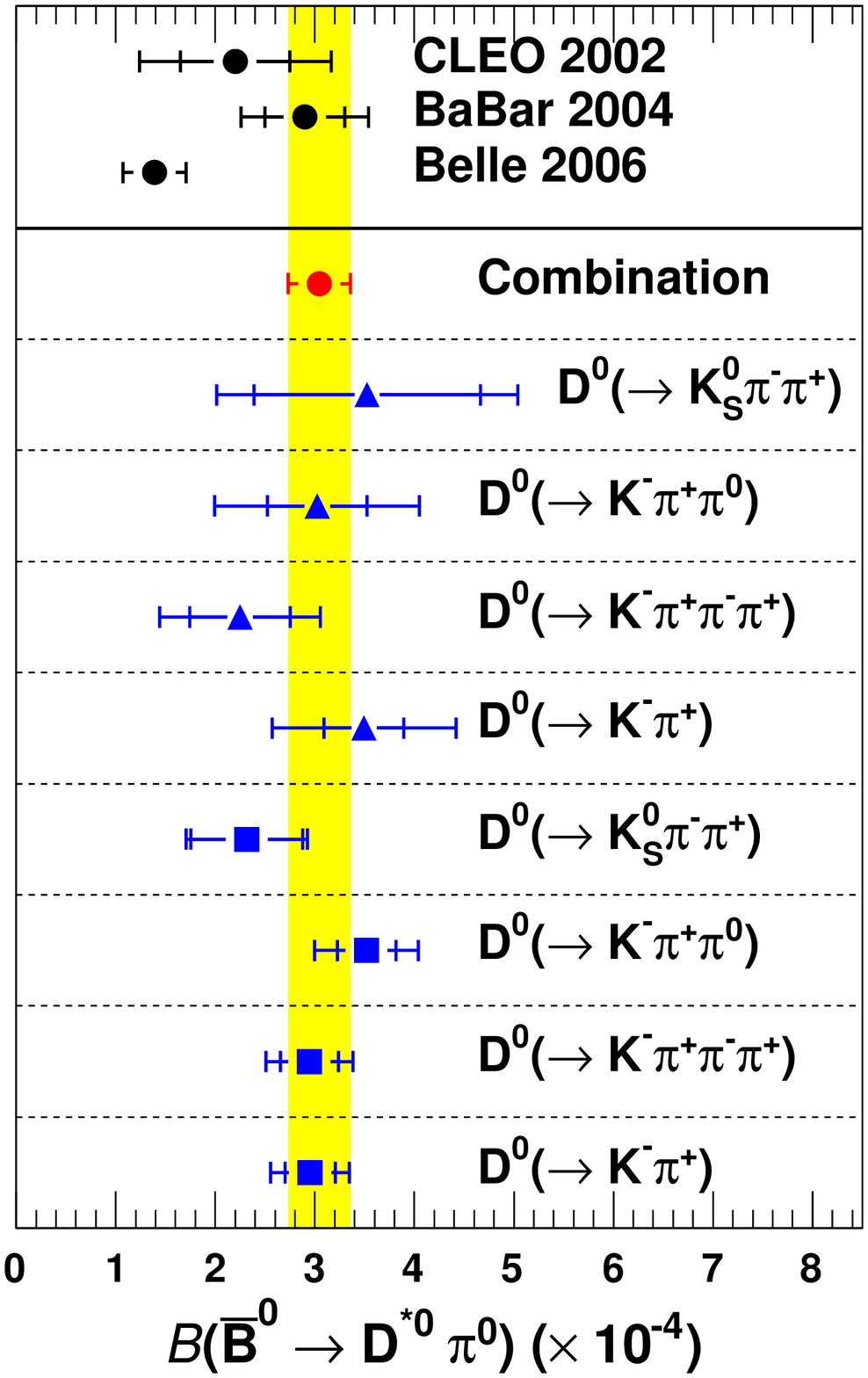}
\includegraphics[width=0.315\linewidth,height=7cm]{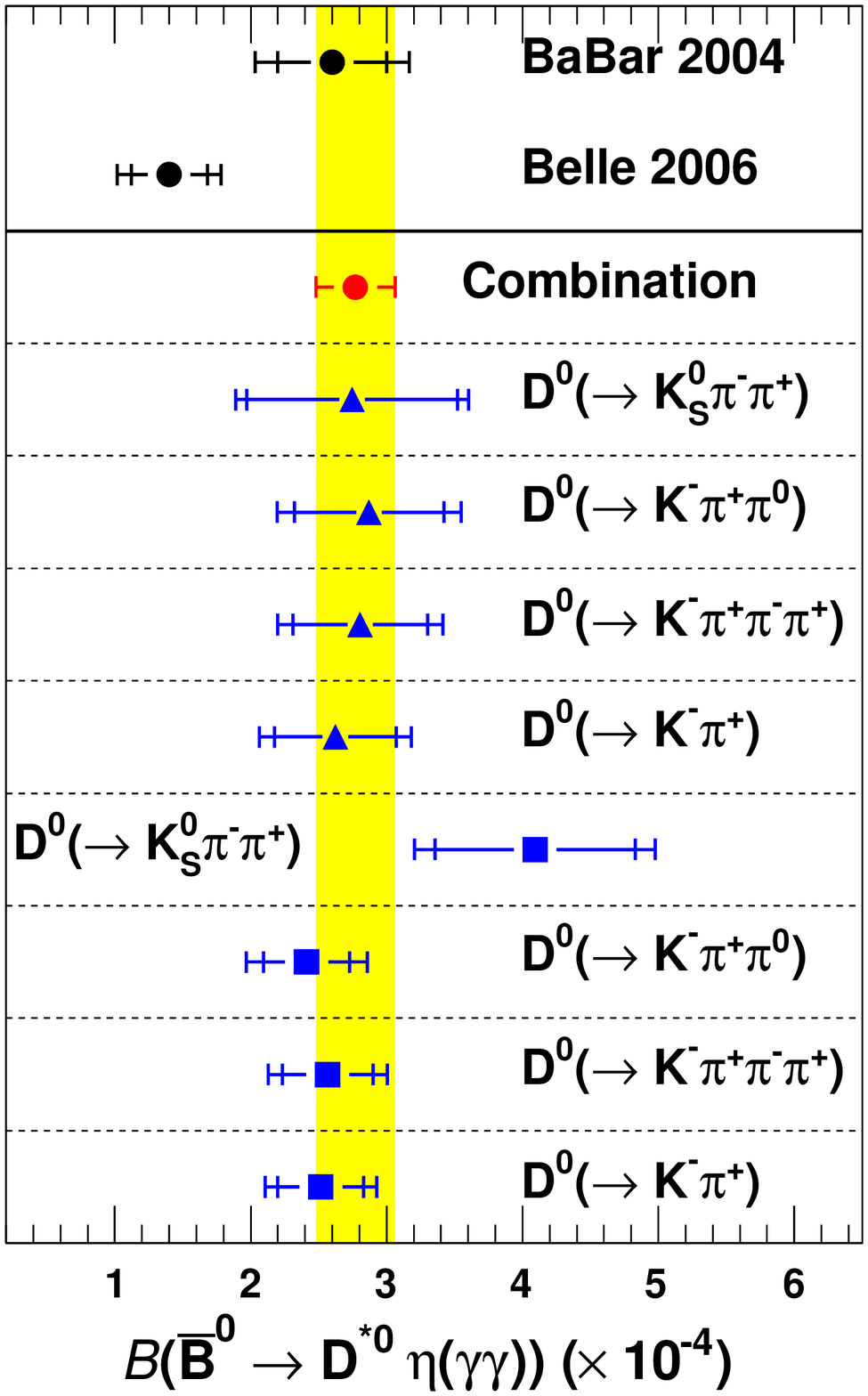}
\includegraphics[width=0.315\linewidth,height=7cm]{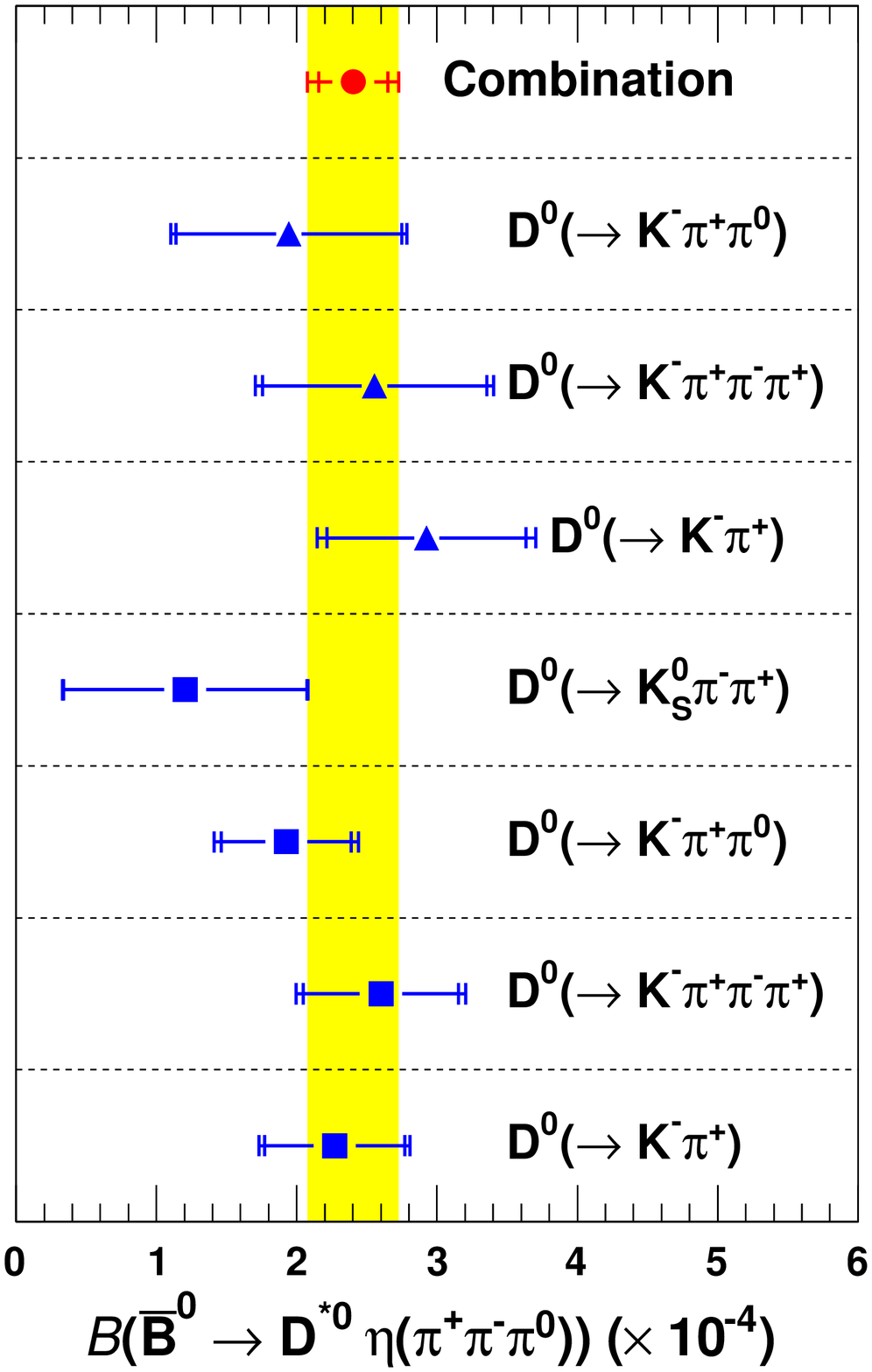}  \\
\includegraphics[width=0.315\linewidth,height=7cm]{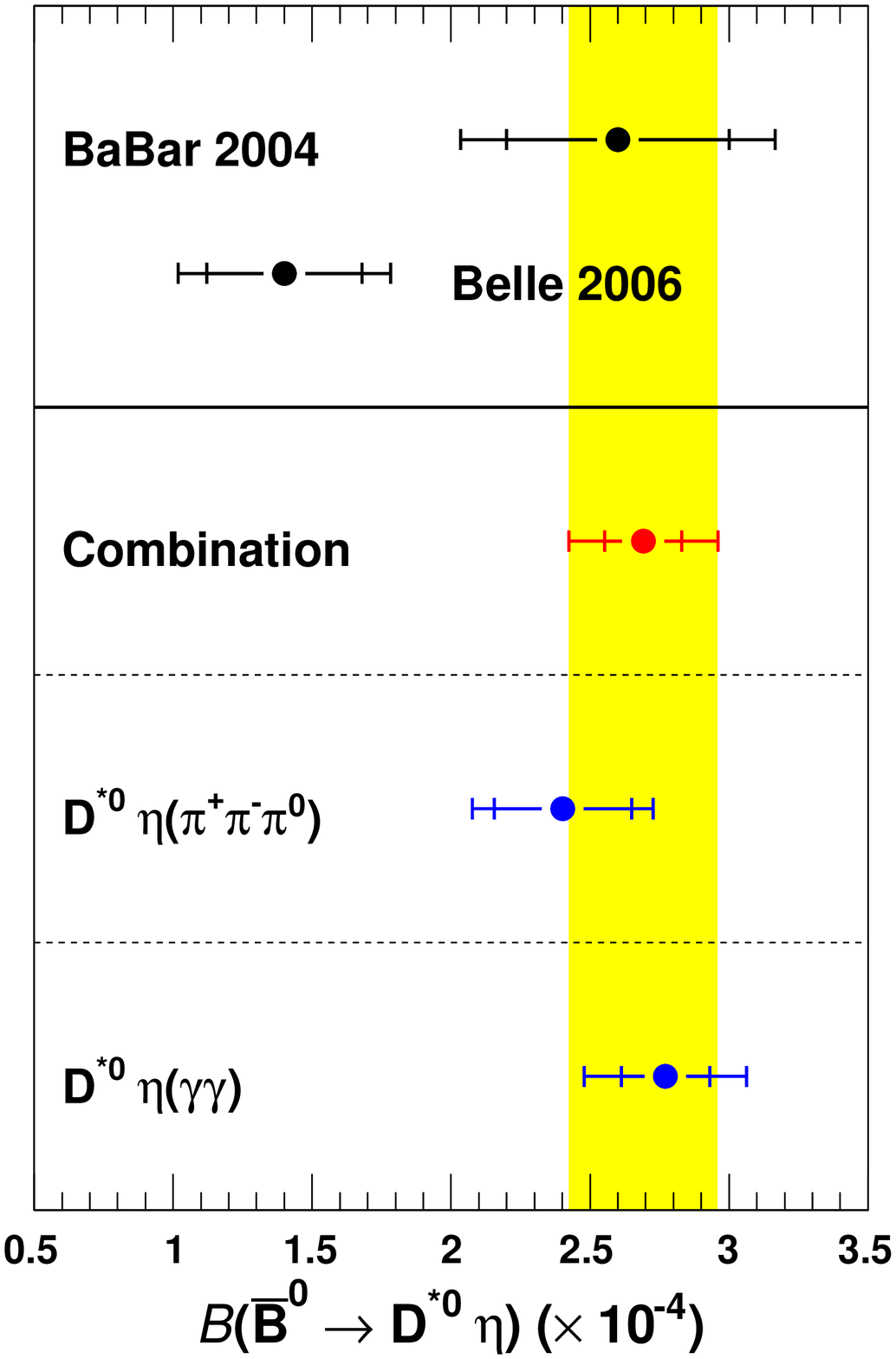}
\includegraphics[width=0.315\linewidth,height=7cm]{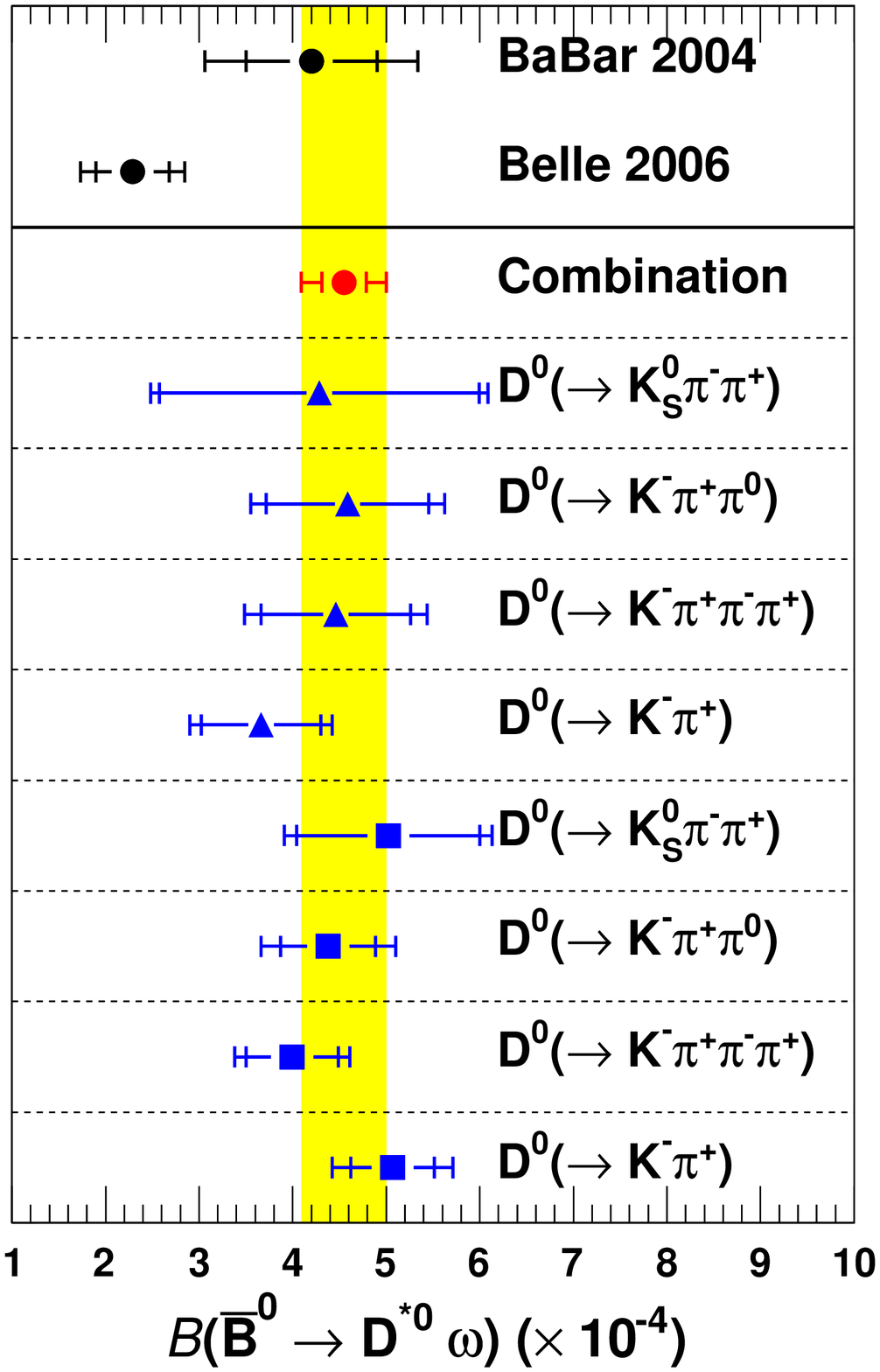}
\includegraphics[width=0.315\linewidth,height=7cm]{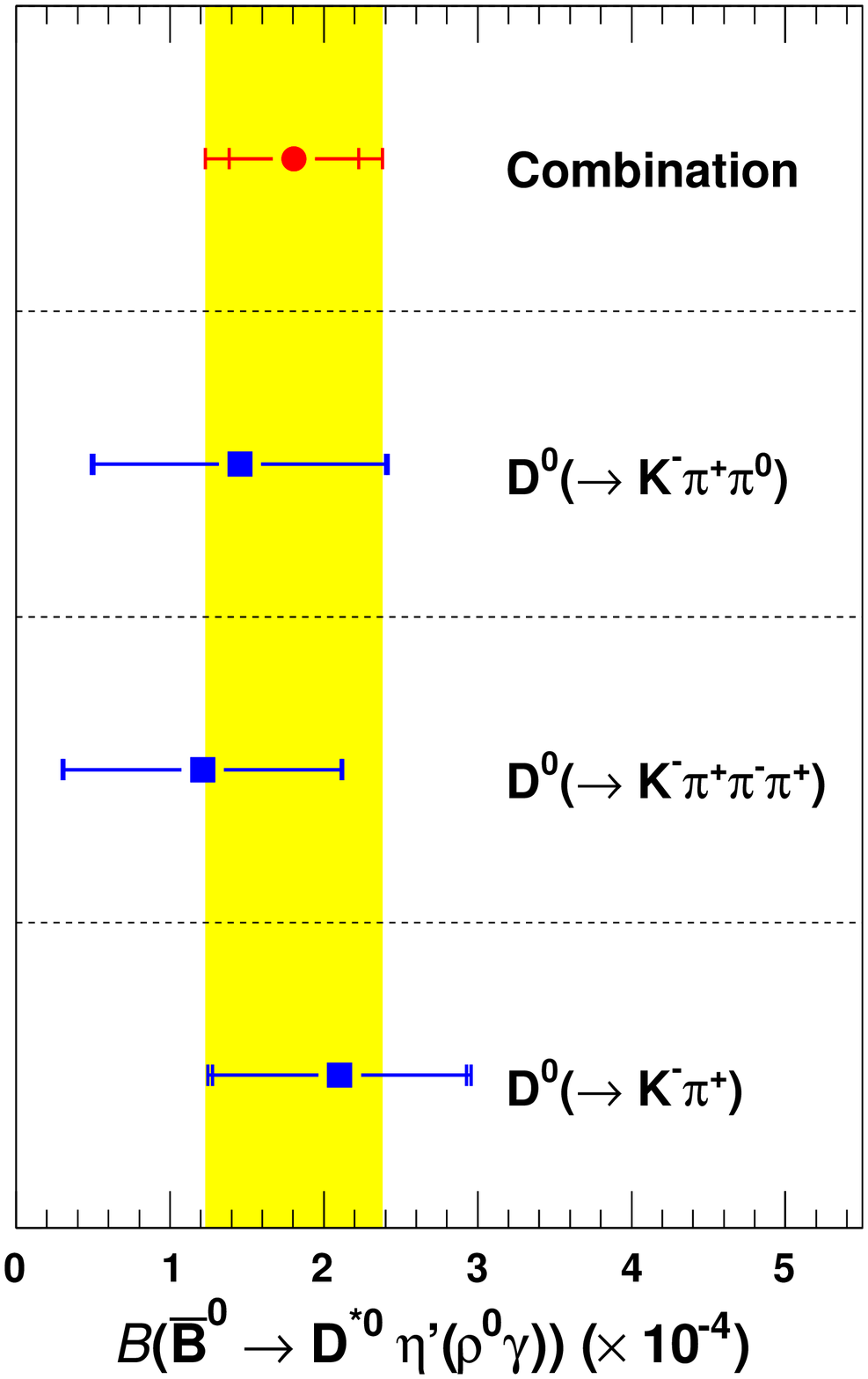} \\
\includegraphics[width=0.315\linewidth,height=7cm]{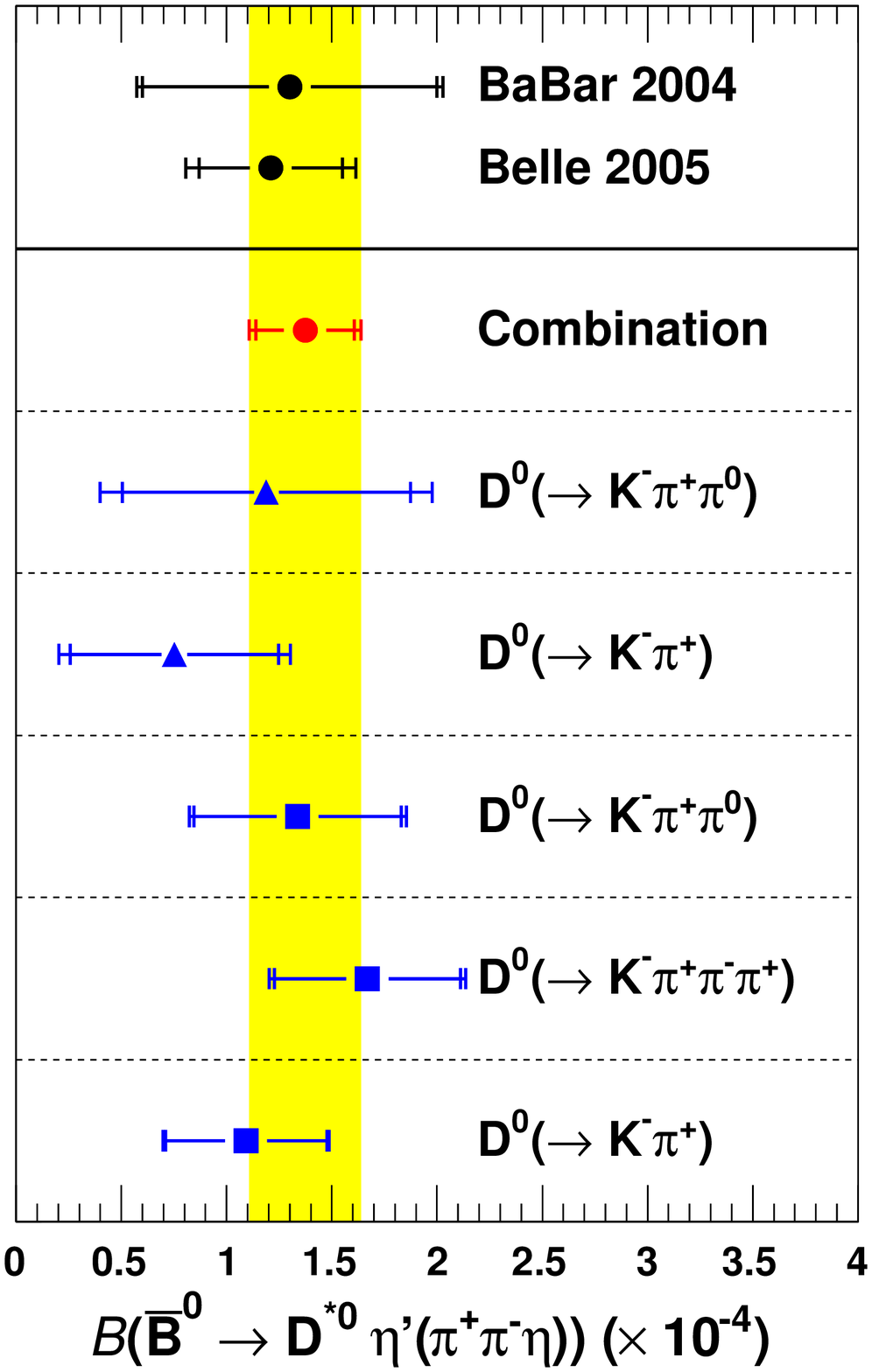}
\includegraphics[width=0.315\linewidth,height=7cm]{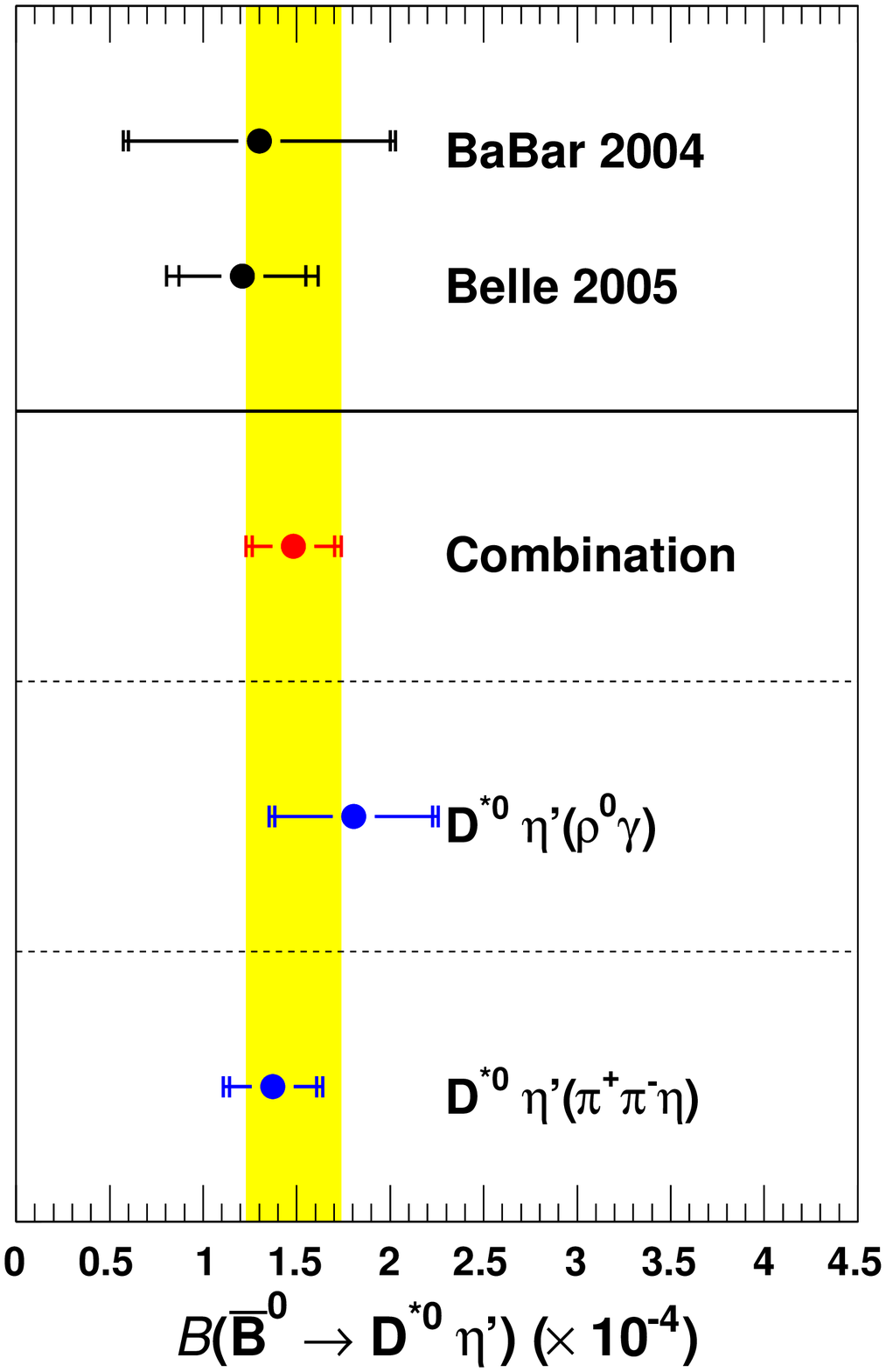}
\caption{\label{fig:ResumefitData_2}
$\BF(\Bzb\ra \Dstarz\hz)$ ($\times10^{-4}$) for the individual
reconstructed $\Dz$, $\Dstarz$, and $\hz$ decay channels together with the {\it BLUE}
combination of this paper measurements (vertical yellow bands and the red points). The blue squares (triangles) are for
measurements with the sub-decay $\Dstarz\ra\Dz\piz$ ($\Dz\g$). The previous experimental results from
$\babar$~\cite{ref:Babar2004,ref:BabarKpipiz}, Belle~\cite{ref:Belle2005,ref:Belle2006}, and
CLEO~\cite{ref:Cleo2002} are also shown (black points). The horizontal bars represent the statistical
contribution alone and the quadratic sum of the  statistical and systematic uncertainty contributions. The width of the vertical
yellow band corresponds to $\pm1 \ \sigma$ of the combined measurement, where the  statistical and systematic uncertainties are summed in quadrature.}
\end{center}
\end{figure*}

\section{RESULTS FOR THE $\BF$ MEASUREMENTS}\label{se:results}

The branching fractions measured in the different secondary decay channels reconstructed in this analysis are given in
Table~\ref{tab:allBF} (for missing entries in the Table; see the discussion on discarded
sub-modes in Sec.~\ref{DataDistAndYield}).

These branching fractions are combined using the so-called  Best Linear Unbiased Estimate ({\it BLUE}) technique~\cite{ref:BLUE},
that accounts for the correlation between the various modes. In the {\it BLUE} method the average value is a linear combination of the individual measurements,
\begin{equation}\label{eq:combi}
\BF = \sum_{i=1}^t(\alpha_i\cdot \BF_i),
\end{equation}
where each coefficient $\alpha_i$ is a constant weight, not necessarily positive, for a given measurement $\BF_i$. The condition $ \sum_{i=1}^t\alpha_i=1$
ensures that the method is unbiased. The set of coefficients $\alpha=(\alpha_1, \ \alpha_2, \ ..., \ \alpha_t)$  is calculated so that the variance of branching fraction is minimal,
\begin{equation}
\alpha = \frac{E^{-1}U}{U^T E^{-1}U},
\end{equation}
where $U$ is a $t$-component vector whose elements are all 1 ($U^T$ is its transpose) and $E$ is the  ($t \times t$) covariance matrix.
The variance of branching fraction is then given by
\begin{equation}\label{eq:erreur}
\sigma^2 = \alpha^T E \alpha.
\end{equation}
The covariance matrix $E$ is evaluated for each source of systematics. Its matrix elements are, for two modes $i$ and $j$,
\begin{equation}
E_{ij} = \rho_{ij}\sigma_i\sigma_j,
\end{equation}
where $\sigma_i$ and $\sigma_j$ are the systematic uncertainties for the modes $i$ and $j$, and $\rho_{ij}$ is
their correlation coefficient. We distinguish several types of systematic uncertainties according to their correlations between the modes:
\begin{itemize}
 \item full correlation, $|\rho_{ij}|\sim 1$: neutrals (but uncertainties for $\piz$ and single $\gamma$ are independent), PID, tracking, number of $B\Bbar$,
 $\BF(\Dstarz)$, $\Dstarz\omega$   polarization in that mode,
 \item medium correlation:  $\BF(\Dz)$, $\BF(\hz)$, whose correlations are taken from the PDG~\cite{ref:PDG} and range from
 $2\%$ to $100\%$, $\Dstze\rho^-$ background in $\Bzb\ra\Dstze\piz$,
\item negligible correlation, $|\rho_{ij}|\sim 0$: statistical uncertainties,
PDF systematics, selection of intermediate particles, MC statistics.
\end{itemize}
The total covariance matrix $E$ is then the sum of the covariance matrices for each source of uncertainty, plus the covariance matrix associated to statistical uncertainties. The systematic
(statistical) uncertainty on the combined value of branching fraction computed by using Eq.~(\ref{eq:erreur}) where the error matrix
includes only the systematic (statistical) uncertainties.

The combined branching fractions in data are given in  Table~\ref{tab:BFDataCombi} with the $\chi^2$ of the combination,
 the number of degrees of freedom of the combination (ndof), and the corresponding probability ($p$-value).
 The individual branching fractions together with the combined value are displayed in Figs.~\ref{fig:ResumefitData_1} and
 \ref{fig:ResumefitData_2} and they are compared to the previous measurements by CLEO~\cite{ref:Cleo2002},
 $\babar$~\cite{ref:Babar2004,ref:BabarKpipiz}, and Belle~\cite{ref:Belle2005,ref:Belle2006}.

The results of this analysis, based on a data sample of 454$\times 10^6~\BB$ pairs, are fully compatible with our previous
measurements~\cite{ref:Babar2004,ref:BabarKpipiz}, and also with those of CLEO~\cite{ref:Cleo2002}.
They are compatible with the measurements by Belle~\cite{ref:Belle2005,ref:Belle2006} for most of the modes,
except for $\Bzb\ra\Dstze\eta$, $\Dstarz\omega$, and $\Dstarz\piz$, where our results are larger. Those four branching fractions
are from 2.5 to 3.7 standard deviations (including systematic uncertainties) away from the latest measurements by Belle~\cite{ref:Belle2005,ref:Belle2006}.
Our measurements are the most precise determinations of the  $\BF(\Bzb\ra\Dstze\hz)$ from a single experiment.
They represent significant improvements with respect to the accuracy of the existing PDG averages~\cite{ref:PDG}.

As a cross check we also perform the branching fractions measurements with  the sub-data set
of 88$\times 10^6~\BB$ pairs that we previously studied~\cite{ref:Babar2004}. When studying the same
decay modes we find values compatible  with that of Ref.~\cite{ref:Babar2004} with both statistical and systematic uncertainties lowered by significant
amounts. In addition to the benefit from improved procedures to reconstruct and analyze the data, this updated analysis
incorporates new decay modes, higher signal efficiency, and better background rejection and modeling.
We use additional control data samples and measure directly in the data the
relative ratio of the $B^-\ra\Dstze\rho^-$ backgrounds. This analysis
employs better fitting techniques and uses more sophisticated methods to combine the results obtained with
the various sub-decay modes.

\section{Polarization of $\Bzb\ra\Dstarz\omega$}
\label{se:fL}

The polarization of the vector-vector ($VV$) decay $\Bzb\ra\Dstarz\omega$ has never been measured.  Until now,
 it was supposed to be similar to that of the decay  $\Bm\ra D^{*0}\rho^-$, based on
 HQET and factorization-based arguments~\cite{ref:CLEOhelicity2}. The angular distributions for the decay $\Bzb\ra\Dstarz\omega$ are described by three helicity amplitudes:
 the longitudinal $H_0$ amplitude and the transverse $H_+$ and $H_-$ amplitudes. In the factorization description of $B\ra VV$ decays, the longitudinal component
 $H_{0}$  is expected to be dominant, leading to the fraction of longitudinal polarization, defined as
\begin{equation}
f_L \equiv \frac{\Gamma_L}{\Gamma} =\frac{|H_0|^2}{|H_0|^2+|H_+|^2+|H_-|^2},
\end{equation}
predicted to be close to unity~\cite{ref:Koerner,ref:Williamson,ref:Neubert,ref:Verkerke}.

Significant transverse polarizations were measured in $B\ra\phi K^*$ (see the review in~\cite{ref:PDG}) and investigated as possible signs of New
Physics~\cite{ref:BPhiKstar}, but could also be the result of non-factorizable  QCD effects~\cite{ref:Cheng}. Similar effects were studied in the context
of SCET~\cite{ref:SCET_2}, and are expected to arise in the $\Bzb\ra\Dstarz\omega$ decay, in particular through enhanced electromagnetic penguin decays~\cite{ref:Beneke},
leading to significative deviation of $f_L$ from unity. It has also been argued in SCET studies that non-trivial long distance contributions
to the $\Bzb\ra\Dstarz\omega$ amplitude may allow a significant amount of transverse polarization of similar size to the longitudinal polarization,
leading to a value $f_L\sim0.5$.

Apart from the motivation of these phenomenological questions, the uncertainty on the angular polarization of $\Bzb \ra \Dstarz\omega$
affects the kinematic acceptance of this decay channel  and therefore would be the dominant contribution to the systematic effects for its $\BF$ measurement.
Hence we measure the fraction of longitudinal polarization for this decay mode. The analysis is performed with $\Bzb\ra\Dstarz\omega$ candidates
selected with the same requirements as for the $\BF$ analysis described in the previous sections. We consider the sub-decays $\Dstarz\ra\Dz\piz$
and $\Dz\ra\Km\pip$, $\Km\pip\piz$, $\Km\pip\pim\pip$, and $\KS\pip\pim$.

\subsection{Description of the method}
\label{sec:DiscussMeth}

The differential decay rate of $\Bzb\ra\Dstarz\omega$ for the sub-decay $\Dstarz\ra\Dz\piz$ is~\cite{ref:Kramer}
\begin{widetext}
\begin{equation}
\begin{array}{rcl}
\label{eq:diffRate}
\frac{\displaystyle{d^3\Gamma}}{\displaystyle{d\cos\theta_{D^*} d\cos\theta_{\omega} d\chi}} & \propto &
4|H_0|^2\cos^2\theta_{D^*}\cos^2\theta_{\omega}+ \\
&  & \left[ |H_+|^2 + |H_-|^2 + 2( Re(H_+H_-^*)\cos2\chi -  Im(H_+H_-^*)\sin2\chi) \right] \sin^2\theta_{D^*}\sin^2\theta_{\omega}+\\
& & ( Re(H_+H_0^*+H_-H_0^*)\cos\chi  -  Im( H_+H_0^*-H_-H_0^* )\sin\chi)\sin2\theta_{D^*}\sin2\theta_{\omega},
\end{array}
\end{equation}
\end{widetext}
where $\theta_{D^*}$ ($\theta_{\omega}$) is the helicity angle of the $D^*$ ($\omega$) meson
(see Sec.~\ref{bruit_BB} for definitions). The angle $\chi$, called the azimuthal angle, is the angle between
the $\Dstarz$ and $\omega$ decay planes in the $\Bzb$ frame. Since the acceptance is nearly independent of $\chi$,
one can integrate over $\chi$ to obtain a simplified expression:

\begin{eqnarray}
\label{eq:integChi}
\frac{\displaystyle{d^2\Gamma}}{\displaystyle{d\cos\theta_{D^*} d\cos\theta_{\omega}}} \propto
4|H_0|^2\cos^2\theta_{D^*}\cos^2\theta_{\omega} \\
+( |H_+|^2 + |H_-|^2 )\sin^2\theta_{D^*}\sin^2\theta_{\omega}. \nonumber
\end{eqnarray}
This differential decay width is proportional to
\begin{equation}
\label{eq:integChi2}
4 f_L \cos^2\theta_{D^*}\cos^2\theta_{\omega} + (1-f_L)\sin^2\theta_{D^*}\sin^2\theta_{\omega},
\end{equation}
which is the weighted sum of purely longitudinal ($f_L=1$) and purely transverse ($f_L=0$) contributions.

We employ high statistics MC simulations of exclusive signal samples of $\Bzb\ra\Dstarz\omega$ decays with the two extreme configurations
$f_L=0$ and $1$ to estimate the ratio of signal acceptance,  $\varepsilon_{0}/\varepsilon_{1}$, of $f_L=0$ events to $f_L=1$ events.
The longitudinal fraction $f_L$, can be expressed in terms of the fraction of background events ($\gamma$)
and the fraction of $f_L=1$ events in the observed data sample ($\alpha$):
\begin{equation}\label{eq:secondDef_fL}
f_L =\frac{\displaystyle{\alpha}}{\displaystyle{\alpha+(1-\alpha-\gamma)\cdot\frac{\varepsilon_{0}}{\varepsilon_{1}}}}.
\end{equation}
The fraction $\gamma$ is taken from the fit of $\Delta E$ for a signal region $|\Delta E|<2.5 \ \sigma_{\Delta E}$ and $\mes>5.27~\gevcc$, where
$\sigma_{\Delta E}$ is the fitted $\Delta E$ width of the signal distribution, ranging from 20.8 to 23.3$\mev$ depending on the mode. The fraction
$\alpha$ is determined from a simultaneous 2-dimensional fit to the distributions of the helicity angles $\cos\theta_{\omega}$ and $\cos\theta_{D^*}$, for
$\Bzb\ra\Dstarz\omega$ candidates selected in the same signal region. The correlation between $\cos\theta_{\omega}$ and $\cos\theta_{D^*}$ is found to be negligible.

The signal shapes are described with parabolas (see Eq.~(\ref{eq:integChi2})), except for the $\cos\theta_{\omega}$ distribution of $f_L=0$ signal
events, which is described by a non-parametric PDF based on the MC simulation. As the signal distribution of $\cos\theta_{\omega}$
is  distorted around zero because of the selection cut on pion momentum and on the $\omega$ boost (see Sec.~\ref{omegaselection}). The signal PDF parameters are fixed
 to those fitted on the $\Dstarz\omega$ simulations. The shape of the $\cos\theta_{\omega}$ and $\cos\theta_{D^*}$ background distributions
 is taken from the data sideband $|\Delta E|<280~\mev$ and $5.235<\mes<5.270~\gevcc$. The consistency of the background shape was checked
 and validated for various regions of the sidebands in data and generic MC simulations.
Possible biases on $f_L$ from the fit are investigated with pseudo-experiment studies for various values of $f_L$ from 5 to 95 \%.
No significant biases are observed.

An additional  study is performed with an embedded signal MC simulation, {\it i.e.} with signal events
modeled from various different fully simulated signal samples and with a generated value
$f_L\simeq90\%$ (as expected from HQET~\cite{ref:CLEOhelicity2,ref:HQET}).
A small bias  on the fitted value of $f_L$ is observed ($\sim 14 \%$ of the statistical uncertainty). This bias
is due to the slight difference on the description of the signal shape  for the $\cos\theta_{\omega}$ distribution and for $f_L=0$,
modeled by a  non-parametric PDF, and that of the actual shape obtained from the embedded signal MC simulation.
This bias is corrected later on and we assign a systematic uncertainty on that correction.

\begin{figure*}[htb]
\begin{center}
\includegraphics[width=0.475\linewidth]{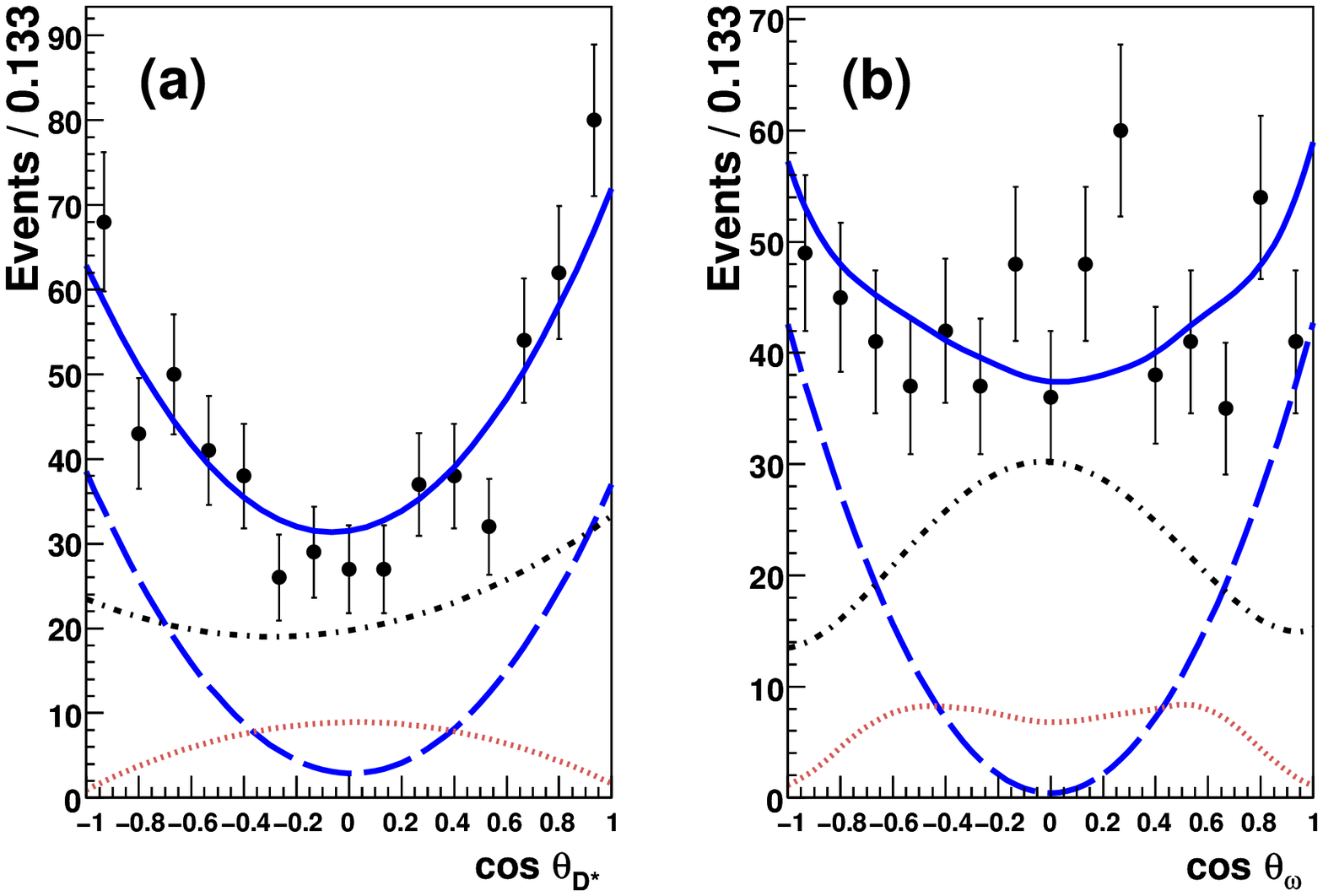}
\includegraphics[width=0.475\linewidth]{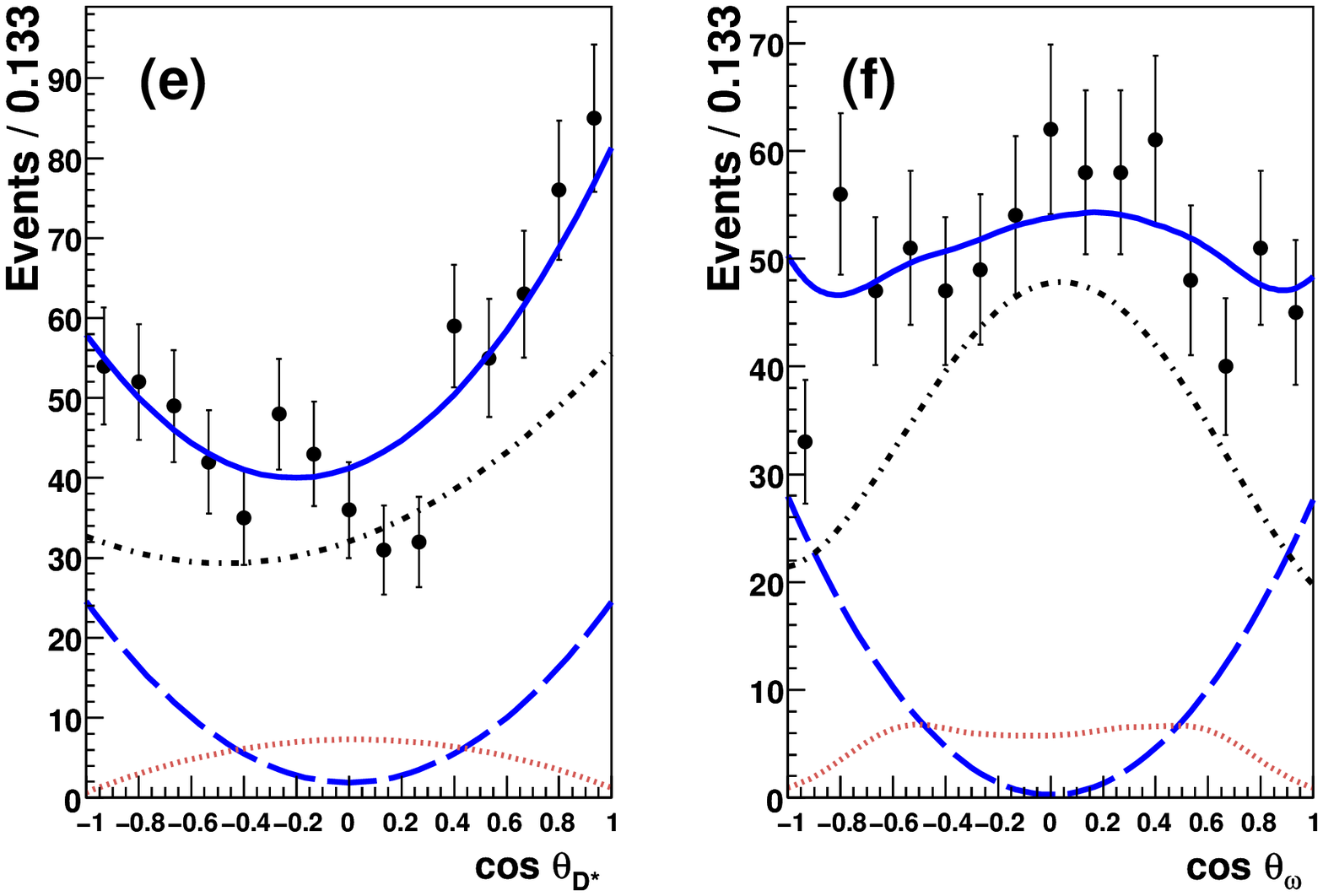}\\
\includegraphics[width=0.475\linewidth]{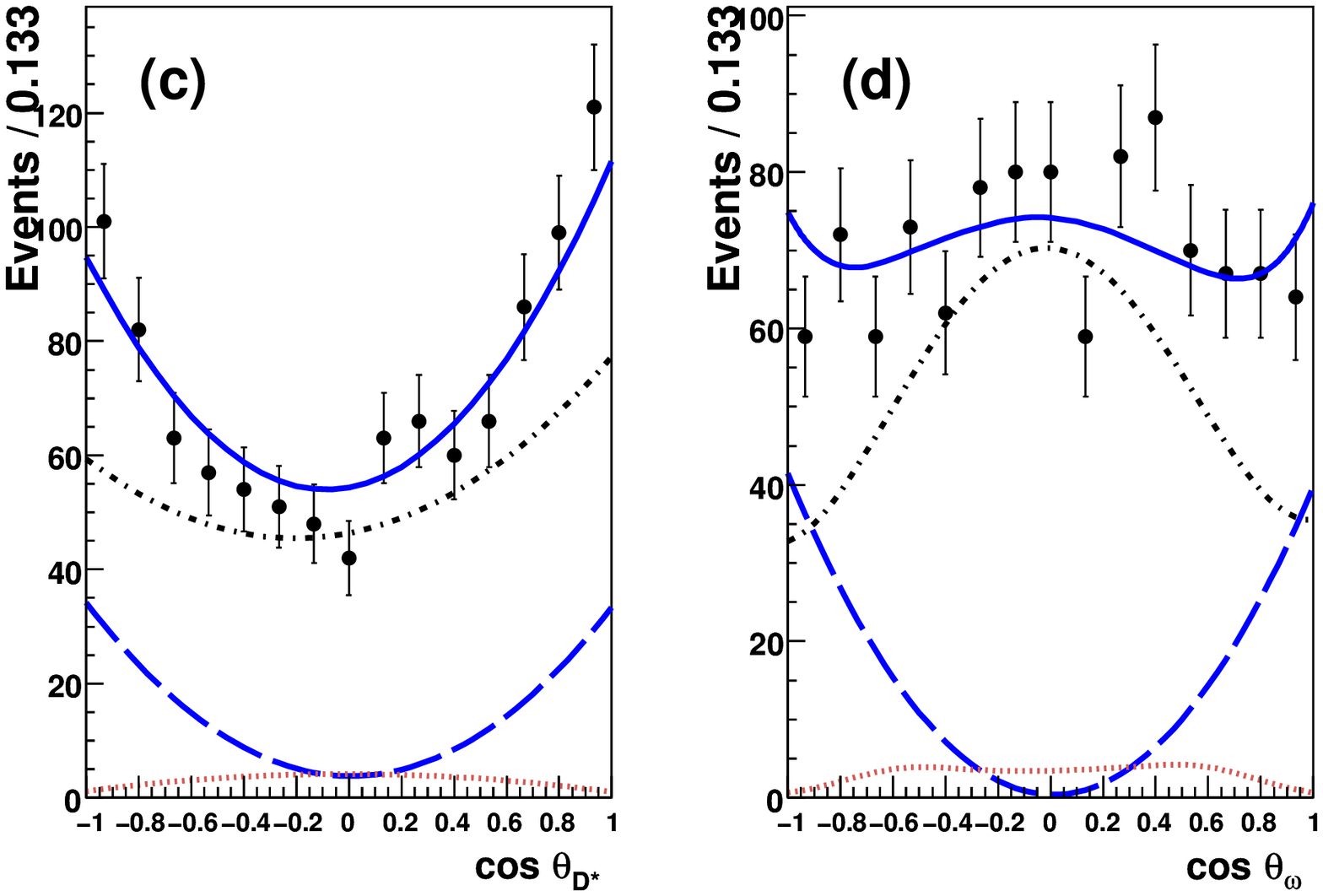}
\includegraphics[width=0.475\linewidth]{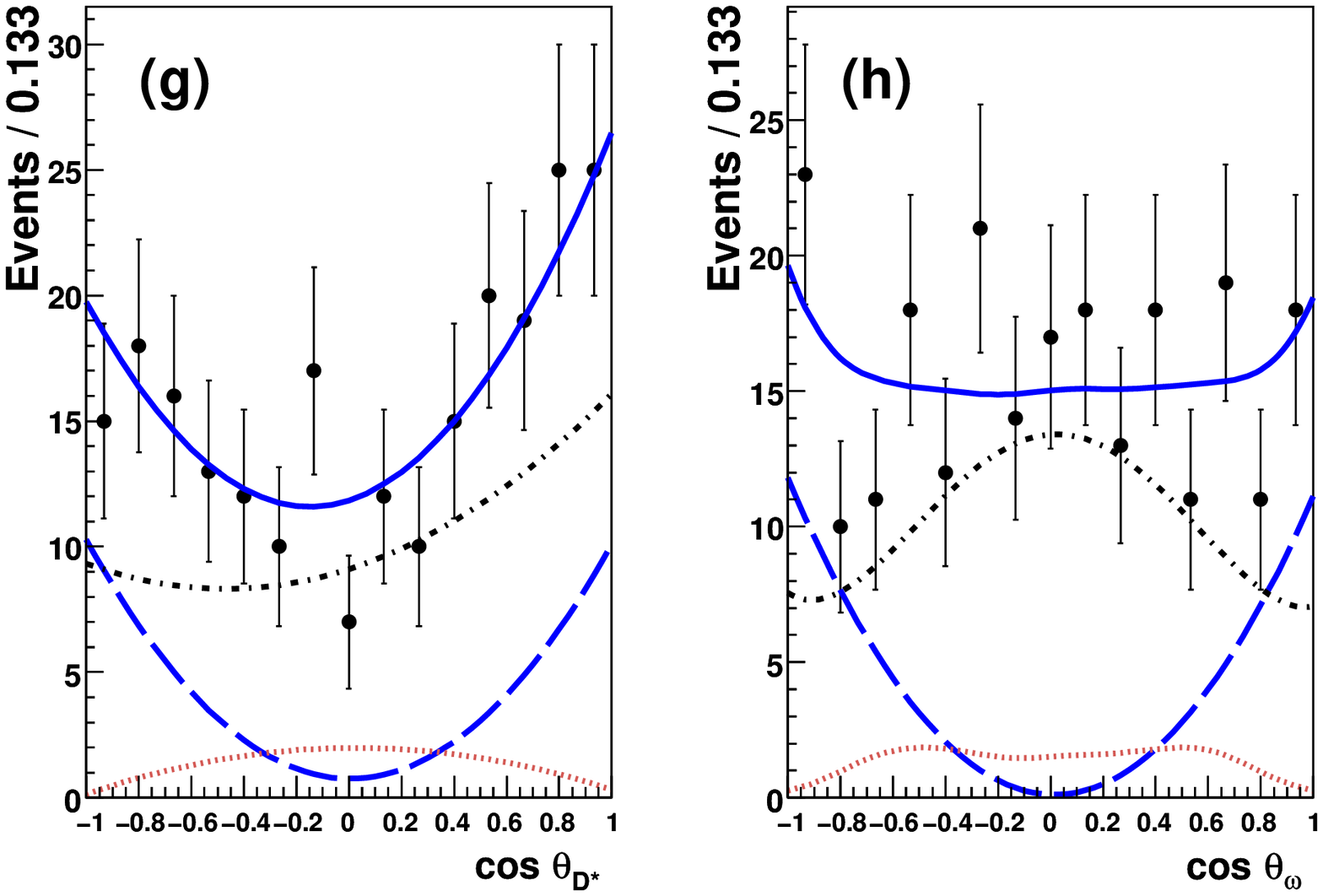}
\caption{Fitted distributions of the helicity $\cos\theta_{D^*}$  and $\cos\theta_{\omega}$ in the decay channel
$\Bzb\ra\Dstarz\omega$ for the $\Dz$ decay modes $K^-\pip$ (a) and (b),  $K^-\pip\piz$ (c) and (d), $K^-\pip\pim\pip$ (e) and (f),and $\KS\pip\pim$ (g) and (f).
The dots with error bars are data, the curves are the various PDF contributions: the solid blue (blue) is the total PDF, the dash-dot (grey) is the background contribution, the
long dash (blue) is the $f_L=1$ signal part and the dots (red) is the $f_L=0$ signal.}\label{fig:fit_cos_data}
\end{center}
\end{figure*}
\begin{table}[htb]
\caption{\label{tab:uncertainties_data}Total relative uncertainties computed
in data on the measurement of $f_L$ in the decay channel $\Bzb\ra\Dstarz\omega$,
with $\Dstarz\ra\Dz\piz$ and $\Dz\ra\Km\pip$, $\Km\pip\piz$, $\Km\pip\pim\pip$, and $\KS\pip\pim$.}
\begin{center}
\begin{tabular}{ l c c c c c }
\hline\hline \\
Sources       & &  \multicolumn{4}{c}{$\Delta f_L/f_L$ ($\%$)} \\
              & &  $K\pi$  &  $K3\pi$  &  $K\pi\piz$  &  $\KS\pi\pi$ \\
\hline \\
Signal PDFs                         & & 2.5 & 2.9 & 2.4 & 2.3\\
Correction of the bias         & & 1.0 & 1.3 & 2.0 & 2.3\\
Background PDF                    &  & 0.3 & 4.2 & 3.6 & 4.0\\
Limited MC statistics               & & 0.1 & 0.2 & 0.3 & 0.3\\
Flat acceptance vs. $\chi$         & & 1.5 & 1.8 & 0.5 & 6.9\\
\hline \\
Total systematic                         & & 3.1 & 5.6 & 4.8 & 8.6\\
\hline \\
Statistical uncertainty & & 9.6 & 16.3 & 16.3 & 25.6\\
\hline \\
Total uncertainty                    &   & 10.0 & 17.2 & 17.0 & 27.0\\
\hline\hline
\end{tabular}
\end{center}
\end{table}

\subsection{Statistical and systematic uncertainties}

The statistical uncertainty on $f_L$ is estimated with
 a conservative approach by varying independently the values of the two fitted parameters $\alpha$ and $\gamma$
by $\pm 1 \ \sigma$ in Eq.~(\ref{eq:secondDef_fL}). An extended study based on MC
 pseudo-experiments accounting for the correlations between $\alpha$ and $\gamma$ gave slightly smaller uncertainty.

The uncertainty due to the signal shape in the simultaneous 2-dimensional  fit to $\cos\theta_{\omega}$ and $\cos\theta_{D^*}$ is
measured using the control sample $B^-\ra D^{*0}\pi^-$, with $\Dstarz\ra\Dz\piz$ and $\Dz\ra K^-\pi^+$. This mode was
chosen for its high purity and for its longitudinal fraction $f_L=1$, which enables us to directly compare its shape to our
 signal $f_L=1$. The distribution of the helicity angle of the $\Dstarz$ is found to be wider in the data than in the MC,
 this difference being parameterized by a parabola. The uncertainty on the signal shape is then measured by refitting
 $\alpha$, with the signal PDF being multiplied by the correction parabola. The relative difference is then taken as the uncertainty.

 We assign a systematic effect due to the correction that we apply for the observed small bias on the
 value for $f_L$, when fitting the embedded signal MC simulation (see the discussion at the end of Sect.~\ref{sec:DiscussMeth}).

The uncertainty due to the background shape is measured by refitting $\alpha$ with the background shape fitted in a lower data
sideband $|\Delta E|<280~\mev$ and $5.200<\mes<5.235~\gevcc$. The relative difference is then taken as the uncertainty.

An uncertainty is assigned to $f_L$ due to the assumption of the acceptance being independent of $\chi$. The acceptance of the MC simulation signal
is measured in bins of $\chi$ and fitted with a Fourier series to account for any deviation from flatness. The resulting fitted function is
used as a parametrization of the acceptance dependency to $\chi$ and multiplied to the decay rate (as written in Eq.~(\ref{eq:integChi})). We then
perform in a study based on pseudo-MC simulation experiments where the events are generated from this new decay rate.  The resulting
$\cos\theta_{\omega}\times\cos\theta_{D^*}$ distributions are fitted with the procedure described above. A small bias
is observed and its value is assigned as  a systematic effect.

The uncertainty on the efficiency ratio $\varepsilon_{0}/\varepsilon_{1}$, from the limited amount of MC
statistics available, is calculated assuming $\varepsilon_{0}$ and $\varepsilon_{1}$ to be uncorrelated, while
individual uncertainties on $\varepsilon_{0}$ and $\varepsilon_{1}$ are calculated assuming a binomial distribution.

The various relative uncertainties are displayed in Table~\ref{tab:uncertainties_data} for the data and are found to
be compatible with those calculated in MC simulations. The dominant uncertainty is statistical. Among the various systematic
sources, the largest contribution comes from the signal and background parametrizations, and for some modes on the assumption of a flat acceptance versus $\chi$.

As a check, the $f_L$ measurement is applied in data first on the high purity and high statistics control sample $B^-\ra D^{*0}\pi^-$,
with $D^{*0}\ra D^0\piz$ and $D^0\ra\Km\pip$. This decay channel is longitudinally polarized, {\it i.e.} $f_L=1$. The fit
of $\cos\theta_{D^*}$ in data yields a value of $f_L$ compatible with one, as expected.

\subsection{Results for the fraction of longitudinal polarization $f_L$}

\begin{figure}[htb]
\begin{center}
\includegraphics[width=0.6\linewidth]{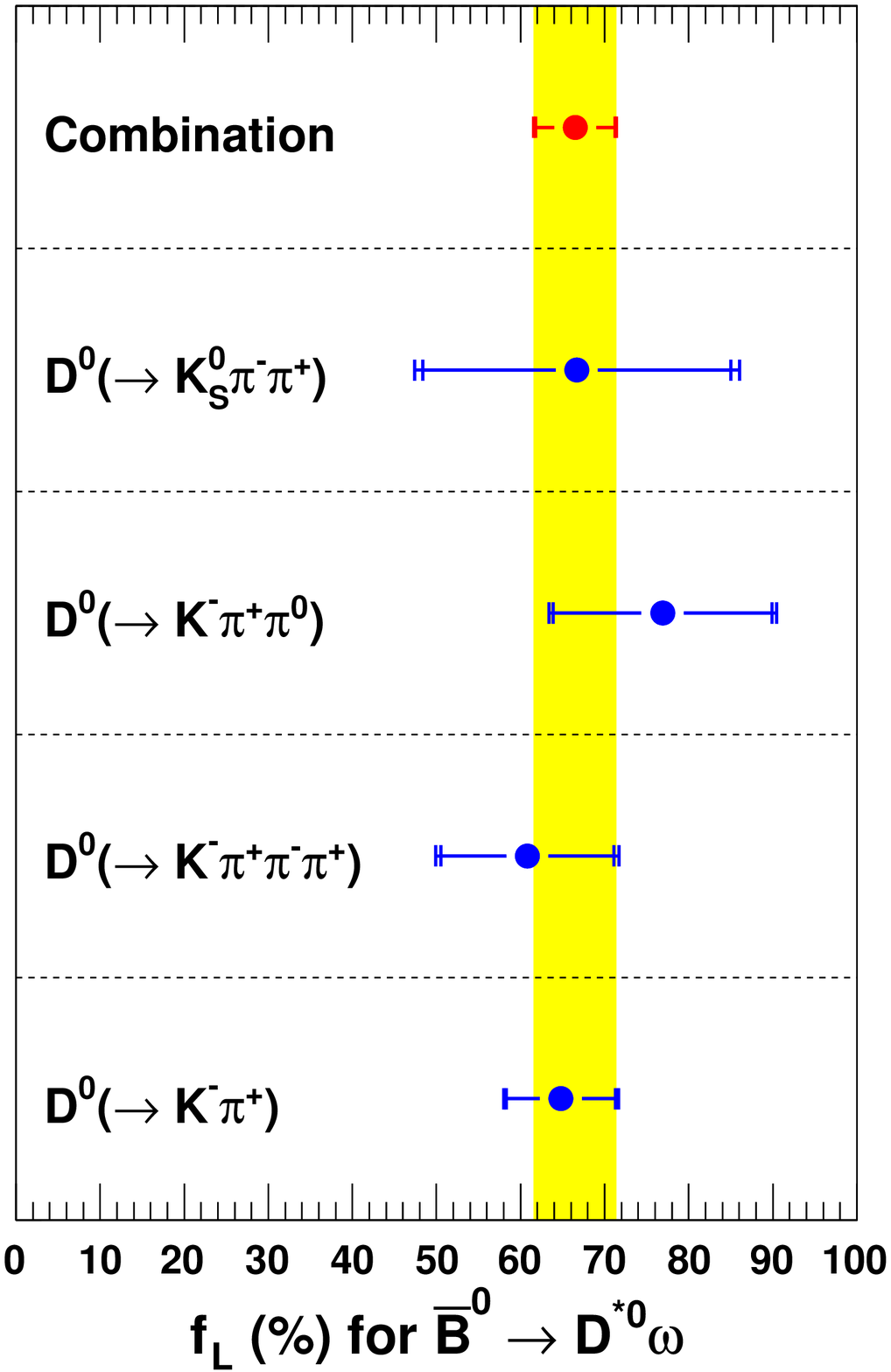}
\caption{Measurements of $f_L$ with the four $\Dz$ modes in data. The yellow band represents the {\it BLUE} combination.}\label{fig:BLUEcosdata}
\end{center}
\end{figure}

The fitted data distributions of the cosine of the helicity angles are given in Fig.~\ref{fig:fit_cos_data}.
The measurements for each $\Dz$ decay channel are then combined
with the {\it BLUE} statistical method~\cite{ref:BLUE} (see Sec.~\ref{se:results}) with
$\chi^2/{\rm ndof}=1.01/3$ ({\it i.e.}: a probability of $79.9\%$).
The measured values of $f_L$, $\alpha$, $\gamma$ and $\varepsilon_{0}/\varepsilon_{1}$ are given with the details of the
combination in Table~\ref{tab:resPol} and in Fig.~\ref{fig:BLUEcosdata}. The final result is $f_L=(66.5\pm 4.7\pm 1.5)\%$,
where the first uncertainty is statistical and the second systematics. This is  the first measurement of the longitudinal fraction of
$\Bzb\ra\Dstarz\omega$, with a relative precision of $7.4\%$.

\begin{table}[htb]
\caption{\label{tab:resPol}Values of $\alpha$ fitted in data, of the background fraction $\gamma$ and of the acceptance ratio
$\varepsilon_{0}/\varepsilon_{1}$, with the corresponding values of the longitudinal fraction $f_L$ after the bias correction.
The first quoted uncertainty is statistical and the second systematic.}
\begin{center}
\begin{tabular}{ l c c c c}
\hline\hline \\
$\Dz$ mode  &  $\alpha$ ($\%$) & $\gamma$ ($\%$) & $\varepsilon_{0}/\varepsilon_{1}$ & $f_L$ ($\%$) \\
\hline	\\
$K\pi$           &  33.4$\pm$2.7 & 52.0$\pm$1.9 & 1.093$\pm$0.012 & 64.8$\pm$\hspace{0.16cm}6.5$\pm$2.1\\
$K3\pi$          &  18.8$\pm$2.3 & 71.2$\pm$2.5 & 1.068$\pm$0.017 & 60.8$\pm$10.3$\pm$3.6\\
$K\pi\piz$       &  19.6$\pm$2.1 & 76.0$\pm$2.3 & 1.109$\pm$0.021 & 76.9$\pm$13.0$\pm$3.8\\
$\KS\pi\pi$      &  24.9$\pm$4.2 & 66.0$\pm$4.9 & 1.092$\pm$0.016 & 66.7$\pm$18.3$\pm$6.2\\
\hline \\
Combi.      & \multicolumn{4}{c}{ $f_L=(66.5\pm 4.7\pm 1.5)\%$}\\
\hline\hline
\end{tabular}
\end{center}
\end{table}

This value differs significantly from the HQET prediction $f_L = (89.5\pm 1.9)\%$~\cite{ref:CLEOhelicity2,ref:HQET}.
This significant transverse amplitude in the $\Bzb\ra\Dstarz\omega$ decay channel may arise from the same
mechanism as the one that is responsible for the transverse polarization observed in $B\ra\phi K^*$. It however supports
the existence of effects from non-trivial long distance contributions to the decay amplitude of $\Bzb\ra\Dstarz\omega$  as predicted  by SCET studies~\cite{ref:SCET_2}.

\section{DISCUSSION}
\subsection{Isospin analysis}

The isospin symmetry relates the amplitudes of the decays
$\Bm\ra D^{(*)0}\pim$, $\Bzb\ra D^{(*)+}\pim$ and $\Bzb\ra
D^{(*)0}\piz$, which can be written as linear combinations of the
isospin eigenstates $\mathcal{A}_{I,D^{(*)}}$, $I=1/2$,
$3/2$~\cite{ref:NeubertPetrov,ref:rosner},

\begin{eqnarray}
&\mathcal{A}(\Dstze\pim) =&
\sqrt{3}\mathcal{A}_{3/2,D^{(*)}},\\
&\mathcal{A}(D^{(*)+}\pim) =& 1/\sqrt{3}\mathcal{A}_{3/2,D^{(*)}}
+
\sqrt{2/3}\mathcal{A}_{1/2,D^{(*)}},\nonumber\\
&\mathcal{A}(D^{(*)0}\piz) =& \sqrt{2/3}\mathcal{A}_{3/2,D^{(*)}}
- \sqrt{1/3}\mathcal{A}_{1/2,D^{(*)}}, \nonumber
\end{eqnarray}
leading to:
\begin{equation}
\mathcal{A}(\Dstze\pim) = \mathcal{A}(D^{(*)+}\pim) +
\sqrt{2}\mathcal{A}(D^{(*)0}\piz).
\end{equation}

The relative strong phase between the amplitudes
$\mathcal{A}_{1/2,D^{(*)}}$ and $\mathcal{A}_{3/2,D^{(*)}}$ is denoted as $\delta$ for the
$D\pi$ system and $\delta^*$ for the $D^*\pi$ system. Final state interactions
between the states $\Dstze\piz$ and $D^{(*)+}\pim$
may lead to a value of $\delta^{(*)}$ different from zero and,
through constructive interference, to a larger value of $\BF$ for
$\Dstze\piz$ than the prediction obtained within the factorization
approximation. One can also define the amplitude ratio $R^{(*)}$,

\begin{equation}
R^{(*)}=\frac{|\mathcal{A}_{1/2,D^{(*)}}|}{\sqrt{2}|\mathcal{A}_{3/2,D^{(*)}}|}.
\end{equation}

In the heavy-quark limit, the factorization model predicts~\cite{ref:BBNS,ref:ChengYang}
$\delta^{(*)}={\cal O}(\Lambda_{\rm QCD}/m_b)$ and $R^{(*)}=1+{\cal O}(\Lambda_{\rm QCD}/m_b)$,
where $m_b$ represents the $b$ quark mass and where the correction to ``1'' is also suppressed by a
power of $1/N_c$, with $N_c$ the number of colors.  SCET~\cite{ref:SCET_1,ref:SCET_2,ref:SCET_3}
predicts that the strong phases $\delta^{(*)}$ (ratios $R^{(*)}$) have the same value in the $D\pi$ and $D^*\pi$
systems and significantly differ from 0 (1).

The strong phase $\delta^{(*)}$ can be computed with an isospin
analysis of the $D^{(*)}\pi$ system. We use the
world average values provided by the PDG~\cite{ref:PDG}
for $\BF(\Bm\ra D^{(*)0}\pim)$, $\BF(\Bzb\ra D^{(*)+}\pim)$  and for
the $B$ lifetime ratio $\tau(\Bp)/\tau(\Bz)$. The values of
$\BF(\Bzb\ra\Dstze\piz)$ are taken from this analysis. We
calculate the values of $\delta^{(*)}$ and $R^{(*)}$ using a
frequentist approach~\cite{ref:ckmfitter}:
\begin{equation}
\delta=(29.0^{+2.1}_{-2.6})^{\circ}, \ \ R=\left( 69.2^{+3.8}_{-3.9} \right) \%,
\end{equation}
for $D\pi$ final states, and
\begin{equation}
\delta^{*}=(29.5^{+3.5}_{-4.5})^{\circ}, \ \ R^*=\left( 67.0^{+4.8}_{-4.7} \right) \%,
\end{equation}
for $D^*\pi$ final states.

In both $D\pi$ and $D^*\pi$ cases, the amplitude ratio is significantly different from the factorization prediction
$R^{(*)}=1$. The strong phases are also significantly different from zero and are equal
in the two systems $D\pi$ and  $D^*\pi$ ($0^{\circ}$ is excluded at $99.998\%$ and $99.750\%$ of confidence level, respectively),
which points out that non-factorizable FSI are indeed not negligible. Those results confirm the SCET predictions.

\subsection{Comparison to theoretical predictions for $\BF(\Bzb\ra\Dstze\hz)$}

Table~\ref{tab:comparaisonTheorie_1} compares the $\BF(\Bzb\ra\Dstze\hz)$ measured with this analysis to the predictions
by factorization~\cite{ref:Chua,ref:Neubert,ref:Deandrea,ref:Deandrea2} and pQCD~\cite{ref:pQCD_1,ref:pQCD_2}.
We confirm the conclusion by the previous $\babar$ analysis~\cite{ref:Babar2004}: the values measured
are higher by a factor of about three to five than the values predicted by factorization. The pQCD predictions are closer to
experimental values but are globally higher, except for the $\Dstze\piz$ modes.

\begin{table*}[htb]
{\footnotesize \caption{\label{tab:comparaisonTheorie_1}
Comparison of the measured branching fractions $\BF$, with the predictions
by factorization \cite{ref:Chua,ref:Neubert,ref:Deandrea,ref:Deandrea2}
and pQCD~\cite{ref:pQCD_1,ref:pQCD_2}. The first quoted uncertainty is
 statistical and the second is systematic.}
\begin{center}
\begin{tabular}{lcccccc}
\hline\hline \\
$\BF(\Bzb\ra)$  $(\times 10^{-4})$ & & This measurement  & & Factorization & & pQCD \\
\hline \\
  $\Dz\piz$  & &  2.69 $\pm$ 0.09 $\pm$ 0.13 &  &0.58~\cite{ref:Chua}; 0.70~\cite{ref:Neubert} & & 2.3-2.6 \\
  $\Dstarz\piz$ & & 3.05 $\pm$ 0.14 $\pm$ 0.28 &  & 0.65~\cite{ref:Chua}; 1.00~\cite{ref:Neubert} & & 2.7-2.9 \\
  $\Dz\eta$ & &  2.53 $\pm$ 0.09 $\pm$ 0.11   &  & 0.34~\cite{ref:Chua}; 0.50~\cite{ref:Neubert} & & 2.4-3.2\\
  $\Dstarz\eta$ & & 2.69 $\pm$ 0.14 $\pm$ 0.23 &  & 0.60~\cite{ref:Neubert} & & 2.8-3.8 \\
  $\Dz\omega$ &  & 2.57 $\pm$ 0.11 $\pm$ 0.14 &  & 0.66~\cite{ref:Chua}; 0.70~\cite{ref:Neubert} & & 5.0-5.6 \\
  $\Dstarz\omega$ & &  4.55 $\pm$ 0.24 $\pm$ 0.39 & &  1.70~\cite{ref:Neubert} & & 4.9-5.8  \\
 $\Dz\etapr$ &  & 1.48 $\pm$ 0.13 $\pm$ 0.07 &  & 0.30-0.32~\cite{ref:Deandrea2}; 1.70-3.30~\cite{ref:Deandrea} & & 1.7-2.6  \\
 $\Dstarz\etapr$ & &  1.48 $\pm$ 0.22 $\pm$ 0.13   & &  0.41-0.47~\cite{ref:Deandrea} & & 2.0-3.2  \\
\hline\hline
\end{tabular}
\end{center}
}
\end{table*}

\begin{figure}
\begin{center}
\includegraphics[width=0.6\linewidth]{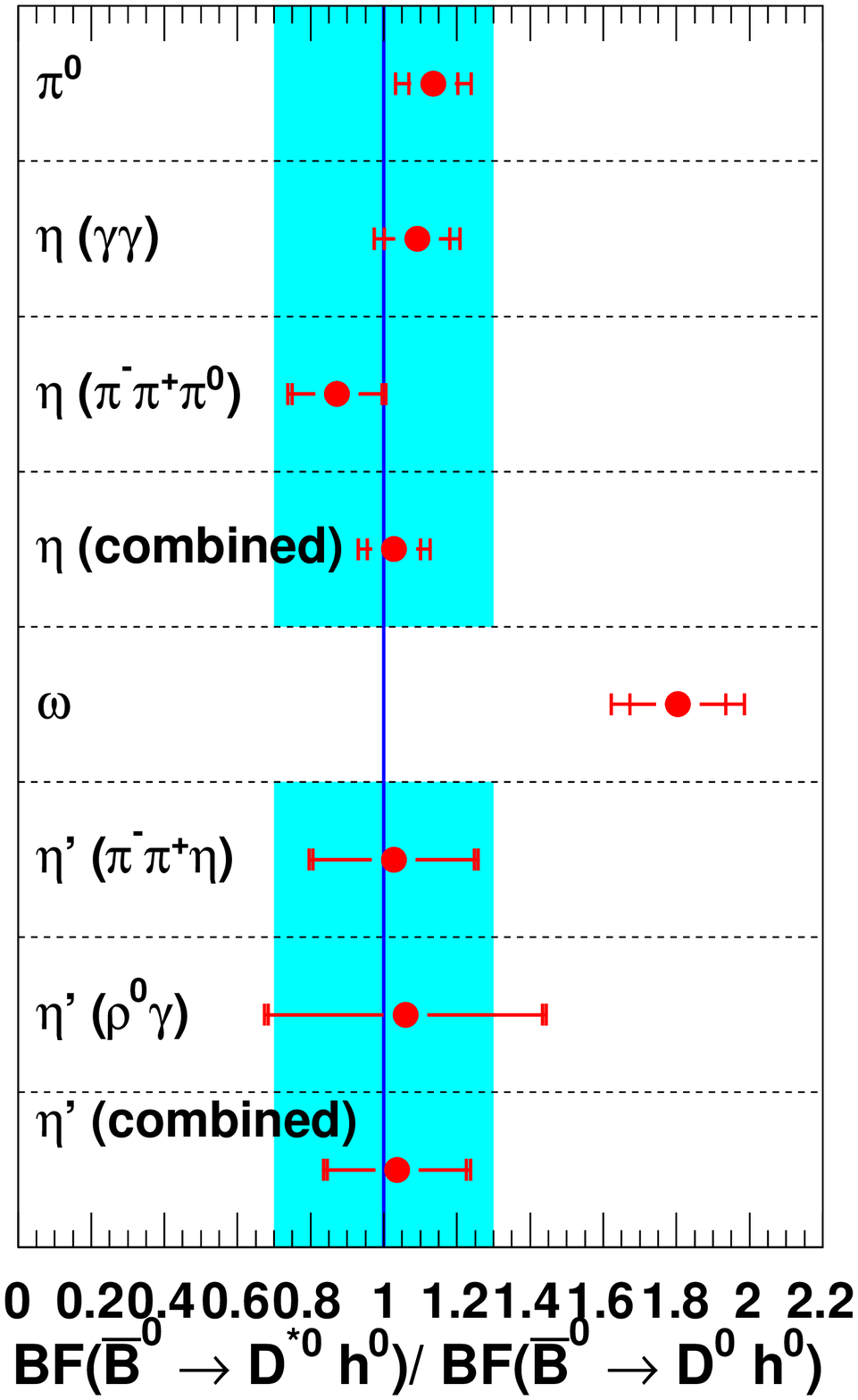}
\caption{\label{fig:BFratio1} Combined ratios $\BF(\Bzb\ra\Dstarz h^0)/\BF(\Bzb\ra\Dz h^0)$
measured in this paper compared to theoretical prediction by SCET~\cite{ref:SCET_2}
(vertical solid line). The vertical band  represent the estimated theoretical uncertainty from SCET (for the case where $h^0$ is
the $\omega$ meson see text).}
\end{center}
\end{figure}

\begin{figure}
\begin{center}
\includegraphics[width=0.6\linewidth]{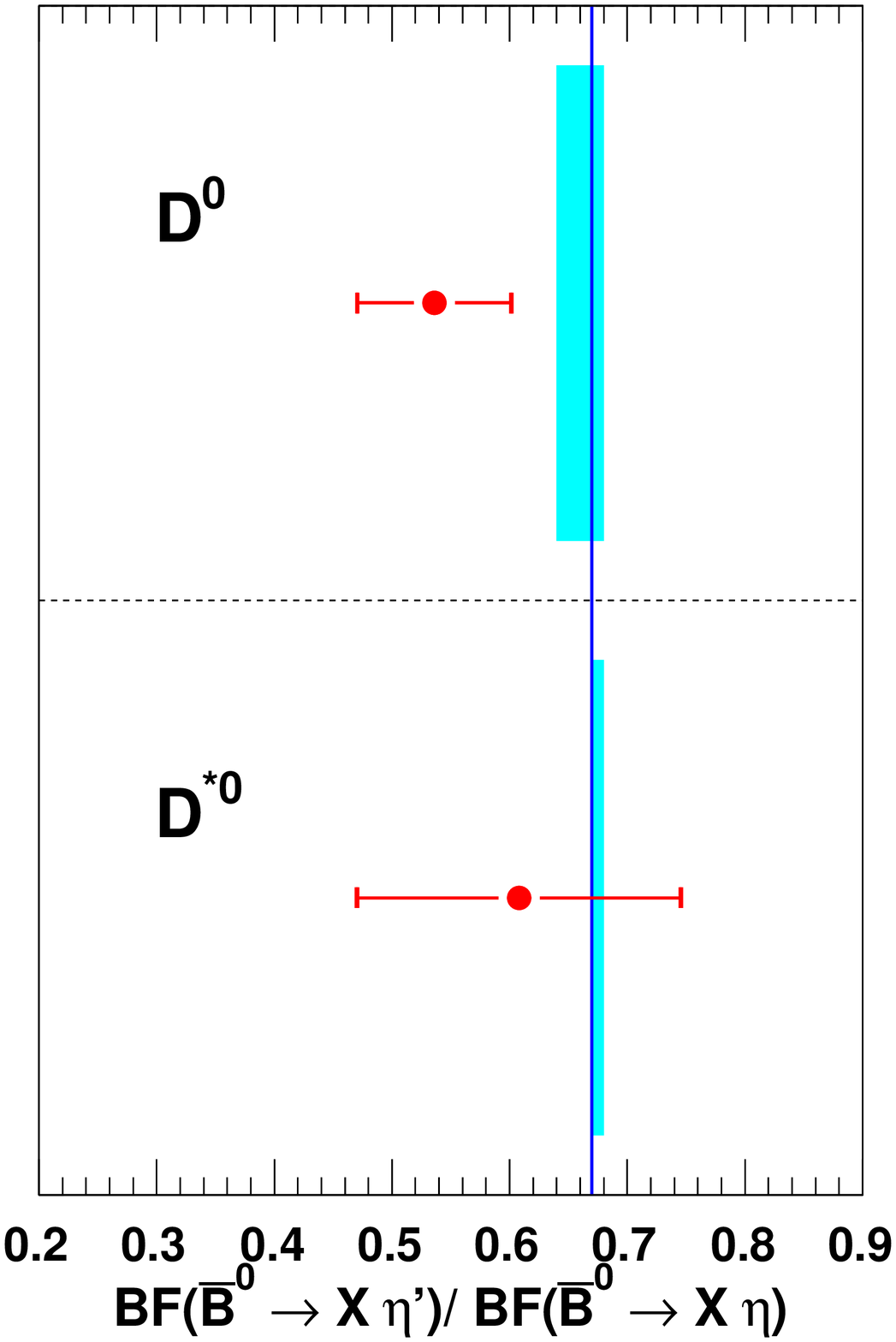}
\caption{\label{fig:BFratio2} Combined ratios $\BF(\Bzb\ra\Dstarz\etapr)/\BF(\Bzb\ra\Dstarz\eta)$
and $\BF(\Bzb\ra\Dz\etapr)/\BF(\Bzb\ra\Dz\eta)$ measured in this paper compared to theoretical
prediction by SCET~\cite{ref:SCET_2} (vertical line) and from factorization~\cite{ref:Deandrea} (vertical bands).}
\end{center}
\end{figure}

The ratios  of the $\BF$ are given in Table~\ref{tab:comparaisonTheorie_2}.
It should be noted that the values of these ratios are not computed directly from those quoted in Table~\ref{tab:BFDataCombi},
as we take advantage of the fact that common systematic uncertainties cancel between $\Dz\hz$ and $\Dstarz\hz$ modes.
Therefore the ratios of the $\BF$ are first calculated for each sub-decays of $\Dz$ and $\hz$, and then
combined with the BLUE method. The ratios $\BF(\Bzb\ra\Dstarz\hz)/\BF(\Bzb\ra\Dz\hz)$ for
$\hz=\piz$, $\eta$, and $\etapr$ are compatible with 1. All are displayed in Fig.~\ref{fig:BFratio1}
together with the theoretical predictions.

Factorization predicts the ratio $\BF(\Bzb\ra\Dstze\etapr)/\BF(\Bzb\ra\Dstze\eta)$
to have a value between 0.64 and 0.68~\cite{ref:Deandrea}, related to the $\eta-\etapr$
mixing. Those ratios are also given in Table~\ref{tab:comparaisonTheorie_2} and Fig.~\ref{fig:BFratio2}
compares the theoretical  predictions with our experimental measurements. The measured ratios
are smaller than the predictions and are compatible at the level of less than two standard deviations.
The SCET gives also a prediction about the ratio
$\BF(\Bzb\ra\Dstze\etapr)/\BF(\Bzb\ra\Dstze\eta)\simeq 0.67$,
which is similar to the prediction by factorization.

SCET~\cite{ref:SCET_1,ref:SCET_2,ref:SCET_3} does not
predict the absolute value of the $\BF$ but it predicts that the ratios
$\BF(\Bzb\ra\Dstarz\hz)/\BF(\Bzb\ra\Dz\hz)$ are about equal to one for $\hz =
\piz$, $\eta$ and $\etapr$. For $\hz=\omega$ that prediction holds
only for the longitudinal component of $\Bzb\ra\Dstarz\omega$, as
non trivial long-distance QCD interactions may increase the transverse
amplitude.  We measure the fraction of longitudinal polarization
to be  $f_L=(66.5\pm 4.7\textrm{(stat.)}\pm 1.5\textrm{(syst.)})\%$  in the
decay mode $\Bzb\ra\Dstarz\omega$, and find that the ratio
$\BF(\Bzb\ra\Dstarz\omega)/\BF(\Bzb\ra\Dz\omega)$
is significantly higher than one, as expected by SCET~\cite{ref:SCET_2}.

\begin{table}[htb]
{\footnotesize \caption{\label{tab:comparaisonTheorie_2}Ratios
of branching fractions $\BF(\Bzb\ra\Dstarz\hz)/$ \\ $\BF(\Bzb\ra\Dz\hz)$ and
$\BF(\Bzb\ra\Dstze\etapr)/\BF(\Bzb\ra\Dstze\eta)$.
The first uncertainty is statistical, the second is systematic.}
\begin{center}
\begin{tabular}{lccc}
\hline\hline \\ $\BF$ ratio & & & This measurement\\ \hline \\
${\Dstarz\piz}/{\Dz\piz}$ &  && 1.14 $\pm$ 0.07 $\pm$ 0.08 \\ \\
${\Dstarz\eta(\gg)}/{\Dz\eta(\gg)}$ & & &  1.09 $\pm$ 0.09 $\pm$ 0.08 \\ \\
${\Dstarz\eta(\pi\pi\piz)}/{\Dz\eta(\pi\pi\piz)}$ & & & 0.87 $\pm$ 0.12 $\pm$ 0.05 \\ \\
${\Dstarz\eta}/{\Dz\eta}$ (Combined) &  & &  1.03 $\pm$ 0.07 $\pm$ 0.07 \\ \\
${\Dstarz\omega}/{\Dz\omega}$ &  & &  1.80 $\pm$ 0.13 $\pm$ 0.13 \\ \\
${\Dstarz\etapr(\pi\pi\eta)}/{\Dz\etapr(\pi\pi\eta)}$ & & &  1.03 $\pm$ 0.22 $\pm$ 0.07 \\ \\
${\Dstarz\etapr(\rho^0\g)}/{\Dz\etapr(\rho^0\g)}$ & & & 1.06 $\pm$ 0.38 $\pm$ 0.09 \\ \\
${\Dstarz\etapr}/{\Dz\etapr}$ (Combined) & & &  1.04 $\pm$ 0.19 $\pm$ 0.07 \\ \\
\hline \\
${\Dz\etapr}/{\Dz\eta}$ &  & &  0.54 $\pm$ 0.07 $\pm$ 0.01 \\ \\
${\Dstarz\etapr}/{\Dstarz\eta}$ & & &  0.61 $\pm$ 0.14 $\pm$ 0.02 \\ \\
\hline\hline
\end{tabular}
\end{center}
}
\end{table} 
\section{CONCLUSIONS}
\label{sec:Conclusions}

We measure the branching fractions of the
color-suppressed decays $\Bzb\ra\Dstze\hz$, where $\hz=\piz$, $\eta$,
$\omega$, and $\etapr$ with 454$\times 10^6\BB$ pairs.
The measurements are mostly in agreement with the previous
results~\cite{ref:Cleo2002,ref:Babar2004,ref:Belle2005,ref:Belle2006,ref:BabarKpipiz}
and are the most precise determinations of the  $\BF(\Bzb\ra\Dstze\hz)$
from a single experiment. They represent significant improvements with respect to the accuracy of the existing
PDG averages~\cite{ref:PDG}.

For the first time we also measure the fraction of longitudinal polarization $f_L$ in the decay mode $\Bzb\ra\Dstarz\omega$
to be  significantly smaller than 1. This reinforces the conclusion drawn from the $\BF$ measurements on the validity of factorisation in color-suppressed
decays and supports expectations from SCET.

We confirm the significant differences from theoretical predictions by factorization
and provide strong constraints on the models of color-suppressed decays.
In particular our results support most of the predictions of SCET on
$\Bzb\ra\Dstze\hz$~\cite{ref:SCET_1,ref:SCET_2,ref:SCET_3}.
\\

\section{ACKNOWLEDGMENTS}
\label{sec:Acknowledgments}


We are grateful for the 
extraordinary contributions of our \pep2\ colleagues in
achieving the excellent luminosity and machine conditions
that have made this work possible.
The success of this project also relies critically on the 
expertise and dedication of the computing organizations that 
support \babar.
The collaborating institutions wish to thank 
SLAC for its support and the kind hospitality extended to them. 
This work is supported by the
US Department of Energy
and National Science Foundation, the
Natural Sciences and Engineering Research Council (Canada),
the Commissariat \`a l'Energie Atomique and
Institut National de Physique Nucl\'eaire et de Physique des Particules
(France), the
Bundesministerium f\"ur Bildung und Forschung and
Deutsche Forschungsgemeinschaft
(Germany), the
Istituto Nazionale di Fisica Nucleare (Italy),
the Foundation for Fundamental Research on Matter (The Netherlands),
the Research Council of Norway, the
Ministry of Education and Science of the Russian Federation, 
Ministerio de Ciencia e Innovaci\'on (Spain), and the
Science and Technology Facilities Council (United Kingdom).
Individuals have received support from 
the Marie-Curie IEF program (European Union), the A. P. Sloan Foundation (USA) 
and the Binational Science Foundation (USA-Israel).


\begin{thebibliography}{99}


\bibitem{ref:ChangChuaSoni}
H-Y. Cheng, C-K. Chua, and A. Soni,  \jprd{71}, 014030 (2005).

\bibitem{ref:exchangeW}
D. Du, \plb{406}, 110 (1997).

\bibitem{ref:Neubert} M. Neubert and B. Stech,
in {\it Heavy Flavours II}, eds. A.J. Buras and M. Lindner (World
Scientific, Singapore, 1998), p. 294.

\bibitem{ref:BauerStechWirbel}
M. Bauer, B. Stech, and M. Wirbel, \zpc{34}, 103 (1987).

\bibitem{ref:NeubertPetrov}
M. Neubert and A.A. Petrov, \plb{519}, 50 (2001).

\bibitem{ref:Deandrea_0}
A. Deandrea, N. Di Bartolomeo, R. Gatto, and G. Nardulli,
\plb{318}, 549 (1993); A. Deandrea \etal, $ibid.$ {\bf 320}, 170
(1994).


\bibitem{ref:Babar2004}
B. Aubert \etal\ ($\babar$ Collaboration), \jprd{69}, 032004 (2004).

\bibitem{ref:HSW}
K. Honscheid, K. R. Schubert, and R. Waldi, \zpc{63}, 117 (1994).


\bibitem{ref:Belle2002}
 K. Abe \etal\ (Belle Collaboration), \jprl{88}, 052002 (2002).

\bibitem{ref:Cleo2002}
T. E. Coan \etal\ (CLEO Collaboration), \jprl{88}, 062001 (2002).


\bibitem{ref:Belle2005}
J. Sch\"umann \etal\ (Belle Collaboration), \jprd{72}, 011103 (2005).

\bibitem{ref:Belle2006}
S. Blyth \etal\ (Belle Collaboration), \jprd{74}, 092002 (2006).

\bibitem{ref:Belle2007}
A. Kuzmin \etal\ (Belle Collaboration), \jprd{76},
012006 (2007).

\bibitem{ref:BabarKpipiz} B. Aubert \etal\ ($\babar$ Collaboration), \jprd{78}, 052005 (2008).

\bibitem{ref:BaBar2010D0pipi} P. del Amo Sanchez  \etal\ ($\babar$ Collaboration), 	PoS ICHEP 2010, 250 (2010), 	
BABAR-CONF-10/004, SLAC-PUB-14203,	arXiv:1007.4464v1 [hep-ex] (2010).

\bibitem{ref:Chua} C-K. Chua, W-S. Hou, and K-C. Yang,
\jprd{65}, 096007 (2002).

\bibitem{ref:LegangerEeg} L.E. Leganger and J.O. Eeg,
\jprd{82}, 074007 (2010).

\bibitem{ref:pQCD_1}
Y.Y Keum, T. Kurimoto, H. Li, C.D~L\"u, and A.I. Sanda, \jprd{69},
094018 (2004).
\bibitem{ref:pQCD_2}
C.D~L\"u, \jprd{68}, 097502 (2003).

\bibitem{ref:SCET_1}
C.W. Bauer, D. Pirjol, and I.W. Stewart, \jprd{65},
054022 (2002).

\bibitem{ref:SCET_3}
S. Mantry, D. Pirjol, and I.W. Stewart, \jprd{68} 114009 (2003).

\bibitem{ref:SCET_2}
A.E. Blechman, S. Mantry, and I.W. Stewart, \plb{608}, 77 (2005).

\bibitem{ref:ChuaHou}  C-K. Chua and W-S. Hou, \jprd{77}, 116001 (2008) ;
R. Fleischer, N. Serra, and N. Tuning, \jprd{83}, 014017 (2011) ; R. Aaij
\etal\ (LHCb Collaboration) arXiv:1106.4435 [hep-ex] (2011), submitted to Phys. Rev. Lett.

\bibitem{ref:GronauFleischer} M. Gronau, \plb{557}, 198 (2003);
 M. Gronau, Y. Grossman, N. Shuhmaher, A. Soffer, and J. Zupan,
 \jprd{69} 113003 (2004); R. Fleischer, \plb{562}, 234 (2003) and \npb{659}, 321 (2003);
B. Aubert \etal\ ($\babar$ Collaboration), \jprl{99}, 081801 (2007).

\bibitem{ref:detector}
B. Aubert \etal\ ($\babar$ Collaboration), \nima{479}, 1 (2002).

\bibitem{ref:PDG}
K. Nakamura \etal (Particle Data Group), J. Phys. G {\bf 37}, 075021 (2010), Mesons section, page 38.

\bibitem{ref:evtgen}
D. Lange, \nima{462}, 152~(2001).

\bibitem{ref:pythia}
T.~Sj\"ostrand, S.~Mrenna, and P.~Skands, \cpc{178}, 852 (2008).

\bibitem{ref:jetset}
T. Sj\"{o}strand, \cpc{82}, 74~(1994).

\bibitem{ref:GEANT4}
 S. Agostinelli \etal\  (Geant4 Collaboration), \nima{506}, 250
(2003).

\bibitem{ref:bukin}
The so called empirically modified Novosibirsk function divides the
fitting region into a peaking region, a low tail region and a high tail
region. For a variable $x$, the modified Novosibirsk function
$f(x) = A_p \times \exp{\left( g(x) \right) }$, where $g(x)$ is defined,
in the peak region $x_1 < x < x_2$, as

\begin{equation}
-\ln2  \times \left( \frac{\ln\left(1+2\tau\sqrt{\tau^2+1}\frac{x-x_p}{\sigma_p\sqrt{2\ln2}}\right)}
{\ln(1+2\tau^2-2\tau\sqrt{\tau^2+1})} \right)^2,  \nonumber
\end{equation}

in the low tail region  $x<x_1$, as

\begin{equation}
\frac{\tau\sqrt{\tau^2+1}(x-x_1)\sqrt{2\ln2}}{\sigma_p(\sqrt{\tau^2+1}-\tau)^2\ln(\sqrt{\tau^2+1}+\tau)}
+ \rho_1\left(\frac{x-x_1}{x_p-x_1}\right)^2 - \ln2, \nonumber
\end{equation}

and in the high tail region $x>x_2$, as

\begin{equation}
-\frac{\tau\sqrt{\tau^2+1}(x-x_2)\sqrt{2\ln2}}{\sigma_p(\sqrt{\tau^2+1}+\tau)^2\ln(\sqrt{\tau^2+1}+\tau)}
+ \rho_2\left(\frac{x-x_2}{x_p-x_2}\right)^2 - \ln2.\nonumber
\end{equation}

The parameters are:
\begin{itemize}
\item $A_p$ is the value at the maximum of the function,

\item $x_p$ is the peak position,

\item $\sigma_p$ is the width of the peak defined as the width at
half-height divided by $2\sqrt{2\ln 2}\simeq2.35$,

\item $\tau$ is an asymmetry parameter.

\item $\rho_{1/2}$ described the width of the respective tails.
\end{itemize}

The positions $x_{1,2}$ are $x_p +  \sigma_p\sqrt{2\ln 2}\left(
\frac{\xi}{\sqrt{\xi^2+1}}\mp 1 \right)$. Left and right tails are attached to the peak with the conditions of continuity and first derivative.


\bibitem{ref:E691}
J.C. Anjos \etal\ (E691 Collaboration), \jprd{48}, 56~(1993).

\bibitem{ref:TMVA}
A. H\"ocker \etal\ (TMVA Group), CERN Report No. CERN-OPEN-2007-007 (2007).

\bibitem{ref:BermanJacob}
S.M. Berman and M. Jacob, \pr{139}, B1023 (1965).

\bibitem{ref:BLUE}
L. Lyons, D. Gibaut, and P. Clifford, \nima{270} 110 (1988);
B. Aubert \etal\ ($\babar$ Collaboration), \jprd{78}, 112003 (2008).

\bibitem{ref:CLEOhelicity2}
S.E. Csorna \etal\ (CLEO Collaboration), \jprd{67}, 112002 (2003).

\bibitem{ref:Williamson}
A.R. Williamson, {\it in the Proceedings of 4th Flavor Physics and CP Violation Conference (FPCP 2006), 
Vancouver, British Columbia, Canada, 9-12 Apr 2006}, eConf C060409 and eprint: arXiv:hep-ph/0605196 (2006).


\bibitem{ref:Koerner}
J.G. Koerner and G.R. Goldstein, \plb{89}, 105 (1979).

\bibitem{ref:Verkerke}
B. Aubert \etal\ ($\babar$ Collaboration), \jprl{92}, 141801 (2004).

\bibitem{ref:BPhiKstar}
P.K. Das and K.C. Yang, \jprd{71}, 094002 (2005). Y.D. Yang;
R. Wang and G.R. Lu, \jprd{72}, 015009 (2005). C.S. Huang;
P. Ko, X.H. Wu, and Y.D. Yang, \jprd{73}, 034026 (2006);
B. Aubert \etal\ ($\babar$ Collaboration), \jprd{78}, 092008 (2008).

\bibitem{ref:Cheng}
H.Y. Cheng, C.K. Chua, and A. Soni, \jprd{71}, 014030 (2005).

\bibitem{ref:Beneke}
M. Beneke, J. Rohrer, and D. Yang, \jprl{96}, 141801 (2006).

\bibitem{ref:Kramer}
G. Kramer and W.F. Palmer, \jprd{45}, 193 (1992).


\bibitem{ref:HQET}
J.L. Rosner, \jprd{42}, 3732 (1990);
M. Neubert, \plb{264}, 455 (1991). G. Kramer, T.
Mannel and W.F. Palmer, Z. Phys. C {\bf 55}, 497 (1992);
J.D. Richman, {\it Heavy Quark and CP Violation, in Probing the Standard Model
of Particle Interactions}, Les Houches Session LXVIII, 1997
(Elsevier, Amsterdam, 1999), eds. R. Gupta, A. Morel, E. De Rafael, and F. David.

\bibitem{ref:rosner}
J.L. Rosner, \jprd{60}, 074029 (1999).

\bibitem{ref:BBNS} M. Beneke, G. Buchalla, M. Neubert, and C.T.
Sachrajda, \npb{591}, 313 (2000).

\bibitem{ref:ChengYang}
H-Y. Cheng and K-C Yang, \jprd{59}, 092004 (1999).

\bibitem{ref:ckmfitter}
J. Charles \etal\ (CKMfitter group), \epjc{41}, 1-131 (2005).

\bibitem{ref:Deandrea} A. Deandrea and A.D. Polosa, \epj{\bf 22}, 677 (2002).

\bibitem{ref:Deandrea2}
J.O. Eeg, A. Hiorth, and A.D. Polosa, \jprd{\bf 65}, 054030 (2002).


\end{thebibliography}
\end{document}